\documentclass[%
 reprint,
 amsmath,amssymb,
 aps,
]{revtex4-2}

\usepackage{graphicx}
\usepackage{dcolumn}
\usepackage{bm}

\newcommand{\hoch}[1]{$\, ^{#1}$}
\newcommand{\be}{\begin{equation}}
	\newcommand{\ee}{\end{equation}}
\newcommand{\bea}{\setlength\arraycolsep{2pt} \begin{eqnarray}}
	\newcommand{\eea}{\end{eqnarray}}

\newcommand{\kgeo}{\kappa_\text{geo}}

\def\fft#1#2{{\frac{#1}{#2}}}

\def\0{{\sst{(0)}}}
\def\1{{\sst{(1)}}}
\def\2{{\sst{(2)}}}
\def\3{{\sst{(3)}}}
\def\4{{\sst{(4)}}}
\def\5{{\sst{(5)}}}
\def\6{{\sst{(6)}}}
\def\7{{\sst{(7)}}}
\def\8{{\sst{(8)}}}
\def\sst#1{{\scriptscriptstyle #1}}

\def\del{{\partial}}

\def\M2{{\bar{\mathcal{M}}}_{(2)}}
\def\Mco2{{\hat{\mathcal{M}}}_{(D-2)}}

\def\Rbar{{ \bar{R} }}

\def\barnab{{\bar{\nabla}}}

\def\ghat{ {\hat{g} }}
\def\Rhat{{ \hat{R} }}

\def\hatnab{{\hat{\nabla}}}

\def\cL{{{\cal L}}}

\def\R{{\mathbb R}}
\def\cK{{\cal{K}}}
\def\Mms{{M_{\text{MS}}}}

\def\ivgcp{{8\pi G_N}}

\def\mms{{m_{\text{MS}}}}
\def\wtot{{w_{\text{tot}}}}

\begin{document}

\preprint{APS/123-QED}

\title{	Novel topological black holes from thermodynamics and deforming horizons}

\author{Jinbo Yang\hoch{1,2,3}}
 \email{yangjinbo@gzhu.edu.cn}
  \affiliation{ \it \hoch{1}Department of Astronomy, School of Physics and Materials Science,\\
 	Guangzhou University, Guangzhou 510006, P.R.China} 
 \altaffiliation{\it \hoch{2}Institute for Theoretical Physics, 
 	Kanazawa University, Kanazawa 920-1192, Japan,\\
 \hoch{3}HKUST Jockey Club Institute for Advanced Study, 
 The Hong Kong University of Science and Technology, Clear Water Bay, Hong Kong, P.R.China}

\date{\today}

\begin{abstract}

Two novel topological black hole exact solutions with unusual shapes of horizons in the simplest holographic axions model, the four-dimensional Einstein-Maxwell-axions theory, are constructed. We draw embedding diagrams in various situations to display unusual shapes of novel black holes. To understand their thermodynamics from the quasi-local aspect, we re-derive the unified first law and the Misner-Sharp mass from the Einstein equations for the spacetime as a warped product $\M2 \times \Mco2$. The Ricci scalar $\Rhat$ of the sub-manifold $\Mco2$ can be a non-constant. We further improve the thermodynamics method based on the unified first law. Such a method simplifies constructing solutions and hints at generalization to higher dimensions. Moreover, we apply the unified first law to discuss black hole thermodynamics.
\end{abstract}

\maketitle

		\section{Introduction}
\renewcommand\theequation{1.\arabic{equation}}
\setcounter{equation}{0}
\renewcommand\theequation{1.\arabic{equation}}

Black hole physics, especially black hole thermodynamics, has brought us deep insights into theoretical physics\cite{Bekenstein:1974ax, Bardeen:1973gs, Hawking:1976de, Maldacena:1997re, Gubser:1998bc, Witten:1998qj, Ryu:2006bv, Wald:1993nt, PhysRevD.50.846}. The widely studied shape of the black hole has a spherical topology, supported by the topological theorem for Einstein gravity\cite{Hawking:1971vc}. According to the theorem, the horizon of a four-dimensional asymptotically flat black hole must be topologically spherical.
Nevertheless, many black objects beyond the spherical horizon in higher dimensional supergravity or string theory have been discovered\cite{Horowitz:1991cd, Vanzo:1997gw, Galloway:1999bp, Emparan:2001wn, Elvang:2007rd, Emparan:2008eg}. One kind of them is the topological black hole in asymptotic Anti-de Sitter (AdS) space. Its horizon shape is not a sphere but rather an Einstein manifold \cite{Mann:1996gj, Birmingham:1998nr, Emparan:1999gf, Birmingham:2007yv}. 
Widely studied topological black holes have planar or hyperbolic horizons \cite{Klemm:1997ea, Morley:2018lwn, Bai:2022obp}.
The hyperbolic black hole can be viewed as a gravitational description of S-brane in string theory \cite{Gutperle:2002ai, Tasinato:2004dy, Lu:2004ye},  
while the planar black hole is widely applied in the context of AdS/CFT duality \cite{Hartnoll:2008vx, Hartnoll:2008kx}. A non-extreme black hole in the AdS background corresponds to a specific boundary phase in finite temperature.
Specifically, the so-called holographic axion model introduces various axions to achieve momentum relaxation, thus implying the finite DC conductivity on the boundary\cite{Andrade:2013gsa, Donos:2014cya, Arias:2017yqj, Jiang:2017imk, Esposito:2017qpj, Baggioli:2018bfa, Esposito:2020wsn, Baggioli:2021xuv, Liu:2022bdu}. In such a model, the planar axionic black hole contains an axionic charge appearing in the first law of black hole thermodynamics \cite{Andrade:2013gsa, Bardoux:2012aw}. The first law satisfies the Gibbs-Duhem relationship hence it has Euler homogeneity.

The Gibbs-Duhem relationship satisfying the Euler homogeneity leads to some insights into the thermodynamics of topological black holes.
Y. Tian, etc firstly suggested to introduce the topological charge for non-planar Reissner-Nordstr\"om(RN)-AdS black hole \cite{Tian:2014goa} which is detailed discussed in Ref\cite{Tian:2018hlw}. 
This topological charge has a similar scaling behavior to the axionic charge such that it preserves the Euler homogeneity. In recent years, Z. Gao, etc have emphasized the importance of Euler homogeneity for understanding the black hole thermodynamics in a unified way with the usual thermodynamics\cite{Zeyuan:2021uol, Wang:2021cmz, Gao:2021xtt, Zhao:2022dgc, Kong:2022gwu, Wang:2022err}. They proposed the restricted phase space formalism and suggested introducing a ``center charge" to the first law of black hole thermodynamics. Such a new thermodynamics quantity indicates degrees of freedom in some sense. It is similar to the color charge introduced by M. Visser in the context of extended phase space formalism\cite{Visser:2021eqk}. These three distinct approaches, topological charge, color charge, and center charge, attach the same issue about adding new quantities to the first law of thermodynamics for topological black holes.

On the other hand, Einstein's gravity has a quasi-local mass called Misner-Sharp (MS) mass\cite{PhysRev.136.B571, Hayward:1994bu} for spherically symmetric spacetime due to the Kodama vector\cite{10.1143/PTP.63.1217}.
It reduces to the Arnowitt-Deser-Misner (ADM) mass when going to the space-like infinity and the Bondi mass when going to the null infinity in the asymptotically flat background. Due to the quasi-local nature, the MS mass is widely applied in the context of primordial black hole formation\cite{Nakao:2018knn, Hutsi:2021nvs, Harada:2021xze, Yoo:2021fxs, Sato:2022yto, Escriva:2022pnz}, detailed study for Hawking evaporation\cite{Ren:2007xw, Ke-Xia:2009tzo, DiCriscienzo:2009kun,  diCriscienzo:2010cp, Vanzo:2011wq} and $P-V$ transition in cosmological background\cite{Kong:2021dqd, Abdusattar:2022bpg}. Moreover, the MS mass is significant for formulating the unified first law which unified the black hole thermodynamics and relativistic hydrodynamics\cite{Hayward:1997jp}. Refs.\cite{Cai:2005ra, Cai:2006rs} further generalized the unified first law to discuss the thermodynamics of the apparent horizon (Hubble horizon) in an expanding universe and not limited to Einstein's gravity. These works also inspire a novel approach to generate spherical black hole solutions of general relativity from thermodynamics\cite{Zhang:2013tca}, and later developed in Refs.\cite{Maeda:2007uu, Zhang:2014goa, Hu:2015xva, Tan:2016wkj, Kinoshita:2021qsv}, including planar and hyperbolic cases in or not in the context of modified gravity.

This article explores the possibility of replacing the spherical part with an unusual shape, not limited to maximally symmetric space or Einstein manifold. 
Suppose the part replacing the sphere is an independent manifold with metric $\hat{g}_{ij}(x)$, where $x$ donate the point in the independent manifold, and $i,j,k$ are the corresponding indexes. If the manifold is maximally symmetric, the Rienman tensor from $\hat{g}_{ij}(x)$ is $\Rhat_{ijkl}=k(\ghat_{ik}\ghat_{jl} - \ghat_{il}\ghat_{jk})$. An Einstein manifold satisfies a weaker condition $\Rhat_{ij}(x)= \lambda \ghat_{ij}(x)$, where $\lambda$ is a constant\cite{Birmingham:1998nr}. Its Ricci scalar $\Rhat$ is naturally a constant. 
Although cases about non-constant curvature are discussed in the context of modified gravity \cite{Dotti:2005rc, Maeda:2010bu, Ray:2015ava, Ohashi:2015xaa}, it is long believed that general relativity demands an Einstein manifold. 
We will show that the simplest holographic axion model contains black hole solutions with unusual shapes of horizons. The spacetime is still a warped product, but the transverse space is not an Einstein manifold but its $\Rhat$ can depend on direction $x$.

In addition, we will generalize the unified first law to these non-constant $\Rhat$ cases. This implies an efficient method for constructing exact solutions inspired by Refs.\cite{Zhang:2013tca, Zhang:2014goa, Zhang:2014ala, Hu:2015xva, Hu:2016hpm, Tan:2016wkj}. We call it the thermodynamics method and use it to justify the ansatzes for obtaining novel solutions. The method simply induces the constraint equation for $\hat{g}_{ij}(x)$. Moreover, the unified first law provides a quasi-local viewpoint to understand the first law of black hole thermodynamics even without precise definitions of global parameters. It is beneficial to deal with those black hole solutions.

The article is organized as follows. In section 2, we will introduce the action of the simplest holographic axion model in $D=4$ and give two novel charged topological black hole solutions. The crucial feature is that the intrinsic metric of the transverse space can have a non-constant Ricci scalar, different from topological RN black holes without axion or planar axionic black holes. 
We will draw embedding diagrams to visualize the shapes of horizons for various situations. 
In section 3, we will re-derive the unified first law of general relativity from Einstein's equation and give an improved thermodynamics method proposed in Ref.\cite{Zhang:2013tca} originally. Such a method simplifies solving Einstein's equation, hence hints at generalizing the novel solutions. We will also apply the quasi-local viewpoint offered by the unified first law to discuss the first law of thermodynamics for these deformed topological black holes. Section 4 will give a conclusive summary.
\section{Action and solutions}
\setcounter{equation}{0}
\renewcommand\theequation{2.\arabic{equation}}
\setcounter{equation}{0}
\renewcommand\theequation{2.\arabic{equation}}

We consider the following action
\be
\begin{split}
	S= \int d^{4}x\,&\sqrt{-g}\,\big(\frac{1}{16\pi G_N}(R-2\Lambda ) \\&-\fft{1}{16\pi}F^{\mu\nu} F_{\mu\nu} 
	- \fft{1}{2}g^{\mu\nu}\partial_{\mu}\psi^I \partial_{\nu}\psi^I  \big) \,,\label{holographicAxions4dim}
\end{split}
\ee
where $F_{\mu\nu}=\partial_{\mu}A_{\nu}-\partial_{\nu}A_{\mu}$ is the strength of the $U(1)$ gauge field $A_{\mu}$ and $\psi^I$ with $I=1,2$ are two massless scalars.
The equations of motion are
\begin{align}
	& R_{\mu\nu} - \frac{1}{2}Rg_{\mu\nu} +\Lambda g_{\mu\nu}= 8\pi G_N \big( T^{(em)}_{\quad\mu\nu} +T^{(\psi)}_{\quad\mu\nu} \big) 
	\,,   	\label{EinsteinEOMs}
	\\  & \nabla_{\mu}F^{\mu\nu} =\frac{1}{\sqrt{-g}}\frac{\partial }{\partial x^{\mu}} \big(\sqrt{-g}\,
	F^{\mu\nu} \big) =0  \,, \label{MaxwellEOMs}
	\\&  \nabla^{\mu} \nabla_{\mu} \psi^I =\frac{1}{\sqrt{-g}}\frac{\partial }{\partial x^{\mu}} \big(\sqrt{-g}\,g^{\mu\nu}\,\frac{\partial \psi^I}{\partial x^{\nu}} \big) =0 \,,
	\label{KGEOMs}
\end{align}
where 
\begin{align}
	&  T^{(em)}_{\quad\mu\nu}= \frac{1}{4\pi} ( F_{\mu\alpha}F_{\nu}^{\;\,\alpha} -\frac{1}{4} F_{\alpha\beta}F^{\alpha\beta}\,g_{\mu\nu} )
	\,,    \label{emTmunus}
	\\  & T^{(\psi)}_{\quad\mu\nu}= \partial_{\mu}\psi^I \partial_{\nu}\psi^I - \frac{1}{2}g_{\mu\nu} \,g^{\lambda\sigma} \partial_{\lambda}\psi^I \partial_{\sigma}\psi^I  
	\,.
	\label{axionsTmunus}
\end{align}
This theory is the same as the 4-dimensional case of the holographic model proposed by Ref.\cite{Andrade:2013gsa} which aims to achieve momentum relaxation. Their model considered $d+1$-dimensional spacetime and introduced $d-1$ massless scalars. The action thus has global shift symmetries, i.e., invariant under transformation $\psi^I\rightarrow\psi^I+c^I$. Hence scalars $\psi^I$ are usually viewed as axions. Later extended studies are called holographic axion models which introduce axions to deduce the momentum relaxation, then obtain a finite DC conductivity for the strong coupling theory living on the boundary of AdS spacetime Refs.\cite{Andrade:2013gsa}.  Distinguishing with the interest of beyond standard model, holographic axion models usually do not introduce the coupling of axions $\psi^I$ and topological term $F\wedge F$. Relevant investigations are well summarized in Ref.\cite{Baggioli:2021xuv}. 

\subsection{Solution I}
The next task is to solve those equations of motion. We take the ansatz as
\be
\begin{split}
	&ds^2= - f(r) dt^2 + \frac{dr^2}{ f(r) } 
	+ r^2(\fft{d\rho^2}{1-k \rho^2-e(\rho)}+\rho^2d\varphi^2) 
	\,,   
	\\ & A_{\mu}dx^{\mu} =-\phi(r) dt  \,,\quad
	\psi^1(\rho) = ~ \alpha\, p(\rho)  \,, \quad
	\psi^2(\varphi)= \alpha  \,\varphi \,,
	\label{usualansatz}
\end{split}
\ee
The Klein-Gorden(KG) equation $\nabla_\mu\nabla^\mu\psi^2=0$ is solved by $\psi^2(\varphi)= \alpha  \,\varphi$. While the Einstein equation gives
\begin{align}
	&k -  f -r\frac{df}{dr} -\Lambda r^2 - G_N (\frac{d\phi}{dr})^2   \nonumber\\&\qquad = - \frac{1}{2\rho}\frac{de}{d\rho} +\frac{4\pi G_N\alpha^2}{\rho^2}\big( 1+ \rho^2(1- k\rho^2-e)(\frac{dp}{d\rho})^2 \big)  
	\,,   \label{traceGab}
	\\ & \frac{1}{2} \frac{d^2f}{dr^2}+\frac{1}{r}\frac{df}{dr} +\Lambda=  G_N (\frac{d\phi}{dr})^2  \label{parrtraceGab} \,,\quad\quad 
	\\&  \rho^2 (1 - k \rho^2 - e) (\frac{dp}{d\rho})^2 -1=0 \label{shapeofgij}\,.
\end{align}
The last equation leads to $dp/d\rho=\pm 1/(\rho\sqrt{1 - k \rho^2 - e})$. It solved $\nabla_\mu\nabla^\mu\psi^1=0$ since the ansatz $\psi^1(\rho) = ~ \alpha\, p(\rho)$ implies
\be
\begin{split}
	\frac{d}{d\rho}\big( \rho (1- k\rho^2-e)\frac{dp}{d\rho}  \big)  +\frac{1}{2}\rho(2k\rho+\frac{de}{d\rho})\frac{dp}{d\rho} =0 \,.
	\label{diffeqsFromAxions}
\end{split}
\ee
Furthermore, the Maxwell equations lead to  
\be
\begin{split}
	\frac{d^2\phi}{dr^2} +\frac{2}{r}\frac{d\phi}{dr} =0 \,,
	\label{diffeqsFromMaxwell}
\end{split}
\ee
such that determinde the electric potential as $\phi = (4\pi Q/\Omega)\,r^{-1} $ up to some freedom for gauge fixing. We have used $\Omega$ to donate the size of the unit surface $\mathcal{S}$ in the transverse space we are interested in. It is an area integral
\be
\Omega = \int_{\mathcal{S}} \frac{\rho}{\sqrt{ 1 - k \rho^2 - e} } d\rho d\varphi \,.
\ee
The $U(1)$ charge inside the region surrounded by such a surface is
\be
Q= \frac{1}{4\pi}\int_{\mathcal{S}} *F =  \int_{\mathcal{S}}(-\frac{d\phi}{dr})\, r^2 \frac{\rho}{\sqrt{ 1 - k \rho^2 - e} } d\rho d\varphi\,.
\ee 
Moreover, the solution $\phi = (4\pi Q/\Omega)\,r^{-1} $ also implies that the derivative of Eq.\eqref{traceGab} with respect to $r$ gives Eq.\eqref{parrtraceGab}.
Noting that the left-hand-side (LHS) of Eq.\eqref{traceGab} only relates to $r$ while the right-hand-side (RHS) of Eq.\eqref{traceGab} only depends on $\rho$, Eq.\eqref{traceGab} should equal to a constant $\eta$. Namely,
\begin{align}
	\eta &=k -  f -r\frac{df}{dr} -\Lambda r^2 -(\frac{4\pi}{\Omega})^2\frac{ G_N Q^2}{r^2} \,, \label{diffeq1FromEin1st}  \\
	\eta &= - \frac{1}{2\rho}\frac{de}{d\rho} +\frac{8\pi G_N\alpha^2}{\rho^2} 
	\,,   \label{diffeq2FromEin1st}
\end{align}
in which the solution $dp/d\rho=\pm 1/(\rho\sqrt{1 - k \rho^2 - e})$ is sustituted. One can introduce $c=k-\eta$ to solve Eq.\eqref{diffeq1FromEin1st}. As for Eq.\eqref{diffeq2FromEin1st}, this equation is solved by $ e(\rho)= -\eta\rho^2 + 16\pi G_N \alpha^2 \log(\beta \rho) $. Thus $\eta$ contributes a $-\eta\rho^2$ term to $e(\rho)$. It hence shifts the $-k\rho^2$ as $-c\rho^2$ in $p(\rho)$ and the metric. 
In summary, the full solution is given by
\be
\begin{split}
	&ds^2= -( c-\frac{8\pi G_NM}{\Omega\,r}  + (\frac{4\pi}{\Omega})^2\frac{ G_NQ^2}{r^{2}}- \fft{\Lambda r^2}{3}    )dt^2 
	\\ &\;\;\quad\quad + ( c-\frac{8\pi G_NM}{\Omega\,r}  + (\frac{4\pi}{\Omega})^2\frac{ G_NQ^2}{r^{2}}- \fft{\Lambda r^2}{3}    )^{-1}dr^2 
	\\ &\;\;\quad\quad+  r^2(\fft{d\rho^2}{1-c \rho^2 - 16\pi G_N \alpha^2 \log(\beta \rho)}+\rho^2d\varphi^2)   \,,   
	\\ &A_{\mu} dx^{\mu} = -\phi(r) dt =- \frac{4\pi Q}{\Omega r} dt  \,,
	\\ &\psi^1 = \alpha \int^{\rho} \fft{ d\tilde{\rho} }{\tilde{\rho} \sqrt{1-c \tilde{\rho}^2- 16\pi G_N\, \alpha^2 \log(\beta \tilde{\rho}) }}  \,, 
	\\ &\psi^2 = \alpha\, \varphi\,.\label{Solu1}
\end{split}
\ee

\subsubsection{Compare with topological RN-(A)dS black holes}
The solution \eqref{Solu1} reduces to three kinds of topological RN black holes up to some scaling of $\rho$ by setting $\alpha=0$ and identifying $\varphi$ with $\varphi+2\pi$. This is because $c=k$ in these cases. Then ($c>0$: genus $g=0$) ($c=0$: genus $g=1$; $c<0$: genus $g>1$), see Refs.\cite{Vanzo:1997gw, Brill:1997mf})   
Meanwhile, such a period condition for $\varphi$ forces the field space to have a cylinder topology though the field space metric $\delta_{IJ}$ is flat.

On the other hand, the $t$-$r$ part of the metric in Eq.\eqref{Solu1} is similar to the $t$-$r$ metric for topological RN solutions even though $\alpha\neq 0$. Hence the parameters space for the metric should include the naked singularity and the extreme black hole. We do not discuss these cases in the following but instead, we are interested in non-extreme black holes. For situations of $\Lambda\le 0$, we will work in the parameter regions of the equation $f(r)=0$ containing two different positive roots. The larger one indicates the black hole horizon location while the smaller one corresponds to the inner Cauchy horizon. As for $\Lambda>0$, the equation $f(r)=0$ may have three positive roots. The largest one should be the cosmic horizon rather than the black hole horizon.

\begin{figure}[h]
	\begin{center}
		\includegraphics[width=8.1cm]{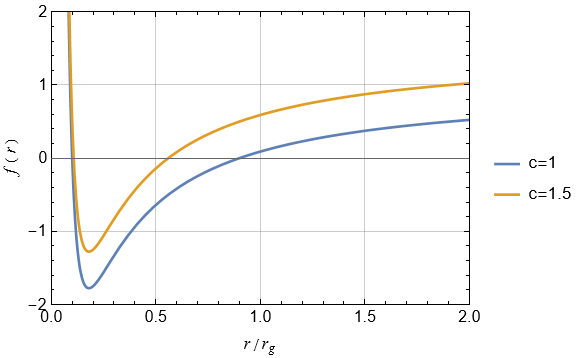} 
	\end{center}
	\caption{\small We set $r_g=8\pi G_NM/\Omega$ and  $ G_N (4\pi Q/\Omega)^2= 0.09\, r_g^2$ to plot $f(r)= c-(8\pi G_NM/\Omega)/r  +G_N (4\pi Q/\Omega)^2/r^2$. The blue curve represents $f(r)$ when $c=1$. It has roots $r=0.9\, r_g$ indicating the black hole horizon location, and $r=0.1\, r_g$ corresponding to the inner Cauthy horizon; While the yellow curve is for the $c=1.5$ case, which has roots  $r=0.5594\,r_g$ (black hole horizon) and $r=0.1073\, r_g$ (inner horizon). } 
	\label{Lam0BH}  
\end{figure}

\begin{figure}[h]
	\begin{center}
		\includegraphics[width=8.1cm]{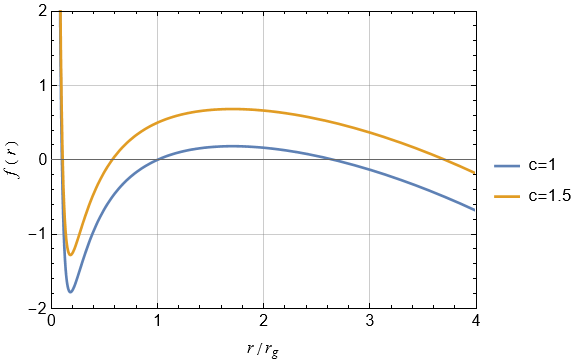}
	\end{center}
	\caption{\small 
		We set $ G_N (4\pi Q/\Omega)^2= 0.09\, r_g^2$ and $\Lambda=0.09\, r_g^{-2}$ to plot $f(r)= c-(8\pi G_NM/\Omega)/r  +G_N (4\pi Q/\Omega)^2/r^2 - \Lambda r^2/3 $ in which $r_g=8\pi G_NM/\Omega$ is the unit of $r$.
		The blue curve is for the $c=1$ case. The largest root of $f(r)=0$ is $r=2.660\, r_g$ which represents the cosmic horizon. While other two roots are $r=1.000\, r_g$ (black hole horizon) and $r=0.1000\, r_g$ (inner horizon); 
		As for the yellow curve, the case of $c=1.5$, roots are $r=3.707\, r_g$ (cosmic horizon), $r=0.5733\, r_g$ (black hole horizon) and $r=0.1072\, r_g$ (inner horizon). }\label{LamPlusBH}
	\begin{center}
		\includegraphics[width=8.1cm]{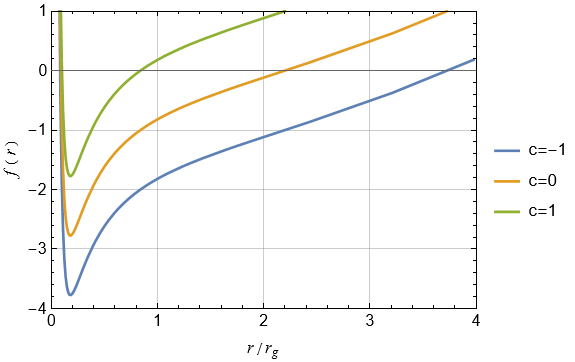}
	\end{center}
	\caption{\small 
		We set $ G_N (4\pi Q/\Omega)^2= 0.09\, r_g^2$ and $\Lambda=-0.09\, r_g^{-2}$ to plot $f(r)= c-(8\pi G_NM/\Omega)/r  +G_N (4\pi Q/\Omega)^2/r^2 - \Lambda r^2/3 $ in which $r_g=8\pi G_NM/\Omega$.
		The blue curve is for $c=-1$. Roots of $f(r)=0$ are $r=3.743\, r_g$ (black hole horizon) and $r=0.08310\, r_g$ (inner horizon); 
		The yellow curve is for $c=0$ with roots $r=2.201\, r_g$ (black hole horizon) and $r=0.09001\, r_g$(inner horizon);
		The green curve is for $c=1$ with roots $r_H=0.8395\, r_g$ (black hole horizon) and $r=0.1000\, r_g$ (inner horizon).}\label{LamMinusBH}
\end{figure}

We plot typical cases for non-extreme black holes in Fig.(\ref{Lam0BH}), (\ref{LamPlusBH}), (\ref{LamMinusBH}). For simplicity, we set $r_g= 8\pi G_N M/\Omega$ as the unit for the $r$ coordinate, and choose the value of $Q$ by requiring $ G_N (4\pi Q/\Omega)^2=0.09\, r_g^2$. The value of $\Lambda$ is taken as $0$ in Fig.(\ref{Lam0BH}); $0.09\, r_g^{-2}$ in Fig.(\ref{LamPlusBH}) and  $-0.09 \, r_g^{-2}$ in Fig.(\ref{LamMinusBH}). It is worth noting that the black hole horizon $r=r_H$ should satisfy $f(r_H)=0$ and $f'(r_H)>0$. While other roots $r_C$ of $f(r)=0$ with negative $f'(r_C)$ should be an inner Cauthy horizon or a cosmic horizon. When $\Lambda\geq 0$ and $c\leq 0$, only the horizon with negative $f'$ exists.  Nevertheless, cases of $\Lambda<0$ and $c\leq 0$ always include a black hole horizon (see Fig.(\ref{LamMinusBH})).

\subsubsection{Shapes of horizons}
Then we will study the geometry of the transverse space which is labeled as $\hat{\cal{M}}_{2}$. Its independent line-element $d\hat{s}^2=\ghat_{ij}dx^idx^j$ seems described by three parameters $c$, $\alpha$ and $\beta$, but one of them can be set to one via a suitable rescaling. We choose $\tilde{\rho}=\beta\rho$, $\tilde{r}=r/\beta$, $\tilde{t}= \beta t$, $\tilde{c}=c/\beta^2$, 
$\tilde{M}=M/\beta^3$, and $\tilde{Q}=Q/\beta^2$, then obtain the line-element for the whole spacetime
\be
\begin{split}
	&ds^2= -(\tilde{c}-\frac{8\pi G_N \tilde{M} }{ \Omega\, \tilde{r} }  + (\frac{4\pi}{\Omega})^2 \frac{G_N \tilde{Q}^2}{\tilde{r}^{2}}- \fft{\Lambda  \tilde{r}^2}{3} )d\tilde{t}^2
	\\&\;\;\quad\quad +(\tilde{c}-\frac{8\pi G_N \tilde{M} }{ \Omega\, \tilde{r} }  + (\frac{4\pi}{\Omega})^2 \frac{G_N \tilde{Q}^2}{\tilde{r}^{2}}- \fft{\Lambda  \tilde{r}^2}{3} )^{-1}d\tilde{r}^2
	\\&\;\;\quad\quad+  \tilde{r}^2(\fft{d\tilde{\rho}^2}{1-c \tilde{\rho}^2 - 16\pi G_N \alpha^2 \log( \tilde{\rho})}+ \tilde{\rho}^2d\varphi^2)    \,.
	\label{rescaleMetric1}
\end{split}
\ee
Particularly, omit the tilde sign, we have the line element of $\hat{\cal{M}}_{2}$:
\be
d\hat{s}^2  =\fft{d\rho^2}{1-c \rho^2-16\pi G_N \alpha^2 \log\rho}+\rho^2d\varphi^2  \,.\label{rescaleddshatsolI}
\ee
Therefore, the Ricci scalar of $\hat{\cal{M}}_{2}$ is
\be
\Rhat(\rho)= 2c+ \fft{16\pi G_N \alpha^2}{\rho^2}  \,, \label{sol1hatRicci}
\ee
which depends on the value of $\rho$ rather than a constant. The intrinstic geometry of $\hat{\cal{M}}_{2}$ is controled by two parameters $c$ and $\alpha$.
$\alpha$ contributes the $\rho$-dependence and probabelly leads to a singularity $\rho=0$. It would be interesting to study the consequence of the existence of a singular direction for quantum gravity, though the entropy of horizon with arbitrary shape is studied in a quantum gravity context\cite{Song:2020arr}.

\begin{figure}[h]
	\begin{center}
		\includegraphics[width=6.3cm]{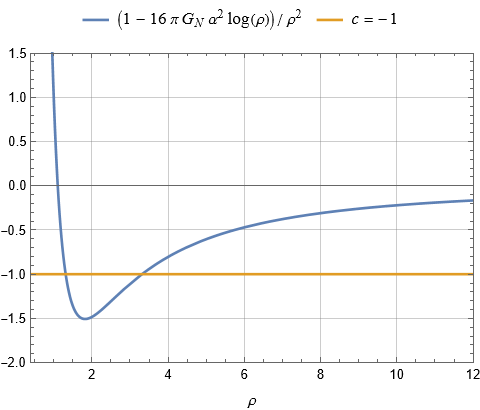}
	\end{center}
	\caption{\small The blue curve describes the function $(1-16\pi G_N \alpha^2 \log\rho )/\rho^2$ when $16\pi G_N \alpha^2$ is taken as 10. It has a minimum value $1$ at $\rho=\,$ 
		The yellow line $c=-1$ intersects the function at $\rho=1.313$ and $\rho=3.314$. Hence we obtain a small branch $0<\rho<1.313$, and a large branch $\rho>3.314$ to ensure $c<(1-16\pi G_N \alpha^2 \log\rho )/\rho^2$, namely, $\ghat_{\rho\rho}>0$ .}  
	\label{loffactor10}
\end{figure}

On the other hand, $\ghat_{\rho\rho}$ should be positive to ensure the correct signature of the metric, such that 
\be
c< \frac{1-16\pi G_N \alpha^2 \log\rho  }{\rho^2}  \,. \label{corsign}
\ee  
The function $(1-16\pi G_N \alpha^2 \log\rho )/\rho^2$ has a minimum at $\rho=e^{1/2 + 1/(16\pi G_N \alpha^2)}$. Thus $c<-8\pi G_N \alpha^2e^{-1-1/(8\pi G_N \alpha^2)}$ never intersects with $(1-16\pi G_N \alpha^2 \log\rho )/\rho^2$.
We take $16\pi G_N \alpha^2= 10$ to plot the function $(1-16\pi G_N \alpha^2 \log\rho )/\rho^2$ in Fig.\ref{loffactor10}.
It shows that a positive $c$ intersects with $(1-16\pi G_N \alpha^2 \log\rho )/\rho^2$ (blue) only once. While a  negative $c$ but larger than $-1.5059$ intersects twice. For instance, the case of $c=-1$ (yellow) has two branches satisfying the inequality \eqref{corsign}. One is $0<\rho<1.313$ called the small branch; while another is the large branch $\rho>3.314$. 
It is worth noting that $\rho=1.313$ and $\rho=3.314$ are roots of $1+ \rho^2-10\log\rho=0$ which make $\ghat_{\rho\rho}$ blowing up. Generally, roots of $1-c \rho^2-16\pi G_N \alpha^2 \log\rho=0$ serve as coordinate singularities which can be removed by choosing the new coordinate $dl=\sqrt{\ghat_{\rho\rho}(\rho)}d\rho$. 
The line-element under the new coordinates $\{l,\varphi\}$ is 
\be
d\hat{s}^2 =dl^2 + \rho^2(l)d\varphi^2  \,,\label{rmcosing}
\ee 
Thus, roots of $1/\ghat_{\rho\rho}$ are somewhere satisfied $d\rho/dl=0$, indicating the minimum or maximum of $\rho$. The integral $l=\int\sqrt{\ghat_{\rho\rho}}d\rho$ will introduce an integral constant. No loss of generality, we require $l=0$ when $\rho$ takes minimum or maximum to determine such a constant.

Then we will draw embed diagrams for several situations to visualize the $\hat{\cal{M}}_{2}$ geometry.
First, we rewrite the line-element \eqref{rescaleddshatsolI} as
\be
d\hat{s}^2 =\fft{c \rho^2+16\pi G_N \alpha^2 \log\rho}{1-c \rho^2-16\pi G_N \alpha^2 \log\rho}d\rho^2+d\rho^2+\rho^2d\varphi^2  \,,\label{preembedsolI}
\ee
which hints at how to embed $\hat{\cal{M}}_{2}$ into a three dimensional flat space.
If $\ghat_{\rho\rho}-1=(c \rho^2+16\pi G_N \alpha^2 \log\rho)/(1-c \rho^2-16\pi G_N \alpha^2 \log\rho)$ is positive, defining $dz_{\text{E}}=\sqrt{\ghat_{\rho\rho}-1}d\rho$ embeds the $\ghat_{\rho\rho}-1>0$ part of $\hat{\cal{M}}_{2}$ into a Euclidean space. While $\ghat_{\rho\rho}-1<0$ should be embeded into a Minkowski space via $dz_{\text{M}}=\sqrt{1-\ghat_{\rho\rho}}d\rho$. Therefore, it would be beneficial to discuss the sign of $\ghat_{\rho\rho}-1$ before drawing the embedding diagram.

\begin{figure}[h]
	\begin{center}
		\includegraphics[width=6.3cm]{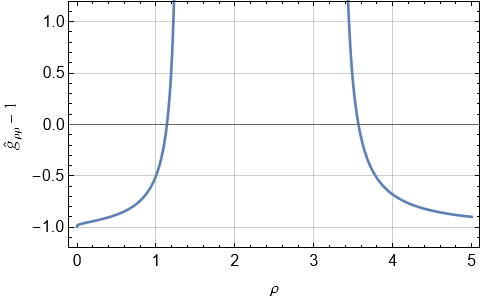} 
	\end{center}
	\caption{\small Plot for $\ghat_{\rho\rho}-1$ in the case of $c=-1$ of $16\pi G_N \alpha^2= 10$:
		$\ghat_{\rho\rho}-1$ changes its sign at $\rho=1.138$ and $\rho=3.566$. It also blows up at $\rho=1.313$ and $\rho=3.314$. Hence the positive $\ghat_{\rho\rho}-1$ regions are $1.138<\rho<1.313$ and $3.314<\rho<3.566$, hence they can be embedded into a three-dimensional flat Euclidean space; the regions $0<\rho<1.313$ and $\rho>3.566$ can be embedded into a Minkowski space due to their negative $\ghat_{\rho\rho}-1$. }\label{c1grhorhom1} 
\end{figure}

Let us back to the case of $c=-1$ and $16\pi G_N \alpha^2= 10$. We plot the coresponding $\ghat_{\rho\rho}-1$ in Fig.\ref{c1grhorhom1}. The function $\ghat_{\rho\rho}-1$ also blows up at $\rho=1.313$ and $\rho=3.314$ which indicate the range of $\rho$ for the small branch and the large branch.  $\ghat_{\rho\rho}-1$ is negative in $0<\rho<1.138$ and positive in $1.138<\rho<1.313$ for the small branch; While for the large branch, $\ghat_{\rho\rho}-1$ changes its sign at $\rho=3.566$ from positve to negative.

\begin{figure}[h]
	\begin{center}
		\includegraphics[width=2.5cm]{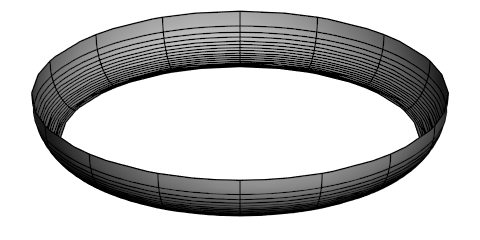} \;
		\includegraphics[width=2.5cm]{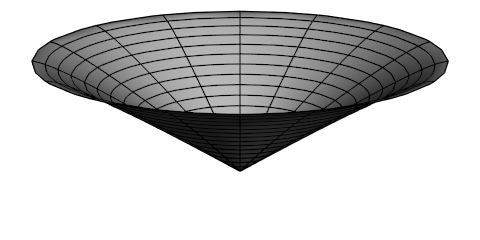} \;
		\includegraphics[width=2.7cm]{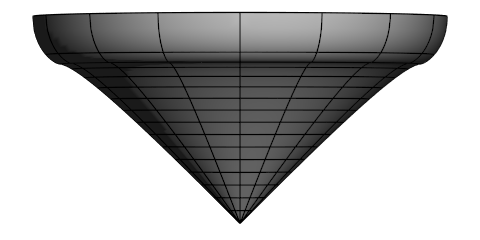}
	\end{center}
	\caption{\small The small branch for the case of $c=-1$ of $16\pi G_N \alpha^2= 10$: The left figure correspondes to the $0<\rho<1.313$ region which is embedded into a Minkowski space; the middle one is for $1.138<\rho<1.313$ embedded in an Euclidean space; the right figure shows two region joined together. }
	\label{sol1cm1Brsmall}
\end{figure}

\begin{figure}[h]
	\centering
	\includegraphics[width=3.6cm]{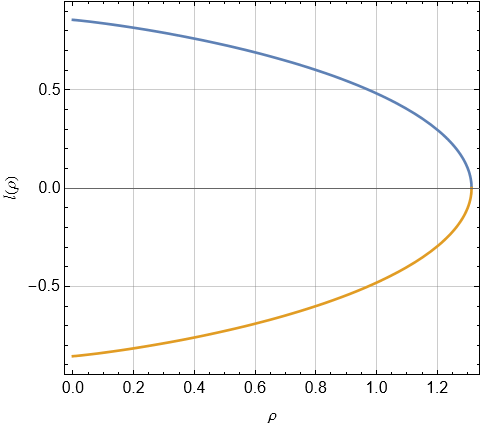} \quad\quad\quad
	\includegraphics[width=3.7cm]{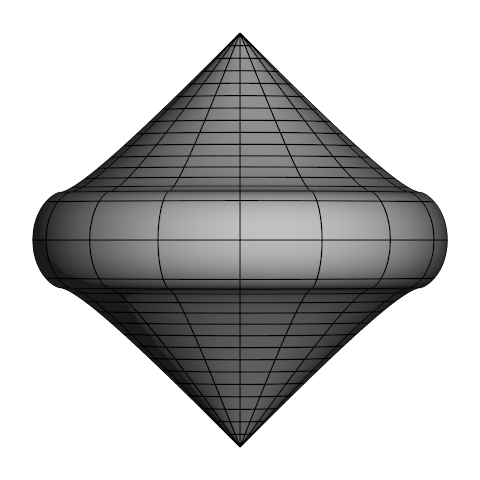} \;
	\caption{\small The small branch for the case of $c=-1$ of $16\pi G_N \alpha^2= 10$: The left figure shows the function $l(\rho)$ in which the yellow curve is for another copy. We set $l=0$ at $\rho=1.313$, i.e. the maximum of $\rho$. The right figure is the complete embedded diagram. }\label{sol1cm1smalemb}
\end{figure}

For the small branch, the $0<\rho<1.313$ region should be embedded into an Euclidean space due to its positive $\ghat_{\rho\rho}-1$. Such a shape is described by the left figure of Fig.\ref{sol1cm1Brsmall}. While the $1.138<\rho<1.313$ region of $\hat{\cal{M}}_{2}$ correspondes to the middle figure of Fig.\ref{sol1cm1Brsmall}. The right figure of Fig.\ref{sol1cm1Brsmall} shows how to join these two parts joined.
Reminding the issue of coordinate singularity $1/\ghat_{\rho\rho}=0$, we tramsform the coordinate $\rho$ to $l$. The relation between $\rho$ and $l$ is described by the left plot in Fig.\ref{sol1cm1smalemb}, in which we have set $l=0$ at the maximum $\rho=1.313$. The yellow curve describes another copy of the blue curve. Hence they represent the full region for the function $\rho(l)$. The right figure of Fig.\ref{sol1cm1smalemb} is the embeding diagram for the whole $\hat{\cal{M}}_{2}$. There are two sharp peaks corresponding to $\rho=0$. They are intrinsic singularities because the independent Ricci scalar is divergent when $\rho$ tends to $0$.

\begin{figure}[h]
	\begin{center}
		\includegraphics[width=2.5cm]{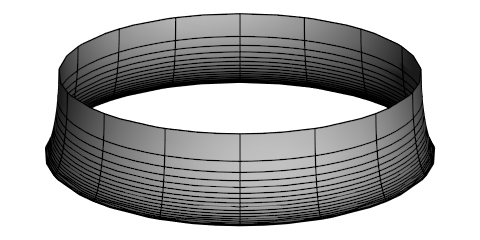} \;
		\includegraphics[width=2.5cm]{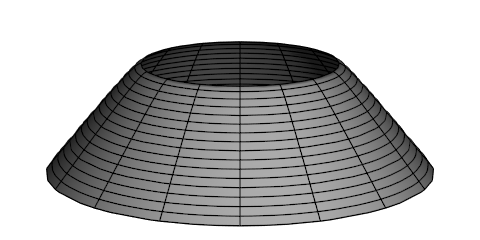} \;
		\includegraphics[width=2.7cm]{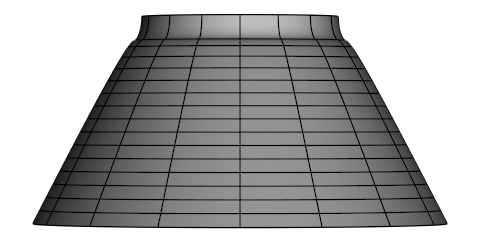}
	\end{center}
	\caption{\small The large branch for the case of $c=-1$ of $16\pi G_N \alpha^2= 10$: The left figure correspondes to the $3.314<\rho<3.566$ region which is embedded into an Euclidean space; the middle one is for the $\rho>3.566$ region embedded in a Minkowski space; the right figure shows two region joined together. }
	\label{sol1cm1Brlarge}
\end{figure}

\begin{figure}[h]
	\centering
	\includegraphics[width=3.6cm]{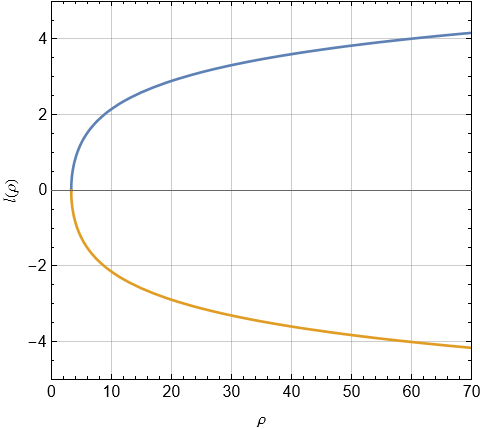} \quad\quad\quad
	\includegraphics[width=3.7cm]{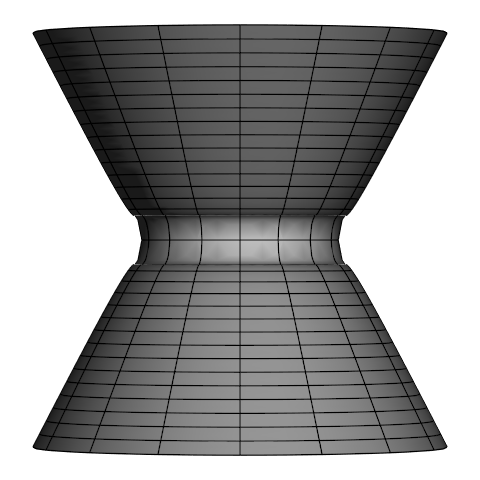} \;
	\caption{\small The large branch for the case of $c=-1$ of $16\pi G_N \alpha^2= 10$: The left figure shows the function $l(\rho)$ in which the yellow curve is for another copy. We set $l=0$ at $\rho=3.566$, i.e. the minumun of $\rho$. The right figure is the complete embedded diagram. }\label{sol1cm1largeemb}
\end{figure}

\begin{figure}[h]
	\begin{center}
		\includegraphics[width=3.6cm]{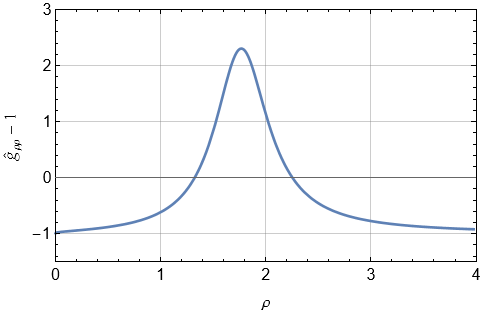}
	\end{center}
	\begin{center}
		\includegraphics[width=2cm]{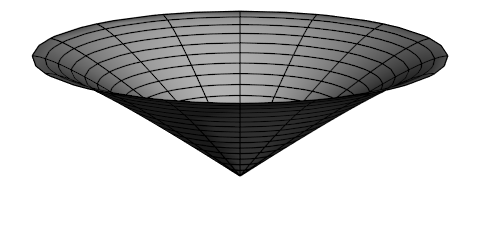} 
		\includegraphics[width=2cm]{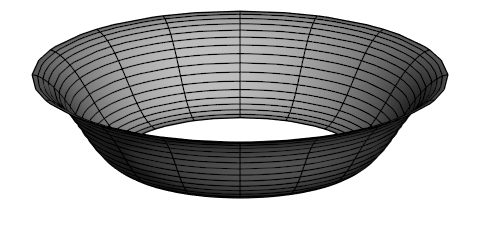} 
		\includegraphics[width=2cm]{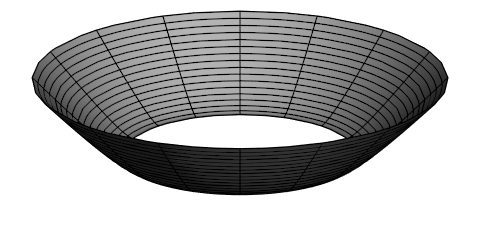} 
		\includegraphics[width=2.3cm]{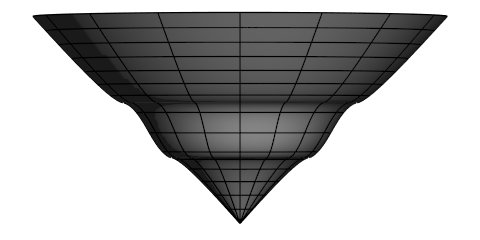}
	\end{center}
	\caption{\small The case of $c=-1.6$ and $16\pi G_N \alpha^2= 10$:
		The upper plot shows that $\ghat_{\rho\rho}-1$ is positive in $1.323<\rho<2.253$ but negative in $0<\rho<1.323$ and $\rho>2.253$.
		The lower figures from left to right are (1) the $0<\rho<1.323$ region embedded into a Minkowski space; (2) the $1.323<\rho<2.253$ region embedded into an Euclidean space; (3) the $\rho>2.253$ region embedded into a Minkowski space again; (4) the whole $\rho>0$ region, respectively. }\label{sol1cm1dot6emb}
\end{figure}

\begin{figure}[h]
	\begin{center}
		\includegraphics[width=3.6cm]{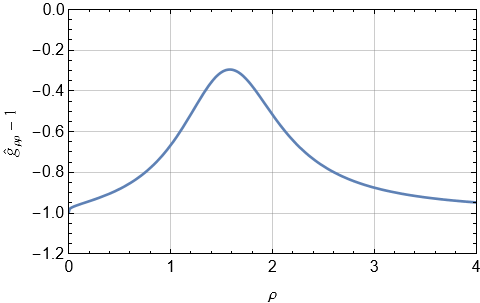}
		\includegraphics[width=4.6cm]{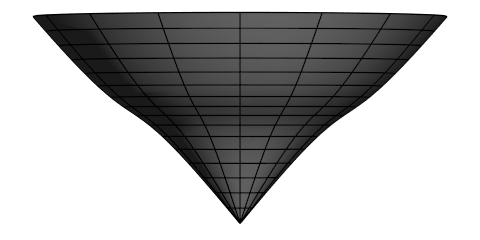}
	\end{center} 
	\caption{\small The case of $c=-2$ and $16\pi G_N \alpha^2= 10$:
		The left plot shows $\ghat_{\rho\rho}-1<0$. Hence the whole $\hat{\cal{M}}_{2}$ should be embedded into a Minkowski space, with the shape as the right figure shows. }\label{sol1cm2emb}
\end{figure}

The large branch is not singular since it starts from the minimum $\rho=3.314$ and then excludes the singularity $\rho=0$. The intrinsic Ricci scalar \eqref{sol1hatRicci} thus has an upper bound. Similarly, the positive $\ghat_{\rho\rho}-1$ region $3.314<\rho<3.566$ indicates that it can be embedded into a flat Euclidean space, as shown in the left figure of Fig.\ref{sol1cm1Brlarge}; While the middle one corresponds to the region $\rho>3.566$ embedded into a Minkowski space. The right figure of Fig.\ref{sol1cm1Brlarge} shows the joined figure. Finally, we plot  Fig.\ref{sol1cm1largeemb} to complete the embedding. The left figure shows the function $l(\rho)$ containing another copy (yellow) in which we have set $l=0$ at the minimum $\rho=3.314$. The right figure shows the entire embedding diagram.
There are other interesting cases when keeping $16\pi G_N \alpha^2=10 $.
If $c$ is smaller than the miminum value $-1.5059$, $\rho$ runs from $0$ to infinity without any point making $\ghat_{\rho\rho}$ blowing up.
We take $c=-1.6$ to plot $\ghat_{\rho\rho}-1$ and draw the embedding digaram in Fig.\ref{sol1cm1dot6emb}. Depende on the sign of $\ghat_{\rho\rho}-1$, the lower four figures from left to right correspond to (1) the $0<\rho<1.323$ region embedded into a Minkowski space; (2) the $1.323<\rho<2.253$ region embedded into an Euclidean space; (3) the $\rho>2.253$ region embedded into a Minkowski space again; (4) the whole $\rho>0$ region, respectively.
Furthermore, a more negative $c$ may imply that no region can be embedded into an Euclidean flat space. For instanse, the left plot of Fig.\ref{sol1cm2emb} shows the case of $c=-2$ which satisfies $0<\ghat_{\rho\rho}<1$ in the whole region of $\rho>0$. Hence $\hat{\cal{M}}_{2}$ in this case should be entirely embedded into a Minkowski space, as shown in the right figure of Fig.\ref{sol1cm2emb}.


\begin{figure}[h]
	\begin{center}
		\includegraphics[width=3.6cm]{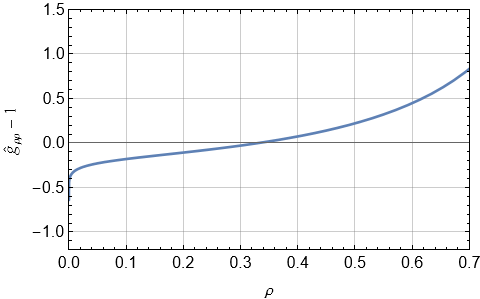} \label{grhorho1stex}
	\end{center}    
	\begin{center} 
		\includegraphics[width=2.5cm]{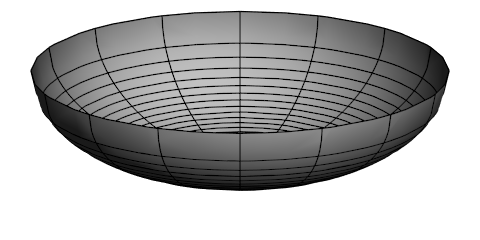} \; 
		\includegraphics[width=2.5cm]{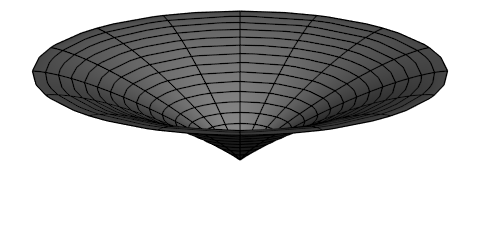} \; 
		\includegraphics[width=2.7cm]{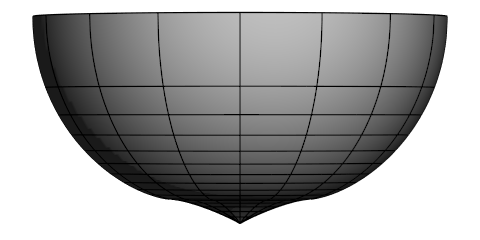}
	\end{center}
	\caption{\small The case of $c=1$ and $16\pi G_N \alpha^2=0.1$: The upper plot shows $\ghat_{\rho\rho}-1$ with a root $\rho=0.3320$.
		The lower figures are embedded diagrams:
		the left figure is for $0.3320<\rho<1$ (Euclidean);
		the middle figure is for $0<\rho<0.3320$ (Minkowski);	the right one is the joined figure.} \label{c1smallalphaemb}
\end{figure}

\begin{figure}[h]
	\centering 
	\includegraphics[width=3.6cm]{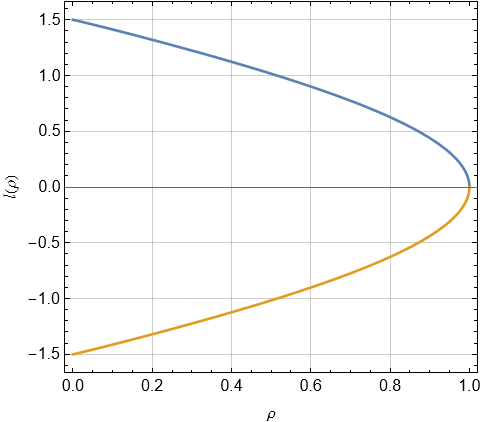} \quad\quad\quad
	\includegraphics[width=3.7cm]{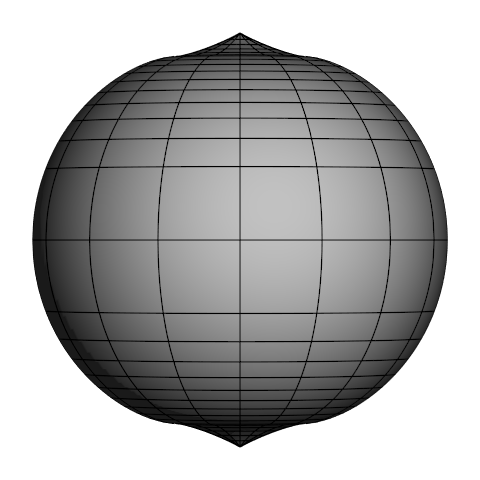} \;
	\caption{\small The case of$c=1$ and $16\pi G_N \alpha^2=0.1$: The left figure shows the function $l(\rho)$. We set $l=0$ at $\rho=1$, i.e. the maximum of $\rho$. The right figure is the complete embedded diagram. }\label{sol1c1smallalphaemb}
\end{figure}

If we still use $16\pi G_N \alpha^2=10 $, a non-negative $c$ does not produce a new embedding diagram but still be similar to the small branch for $c=-1$. They will have a very small maximum $\rho$.
Instead, we take $c=1$ and $16\pi G_N \alpha^2=0.1$ to show a case with positive $c$. The equation $1-\rho^2 - 0.1 \log\rho=0$ has a root $\rho= 1$, so the range of $\rho$ is $0<\rho<1$. 
The upper plot in Fig.\ref{c1smallalphaemb} shows that $\ghat_{\rho\rho}-1$ has a root $\rho=0.3320$. Then the lower figures show several parts of the embedded diagram. The left plot describes the $0<\rho<0.3320$ region embedded in Euclidean space; the middle one describes the $0.3320<\rho<1$ region embedded in a Minkowski space; the right one shows the joined embedding diagram.  
The left plot in Fig.\ref{sol1c1smallalphaemb} shows the numerical results of $l(\rho)$ in which $l=0$ at the maximum $\rho=1$. The yellow curve shows another copy. Then we obtain the whole embedding diagram, the right figure in Fig.\ref{sol1c1smallalphaemb}, which serves as a deformed sphere.

\subsection{Solution II}
There is an alternative ansatz without posing period conditions in the field space. It is
\be
\begin{split}
	&ds^2= - f(r) dt^2 + \frac{dr^2}{ f(r) } 
	+ r^2\,e^{\lambda(\rho)} (d\rho^2+\rho^2d\varphi^2)  \,,   
	\\ & A_{\mu}dx^{\mu} =-\phi(r) dt  \,,\quad
	\psi^1 = \alpha \rho\cos\varphi \,, \quad
	\psi^2 = \alpha \rho\sin\varphi \,,\label{usualansatzalt}
\end{split}
\ee
in which the configuration of $\psi^I$ has already solved two KG equations. The Maxwell equations still give Eq.\eqref{diffeqsFromMaxwell}, while  the Einstein equations reduce to
\begin{align}
	&-f -r\frac{df}{dr}-\Lambda r^2 - G_N (\frac{d\phi}{dr})^2 \nonumber \\&= \frac{e^{-\lambda}}{2\rho} \big( \frac{d\lambda}{d\rho}+ \rho\,(16\pi G_N \alpha^2 +\frac{d^2\lambda}{d\rho^2}) \big)
	\,,    \label{ansdiffeq1Alt}
	\\ &\frac{1}{2}\frac{d^2f}{dr^2}+\frac{1}{r}\frac{df}{dr} + \Lambda =  G_N (\frac{d\phi}{dr})^2   \,.
	\label{ansdiffeq2Alt}
\end{align}
Thus the same $f(r)$ and $\phi(r)$ with Eq.\eqref{Solu1} solve Eq.\eqref{diffeqsFromMaxwell} and Eq.\eqref{ansdiffeq2Alt}, and imply the LHS of Eq.\eqref{ansdiffeq1Alt} becomes a constant $-c$. Hence the the RHS of Eq.\eqref{ansdiffeq1Alt} leads to 
\be
\frac{d\lambda}{d\rho}+ \rho\,(16\pi G_N \alpha^2 +\frac{d^2\lambda}{d\rho^2})  =-2c\rho e^{\lambda}\,, \label{altshapeEq}
\ee
which is a nonlinear differential equation for $\lambda(\rho)$  if $c\neq 0$. It is hard to find an exact general solution. However, in the case of $c=0$, Eq.\eqref{altshapeEq} reduces to a linear equation which can be exactly solved. The solution is
\be
\begin{split}
	\lambda(\rho)&= -4\pi G_N \alpha^2 \rho^2 +n\log (\beta\rho) \,.  \label{deformedPlane}
\end{split}
\ee
Therefore, the full solution for $c=0$ is given by 
\be
\begin{split}
	&ds^2= -(-\frac{8\pi G_N M}{\Omega\,r}  + (\frac{4\pi}{\Omega})^2 \frac{G_N Q^2}{r^{2}}- \fft{\Lambda r^2}{3})dt^2 
	\\&\;\;\quad\quad +(-\frac{8\pi G_N M}{\Omega\,r}  + (\frac{4\pi}{\Omega})^2 \frac{G_N Q^2}{r^{2}}- \fft{\Lambda r^2}{3})^{-1} dr^2
	\\&\;\;\quad\quad + r^2 (\beta\rho)^{n} e^{-4\pi G_N \alpha^2 \rho^2}  (d\rho^2+\rho^2d\varphi^2)  
	\,,
	\\ &A_{\mu}dx^{\mu} =- \frac{4\pi Q}{\Omega r} dt  \,,\quad
	\psi^1 = \alpha \rho\cos\varphi \,, \quad
	\psi^2 = \alpha \rho\sin\varphi\,.\label{Solu2}
\end{split}
\ee

\subsubsection{Compare with the planar axionic RN-AdS black hole}
On the other hand, simply demanding $\lambda$ as a constant also solves Eq.\eqref{altshapeEq}. Then the constant $c$ should be $c=- 8\pi  G_N \alpha^2 \,e^{-\lambda}$. We re-scale $\alpha$ and introduce a coordinates transformation
\be
\tilde{\alpha}=\alpha\,e^{-\lambda/2} \,,\quad
x=e^{\lambda/2}\rho\cos\varphi \,,\quad
y=e^{\lambda/2}\rho \sin\varphi \,,
\ee
such that the solution is given by
\be
\begin{split}
	&ds^2= -f(r)dt^2 + \frac{dr^2}{ f(r) }+ r^2 (dx^2+dy^2)  \,,   
	\\& f(r)= - 8\pi  G_N \tilde{\alpha}^2 -\frac{8\pi G_N M}{\Omega\,r}  + (\frac{4\pi}{\Omega})^2\frac{G_N Q^2}{r^{2}}- \fft{\Lambda r^2}{3} \,,
	\\ &A_{\mu}dx^{\mu} =- \frac{4\pi Q}{\Omega r} dt  \,,\quad
	\psi^1 =\tilde{\alpha}\,x \,, \quad
	\psi^2
	= \tilde{\alpha}\,y \,,\label{planarAxionic}
\end{split}
\ee
which the metric has a planar transverse space $dx^2+dy^2$. It is worth noting that the planar $D$-dimensional AdS black hole in Ref.\cite{Andrade:2013gsa} reduces to Eq.\eqref{planarAxionic} if $D=4$ and $\Lambda<0$. 
The remarkable feature of such kind of planar black hole achieves momentum relaxation through the configuration of scalars to break the transition symmetries called the holographic axion model\cite{Andrade:2013gsa, Donos:2014cya, Baggioli:2021xuv}. In the holographic context, it would be convenient to redefine parameters $M/\Omega$, $Q/\Omega$, and $\Lambda$ to explicitize the horizon location $r_H$. First, we rewrite $\Lambda$ as $\Lambda=-3/L^2$ and then adjust the gauge condition for $A_{\mu}$ to make $\phi=-\mu(1-r_H/r)$ which vanishes on the horizon but has a finite value on the AdS boundary. Hence we have relation $(4\pi Q)/\Omega = \mu r_H$. Moreover, investigations of holographic models usually apply the coordinate $u=L/r$ such that the flat boundary is at $u=0$, but we still use the coordinate $r$ in this paper since using $r$ is beneficial to formulate the unified first law, discussed in the next section. Therefore, $r=\infty$ is the AdS boundary.  As for $M/\Omega$, the requirment $f(r_H)$ gives $M/\Omega= -\alpha r_H^2+\mu^2 r_H/(8\pi) + r_H^3/(8\pi G_N L^2) $ in which we omit the tilde sign of $\alpha$. Therefore, the solution \eqref{planarAxionic} is re-formulated as 
\be
\begin{split}
	&ds^2= -f(r)dt^2 + \frac{dr^2}{ f(r) }+ r^2 \delta_{ij} dx^idx^j  \,, \quad 
	\\&  f(r)=-8\pi G_N\alpha^2- \frac{r_H}{r}(-8\pi G_N\alpha^2+G_N\mu^2+\frac{r_H^2}{L^2})
	\\&\quad\qquad + \frac{G_N\mu^2 r_H^2}{r^2} + \frac{r^2}{L^2} \,,
	\\& A_{\mu}dx^{\mu} =\mu (1-\frac{r_H}{r}) dt  \,,\quad
	\psi^I =\alpha \delta^I_i x^i \,,  \,.
	\label{planarAxholgauge}
\end{split}
\ee
Following the method summarized in Ref.\cite{Baggioli:2021xuv}, we consider the following perturbation around the solution \eqref{planarAxholgauge} to calculate the DC conductivity:
\be
\begin{split}
	&\delta g_{tx}=r^2 H_{tx}(r)  \,,\; \delta g_{rx}= r^2 H_{rx}(r)\,,\; 
	\delta A_{x}= -tE + a(r) \,.
\end{split}
\ee
No $\zeta$ and $\delta \psi^I$ are considered here because we only focus on the electric DC conductivity in this paper but left the thermo-conductivity for future works.
The perturbative Maxwell equation leads to a conserved current
\be
\mathcal{J} = -f(r) \delta A_x'(r) + \phi'(r) r^2 H_{tx}(r)  \,, \label{electriccurrent}
\ee
where $'$ is $d/dr$ for short. We have considered that $d/dr=-(u^2/L)d/du$ will introduce a ``$-$'' sign for Eq.\eqref{electriccurrent}, different from Ref.\cite{Baggioli:2021xuv}. In addition, the $rx$ component of the linearized Einstein equation implies a constraint for $H_{rx}(r)$, such that
\be
\begin{split}
	H_{rx}(r) &=  E \frac{ r_H\mu}{4\pi \alpha^2 r^2f}	\,. \label{perEincons}
\end{split}
\ee
Then one can obtain the boundary DC conductivity by the horizon data because values of $\mathcal{J}$ are independent of the location $r$. According to Refs.\cite{Baggioli:2021xuv}, the perturbation should satisfy the boundary condition near the horizon:
\be
\delta A'_x \sim -\frac{E}{f}  \;,\quad  H_{tx}(r) \sim  fH_{rx}(r) \,, \label{perNHor4DCcond}
\ee
Hence we obtain $\mathcal{J}=(1+\mu^2/(4\pi\alpha^2)E)$. The $4\pi$ factor difference with Ref.\cite{Baggioli:2021xuv} is due to the unit selection for the Maxwell field. We use $-F^2/(16\pi)$ rather than $-F^2/(4\pi)$ as its Lagrangian in the action \eqref{holographicAxions4dim}. Notice that the Ohm's law is $\mathcal{J}= \sigma_{\text{DC}} E$, we obtain the finite DC conductivity,
\be
\sigma_{\text{DC}}= 1+\frac{\mu^2}{4\pi \alpha^2}\,. \label{holaxDCcond}
\ee

It is worth noting that the metric \eqref{planarAxionic} has the same $t$-$r$ part with the hyperbolic RN-AdS black hole, though the transverse space describes a plane. Such an observation is also one motivation in the Ref.\cite{Ren:2019lgw} to construct a hyperbolic black hole in an
Einstein-Maxwell-Dilation(EMD) theory shares the same $t$-$r$ geometry with a planar black hole containing axionic charges. Here, solution II given by Eq.\eqref{Solu2} hints at a continuing family of transverse shapes between the hyperbolic solution without axions and the planar solution with axions.

We will further calculate the DC conductivity for a general metric describing a deformed topological black hole to end the comparison. Replacing the $-8\pi G_N\alpha^2$ and $\delta_{ij}$ in Eq.\eqref{planarAxholgauge} to $c$ and $e^{\lambda(x)}\delta_{ij}$ in which the function $\lambda(x)$ satisfy
\be
\partial^2 \lambda + 16\pi G_N \alpha^2 = -2c\, e^{\lambda}\,, \label{genshapeEq}
\ee
where $\partial^2 \lambda$ is $\delta^{ij} (\partial^2 \lambda/\partial x^i \partial x^j) $ for short. Hence the following metric, gauge field, and axions also solve the equations of motion: 
\be
\begin{split}
	&ds^2= -f(r)dt^2 + \frac{dr^2}{ f(r) }+ r^2 e^{\lambda(x)} \delta_{ij} dx^idx^j  \,,   
	\\&  f(r)= c- \frac{r_H}{r}(c+G_N\mu^2+\frac{r_H^2}{L^2}) + \frac{G_N\mu^2 r_H^2}{r^2} + \frac{r^2}{L^2} \,,
	\\ &A_{\mu}dx^{\mu} =\mu (1-\frac{r_H}{r}) dt  \,,\quad
	\psi^I =\alpha \delta^I_i x^i \,. \label{dfplanarAxholgauge}
\end{split}
\ee
Then consider the perturbation
\be
\begin{split}
	&\delta g_{ti}= (r^2 H_{t}(r) ) \partial_iX \,,\;
	\\&\delta g_{rx}= (r^2 H_{r}(r) ) \partial_iX \,,\;
	\\&\delta A_{i}= (-E t + a(r) )\partial_i X \,.
\end{split}
\ee
Requiring $X$ be a harmonic scalar on $\hat{\cal{M}}_{2}$, i.e., $\partial^2X=0$, will solve most equations of motion but left three independent equations which are similar to the simplest holographic axion model discussed above. Those equations ensure the validity of the current \eqref{electriccurrent} and the constraint \eqref{perEincons} by simply replacing $\alpha^2$ by $-c/(8\pi G_N)$. Hence they lead to the explicit result of the DC conductivity
\be
\sigma_{\text{DC}}= 1 -\frac{ 2 G_N \mu^2}{c}\,. \label{dfholaxDCcond}
\ee

\subsubsection{Shapes of horizons}
We will then study the horizon shapes of Solution II. Even if we omit cases of $c\neq$ but only focus on cases of $c=0$ in this article, there are various shapes of the transverse space $\hat{\cal{M}}_{2}$.
Again, one can introduce a suitable rescaling to reduce the parameters in the line-element \eqref{Solu2}. Hence the $\hat{\cal{M}}_{2}$ part becomes 
\be
d\hat{s}^2 = \rho^n e^{-\rho^2} (d\rho^2+\rho^2d\varphi^2) \,. \label{sol2Mhatrescaled}
\ee
Thus only the parameter $n$ controls the $\hat{\cal{M}}_{2}$ geometry. The independent Ricci scalar is 
\be
\Rhat = 4 \, \fft{e^{\rho^2} }{\rho^{n}} \,, \label{sol2Rhat}
\ee
which obviously blows up at $\rho=0$ if $n>0$. 
When $n=0$, the point $\rho=0$ obviously becomes a regular center, but it becomes subtle for cases of $n<0$. Suppose we start from a finite value $\rho$ and go along a direction with a fixed $\varphi$. Such a path is no doubt a spacelike geodesic, and its affine parameter is the proper distant $l$, which is determined by $dl= \sqrt{\rho^n e^{-\rho^2}}d\rho$ according to Eq.\eqref{sol2Mhatrescaled}.
When we get close to $\rho=0$, $dl/d\rho$ behaves as $\rho^{n/2}$ then $l\sim \rho^{n/2+1}$ if $n\neq -2$ and $l\sim\log\rho$ if $n=-2$. Therefore, for cases of $n\le -2$, $l$ tends to negative infinity as $\rho$ tends to $0$ even though $\Rhat$ keeps finite; if $-2<n\le 0$, both of $l$ and $\Rhat$ have finite values. for cases of $n>0$, the finite limit of $l$ surpports that $\rho=0$ is the intrinsic singularity.
In addition, when $\rho$ tends to infinity, $dl/d\rho$ will rapidly decrease due to the exponential factor $e^{-\rho^2/2}$. Thus, $l$ will converge to a finite limit, but $\Rhat$ will blow up within such a finite affine parameter because of Eq.\eqref{sol2Rhat}. Therefore, infinity far $\rho$ is the intrinsic singularity despite the value of $n$.

\begin{figure}[h]
	\begin{center}
		\includegraphics[width=2cm]{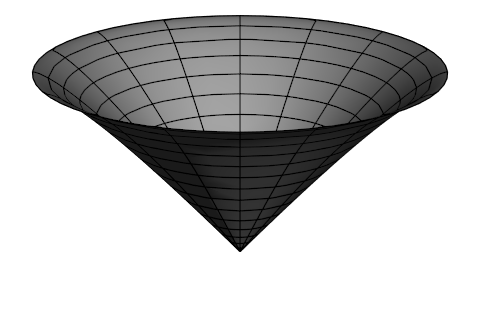} 
		\includegraphics[width=2cm]{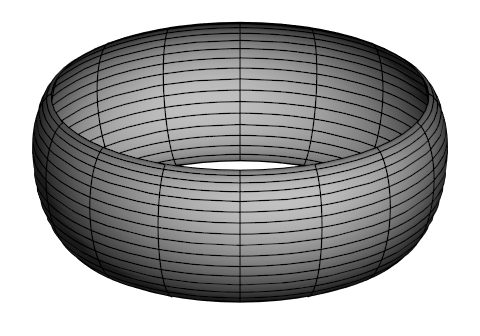} 
		\includegraphics[width=2cm]{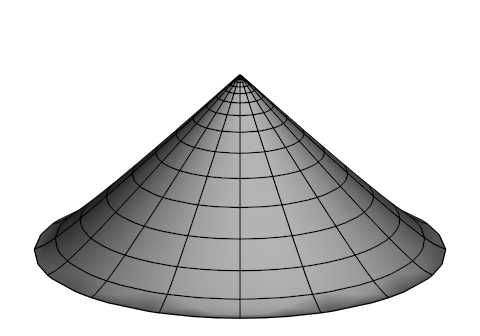} 
		\includegraphics[width=2.3cm]{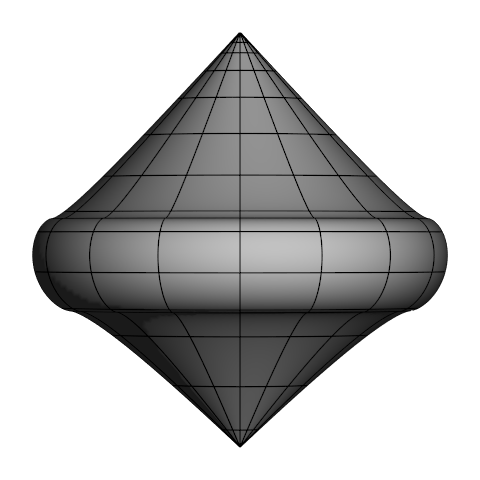}
	\end{center} 
	\caption{\small Embedding diagrams for $n=2$: from left to right, they are (1) $0<\rho <1$, the Minkowski region; (2) $1<\rho <\sqrt{3}$, the Euclidean region; (3) $ \rho >\sqrt{3}$, the Minkowski region; (4) the whole embedding diagram.  }\label{sol2n2emb}
\end{figure}

We will draw the embedding diagram for typical cases to visualize the above features. 
Hence we should define $\tilde{r}= \rho^{1+n/2} e^{-\rho^2/2} $ and re-write Eq.\eqref{sol2Mhatrescaled} as
\be
d\hat{s}^2 =-\rho^n e^{-\rho^2} (\rho^2-\fft{n}{2})(\rho^2-\fft{n+4}{2}) d\rho^2 +d\tilde{r}^2 +\tilde{r}^2d\varphi^2 \,,
\ee
in which the sign of the factor $(\rho^2-n/2)(\rho^2-(n+4)/2)$ determines the signature of the higher dimensional flat space.
For the region $\sqrt{n/2}<\rho<\sqrt{(n+4)/2}$, we introduce
\be
\fft{dz_{\text{E}}}{d\rho} =  \rho^{n/2} e^{-\rho^2/2} \sqrt{-(\rho^2-\fft{n}{2})(\rho^2-\fft{n+4}{2})}  \,,
\ee
which specifies the embedding into an Euclidean space. While $\rho<\sqrt{n/2}$ and $\rho>\sqrt{(n+4)/2}$ should be the Minkowski regions. They lead to the following embedding:
\be
\fft{dz_{\text{M}} }{d\rho} =  \rho^{n/2} e^{-\rho^2/2} \sqrt{(\rho^2-\fft{n}{2})(\rho^2-\fft{n+4}{2})}  \,.
\ee
We will then pick up some typical values of $n$ to draw the embedded diagram.
Fig.\ref{sol2n2emb} shows the case of $n=2$, which is typical for $n>0$. Figures from left to right represent (1) the $0<\rho <1$ region embedded into a Minkowski space; (2) the $1<\rho <\sqrt{3}$ region embedded into an Euclidean space; (3) the $ \rho >\sqrt{3}$ region embedded into a Minkowski space; (4) the full embedding diagram of $\hat{\cal{M}}_{2}$. 
Points for divergent $\Rhat$, $\rho=0$ and $\rho=\infty$, appear in two Minkowski regions. While the geometry of the Euclidean part is smooth.

\begin{figure}[h]
	\centering
	\includegraphics[width=2.5cm]{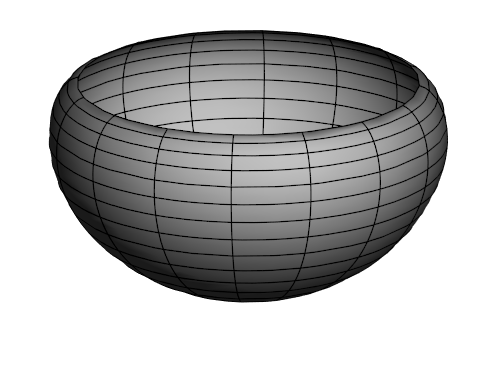} \;
	\includegraphics[width=2.5cm]{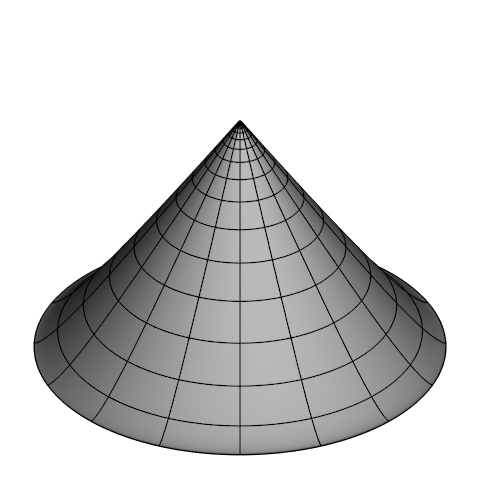} \;
	\includegraphics[width=2.7cm]{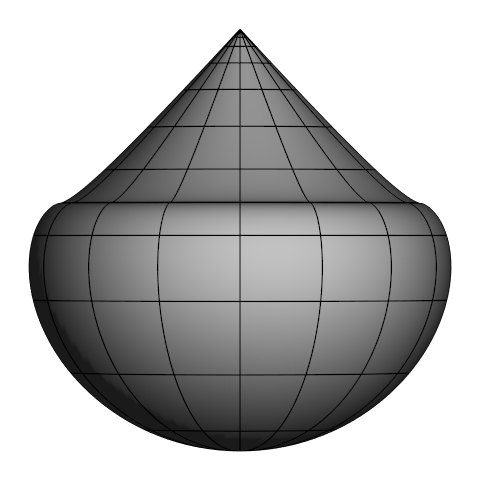}
	\caption{\small Embedding diagrams for $n=0$: the left figure is for the  Euclidean region $0\le \rho < \sqrt{2} $ containning the regular center $\rho=0$; the middle figure shows the $ \rho > \sqrt{2}$ region embedded into a Minkowski space in which the $\rho=\infty$ is a singular direction. The right one is the whole embedding diagram. }\label{sol2n0emb}
\end{figure}

The case of $n=0$ is shown in Fig.\ref{sol2n0emb}. The left figure represents the  $0\le \rho < \sqrt{2} $ region embedded into Euclidean space. While the middle one 
is for the $ \rho > \sqrt{2}$ Minkowski region. The right figure is for the whole $\hat{\cal{M}}_{2}$. The regular center $\rho=0$ is in the Euclidean region, and the singular $\rho=\infty$ is in the Minkowski region.

\begin{figure}[h]
	\centering
	\includegraphics[width=2.5cm]{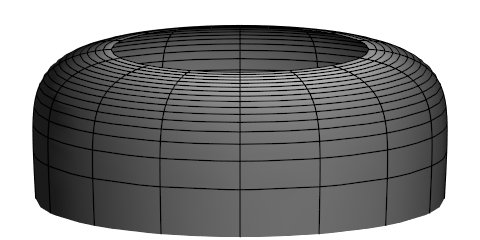} \;
	\includegraphics[width=2.5cm]{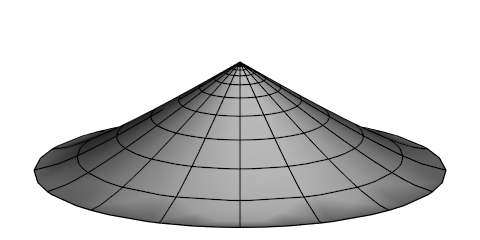} \;
	\includegraphics[width=2.7cm]{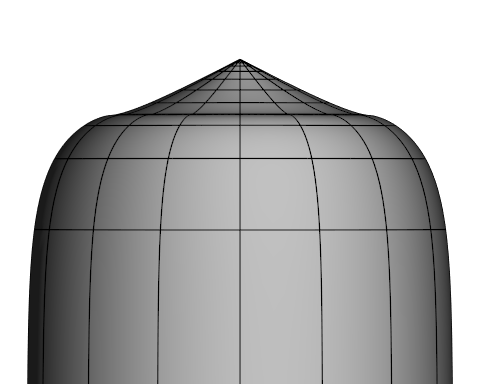}
	\caption{\small Embedding diagrams for the $n=-2$ case: the left figure is for $0\le \rho < 1$ (Euclidean);  the middle one is for $\rho> 1$ (Minkowski); the right figure is the whole embedding diagram. }\label{sol2nm2emb}
\end{figure}

\begin{figure}[h]
	\centering
	\includegraphics[width=3.5cm]{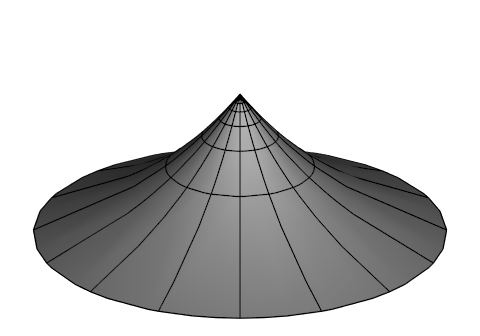} \quad\qquad\quad
	\includegraphics[width=3.5cm]{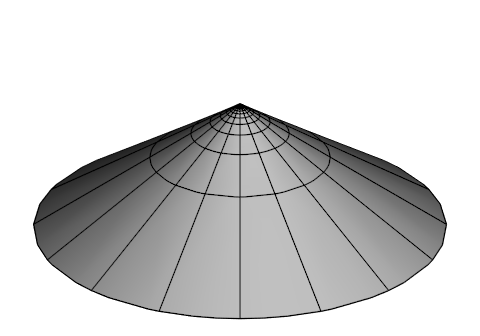} \;
	\caption{\small 
		The left embedding diagram is for $n=-4$ while the right one is for $n=-6$.
		There is no Euclidean region in these cases. }\label{sol2nm4nm6emb}
\end{figure}

The case of $n=-2$ also contains one Euclidean region and one Minkowski region, as shown in Fig.\ref{sol2nm2emb}. The left and middle figures are for the $0\le \rho < 1$ Euclidean region and the $\rho> 1$ region respectively. The whole embedding diagram is the right figure. The infinity $\rho$ also appears in the Minkowski region, but the region near $\rho=0$ is enlarged as a lone tube in the Euclidean region, distinguished with the $n=0$ case. A heavier enlargement happens when $n<-2$. Fig.\ref{sol2nm4nm6emb} includes the cases for $n=-4$ and $n=-6$, entirely embedded into a Minkowski space. They contain $\rho=\infty$ as the center peaks, and the $\rho\sim 0$ regions extended to extreme far-away proper distance. 

Solutions I and II show how suitable profiles for axions deform transverse spaces even in $D=4$. The non-trivial shape makes the physical meaning of parameters $M$ and $Q$ hard to understand. Consider turning off the cosmological constant $\Lambda$. Both solutions are not asymptotically flat since the non-trivial shape extends to infinity. Taking $\Lambda\neq 0$ does not make it better. Lacking a well-defined asymptotically structure like Minkowski or AdS indicates the conceptual difficulties for formulating the black hole thermodynamics via the global parameters $M$ and $Q$. To overcome such a difficulty, we will develop a quasi-local viewpoint based on the MS mass and the unified first law in the next section. 

\section{The generalized unified first law}

\setcounter{equation}{0}
\renewcommand\theequation{3.\arabic{equation}}

This section will briefly introduce the MS mass and the unified first law, and then discuss the thermodynamic method for generating solutions based on them. The method was originally proposed in Refs.\cite{Zhang:2013tca} for spherically symmetric spacetime. We aim to modify this method to adapt other shapes with non-constant $\Rhat(x)$ instead of a sphere.  
The modified thermodynamic method justifies some ansatzs in Eq.\eqref{usualansatz} and Eq.\eqref{usualansatzalt}.
Another goal of this section is to show how the unified first law offers a quasi-local viewpoint for black hole thermodynamics.

\subsection{Derived from Einstein's equation}
This subsection will re-derive the unified first law from the Einstein equations. To keep generality, we consider a $D$-dimensional warped product spacetimes $\M2\times\Mco2$. Its line element is
\begin{equation}
	\begin{split}
		ds^2&= g_{\mu\nu}(X)dX^{\mu}dX^{\nu} 
		\\&=I_{ab}(u)du^adu^b+r^2(u)\hat{g}_{ij}(x)dx^idx^j \,. \label{decom}
	\end{split}
\end{equation}
Coordinates frame $\{X^{\mu}\}$ of the whole spacetime is specified as $\{u^a, x^i\}$. It is beneficial to choose the following viewpoint. Coordinates $u^a$ label the point in the 2-dimensional manifold $\M2$, while $x^i$ label the point in the $(D-2)$-dimensional manifold $\Mco2$. Both manifolds have its independent metric $I_{ab}(u)$ and $\ghat_{ij}(x)$. The areal radius $r(u)$ is a scalar function in spacetime, also a function in $\M2$. Its value means enlarging the unit $\Mco2$ in $r$ times.
It is worth emphasizing again that the manifold $\Mco2$ is not limited to the maximally symmetric space investigated in Ref~\cite{Hayward:1997jp, Maeda:2007uu}. We do not presume a constant Ricci scalar $\Rhat(x)$ of $\Mco2$. Meanwhile, we assume that the manifold $\Mco2$ has a finite $(D-2)$-dimensional ``unit area" $\Omega$, or just the finite part of the $\Mco2$ with volume ``unit area" is concerned.

We then decompose the Einstein equations $G_{\mu\nu}+\Lambda g_{\mu\nu}=8\pi G_N T_{\mu\nu}\,$ via applying results in Appendix A. Results of $ab$ components are
\begin{align}
	&~ -\frac{D-2}{r} (\barnab_a \barnab_b r - \frac{1}{2}I_{ab}\, \barnab^2 r) \nonumber \\ &~
	+\frac{D-2}{2r} I_{ab}\,\big(\barnab^2r -\frac{D-3}{r}(\cK-I^{rr}) \big) 
	+\Lambda I_{ab} \nonumber \\&=8\pi G_N T_{ab}\,,\label{radiueom} 
\end{align}
while $ij$ components are
\begin{align}
	&~\hat{R}_{ij}-(D-3)\cK\hat{g}_{ij}
	-\hat{g}_{ij}\big( \frac{r^2}{2} \Rbar-(D-3)\, r\barnab^2 r  \nonumber
	\\&\qquad \;+\frac{(D-3)(D-4)}{2}(\cK-I^{rr}) -\Lambda r^2 \big) =8\pi G_N T_{ij}\,,\label{angulareom} 
\end{align}
where $I^{rr}$ is $\barnab_ar \barnab^ar$ for short; $\barnab$ and $\hatnab$ is the Levi-Civita connection of $\M2$ and $\Mco2$ respectively; $\cK$ is the reduced Ricci scalar of $\Mco2$ given by $\cK=\Rhat/((D-2)(D-3))$. 
The so-called unified first law is directly derived from Eq.\eqref{radiueom} and hints at suitable definitions for the MS mass. To show this, we need to define the energy supply vector $\Psi^a$ and the work term $W$ first. Namely,
\be
\Psi^a= \big(T^{ab} -\frac{1}{2} T_{cd}I^{cd}\, I^{ab} \big)\,\barnab_b r \,,\qquad
W= -\frac{1}{2} T_{ab}I^{ab} \,. \label{DefinitionPsiW} 
\ee
It is worth noting that the energy supply vector is constructed by projecting the traceless tensor $T^{ab} - T_{cd}I^{cd}\, I^{ab}/2$ on the direction $\barnab^a r$. With the Einstein's equation Eq.\eqref{radiueom}, $\Psi^a$ and $W$ should satisfy
\begin{align}
	\frac{D-2}{2r} ((\barnab^2 r) \barnab_a r -\barnab_a I^{rr} ) &=8\pi G_N \Psi_a \,, \label{radiueomTrless} 
	\\ \frac{D-2}{2r} \,\big(\frac{D-3}{r}(\cK-I^{rr} ) -\barnab^2r 
	\big)  -\Lambda  &=8\pi G_N W \,, \label{radiueomTr} 
\end{align}
where we have lowered the index of the energy supply vector $\Psi^a$.
Obviously, the term with $\barnab^2 r$ factor would disppear in the combination $\Psi_a+W\barnab_a r$ due to the common factor $(D-2)/(2r)$ in Eq.\eqref{radiueomTrless} and Eq.\eqref{radiueomTr}. Moreover, even considering the situation of non-constant $\cK$, it should only depend on $x$, i.e., $\barnab_a\cK=\partial\cK/\partial u^a=0$. Therefore we have $-\barnab_a I^{rr}=\barnab_a (\cK- I^{rr})$ such that 
\bea
&\frac{D-2}{2r} (\barnab_a (\cK-I^{rr}) +\frac{D-3}{r}(\cK-I^{rr})\,\barnab_a r  )
-\Lambda\,\barnab_a r \nonumber\\&=8\pi G_N ( \Psi_a + W \barnab_a r) \,. \label{radiueomComb} 
\eea
The LHS of \eqref{radiueomComb} further hints at the following simplification:
\bea
&\barnab_a \big(\frac{D-2}{16\pi G_N}r^{D-3}(\cK-I^{rr}) -\frac{\Lambda}{8\pi G_N}\,\frac{r^{D-1}}{D-1}\big)\nonumber
\\&= r^{D-2} ( \Psi_a + W \barnab_a r) \,, \label{radiueomCombSim} 
\eea
which indicates that $r^{D-2} ( \Psi_a + W \barnab_a r)$ equals to a total derivative of some scalar function. on $\M2$. 
Multiply the size $\Omega$ of the unit $\Mco2$. The LHS of Eq.\eqref{radiueomCombSim} hints
\be
\Mms(u,x) = \frac{(D-2)\,\Omega}{16\pi G_N}r^{D-3} \big(\cK(x)-I^{rr}\big) -\frac{\Lambda}{8\pi G_N}\,\frac{\Omega\, r^{D-1}}{D-1} \,, \label{xdepMSmass} 
\ee
which defines the MS mass with $x$-dependence. Introduce the area $A=\Omega r^{D-2}$ and ``volume" $V=\Omega r^{D-1}/(D-1)$ for Eq.\eqref{radiueomCombSim}, such that
\be
\barnab_a \Mms
= A\,\Psi_a+ W \barnab_a V \,, \label{x1stlaw}
\ee
which serves as the unified first law with $x$-dependence.

An alternative expression for the MS mass and the unified first law is integrating out $x$ to define the average MS mass, concretely,
\be
\begin{split}
	\mms(u) =&~  \frac{(D-2)\,\Omega}{16\pi G_N}r^{D-3} (k-I^{rr}) -\frac{\Lambda}{8\pi G_N}\,\frac{\Omega\, r^{D-1}}{D-1}
	\,, \label{avgledMSmass}
\end{split}
\ee
in which $k$ is the average $\cK$ in the sense of 
\be
k=\Omega^{-1} \int_{\Mco2} \cK(x)\sqrt{\ghat(x)}\, d^{D-2}x \,,
\ee
and define the average work term
\be
w(u) = \Omega^{-1} \int_{\Mco2} W(u,x)\sqrt{\ghat(x)}\, d^{D-2}x \,.
\ee  
On the other hand, the energy supply vector in GR does not depend on $x$ according to Eq.\eqref{radiueomTrless}. There is no need to define the average energy supply vector $\psi^a$ since $\psi^a$ should be the same as $\Psi^a$. Therefore, the unified first law has the following average version
\be
\barnab_a \mms
= A\,\psi_a+ w \barnab_a V \,. \label{1stlaw}
\ee
We conclude that two versions of the unified first law are needed to include shapes for $\Mco2$ with a non-constant Ricci scalar. The non-average first law \eqref{x1stlaw} and the average one \eqref{1stlaw} share the same $A\psi$ term. It would be interesting to compare the holographic viewpoint. If we treat $\Mco2$ with fixed $r$ as the holographic screen with a fixed $r$, which is similar to the screen defined in Refs.\cite{Tian:2014goa, Tian:2018hlw}, then it is natrual to view $\Mms/\Omega$, $A/\Omega$ and $V/\Omega$ as some screen densities but the $A\psi_a$ term in Eq. \eqref{x1stlaw} will be replaced by the $r^{D-2}\psi_a$ term.

\subsection{Thermodynamics Method}  

If all matter sources contribute an energy-momentum tensor satisfying the following two conditions: 
(i) the sum of energy supply vectors vanishes, i.e., $\sum_{(i)}\psi_{(i)}^a=0$\,;
(ii) the sum of average work terms $\wtot \equiv\sum_{(i)} w_{(i)}$ only depends on $r$ except the situation of $\wtot(r)=(D-2)(D-3)k/(16\pi G_N\,r^2)-\Lambda/(8\pi G_N)$\,, 
then the line-element $d\bar{s}^2=I_{ab}du^adu^b$ is determined as 
\be
d\bar{s}^2=-I^{rr}dt^2+ \frac{dr^2}{I^{rr} } \,, \label{gttgrr=-1}
\ee
where the function $I^{rr}$ should be
\be
\begin{split}
I^{rr}=&~ k-\frac{16\pi G_N }{(D-2)\,r^{D-3}}\big( M/\Omega 
\\&\qquad\quad +\int^{r} \wtot(\xi) \xi^{D-2}d\xi \big)
-\frac{2\Lambda r^2}{(D-1)(D-2)} \,.\label{generalIrr}
\end{split}
\ee

It is straightforward to obtain Eq.\eqref{generalIrr} via the average unified first law. The average unified first law under conditions (i) and (ii) gives
\be
\barnab_a\mms = \wtot(r)\,\Omega\, r^{D-2}\barnab_ar \,. \label{mmspurer}
\ee
Since the sum of average work terms serves as a function of $r$, directly integrating $r$ implies 
\be
\mms(r)= M+ \Omega\int^{r} \wtot(\xi) \xi^{D-2}d\xi \,, \label{onlyworktermmms}
\ee
where $M$ is the mass parameter that can absorb the integral constant from the second term. Therefore Eq.\eqref{avgledMSmass} implies that $I^{rr}$ should be Eq.\eqref{generalIrr}.

The next task is to confirm Eq.\eqref{gttgrr=-1}. A concrete calculation under the Eddinton-Finkelstein-like coordinates makes it explicit. Appendix B gives some useful results. Firstly, the line-element $I_{ab}du^adu^b$ can be generally written as
\be
d\bar{s}^2=-\frac{f(v,r)}{\sigma^2(v,r)} dv^2 +\frac{2dvdr}{\sigma(v,r)} \,,\label{dsbarEF}
\ee
where the function $f(v,r)$ is exactly $I^{rr}$.
Such coordinates frame can be always chosen on $\M2$ thus respecting the generality. 
In addition, the Laplacian of $r$ on $\M2$ is $\barnab^2 r=f'-f\sigma'/\sigma $. Thus, the Eq.\eqref{radiueomTrless} forces the total energy supply vector to become as
\be
\Psi_a=  - \frac{D-2}{16\pi G_N} \frac{f}{r}\frac{\sigma'}{\sigma}   \barnab_a r  \,. 
\ee
Obviously, $\Psi_a=0$ if $\sigma'=0$. In this case, $\sigma$ is at least a non-vanishing function of $v$.The Appendix B also explains why we should have $\sigma \neq 0$. Despite the concrete expression, the coodinate trasformation $t\equiv \int^v d\eta/\sigma(\eta) +\int^r d\xi/f(\xi) $ changes Eq.\eqref{dsbarEF} as Eq.\eqref{gttgrr=-1}. On the other hand, it is worth noting that the case of $I^{rr}=0$ may ruin such a proof. Fortunately, the condition $I^{rr}=0$ is too strong such that the average MS mass is fixed as $\mms=(D-2)k\Omega r^{D-3}/(16\pi G_N) -\Lambda\Omega r^{D-1}/(8(D-1)\pi G_N) $. Therefore the work term becomes $\wtot=(D-2)(D-3)k/(16\pi G_N\,r^2)-\Lambda/(8\pi G_N)$ which we have excluded in the condition (ii). 

This method simplifies solving the $ab$ components of Einstein equations. 
Hence it simplifies the proof of Birkhoff's theorem. A spherically symmetric $D=4$ spacetime is a warped product $\M2\times S^2$ because the spherical symmetry indicates the spacetime can be foliated by a set of orbits of the $SO(3)$ rotation group, i.e., a set of spheres  (see Ref.\cite{HawkingEllis}). Thus the metric should be Eq.\eqref{decom} while the $\ghat_{ij}$ serves as the metric of a unit sphere with $k=1$.
The vacuum condition implies $\psi_a=0$ and $W=0$ hence \emph{the MS mass is a constant} $M$. Therefore the above proof forces the line-element $d\bar{s}^2$ becoming $d\bar{s}^2=-fdt^2+dr^2/f$ where the function $f$ is $1-2G_N M/r$. Combinding with the line element of a unit sphere $d\Omega^2=\ghat_{ij}dx^idx^j$, the whole metric for the solution is $ds^2=-(1-2G_N M/r)dt^2+dr^2/(1-2G_N M/r)+r^2d\Omega^2$, namely, the Schwarzchild metric. Finally, we should substitute this result to the constraint equations Eq.\eqref{angulareom} to ensure it is satisfied. Such a proof for Birkhoff's theorem can be easily generalized to higher dimensions and the situation with a cosmological constant.
Moreover, this method hints at a simplified construction and a probably higher dimensional generalization for our solutions. 
We will then check the Maxwell equation for gauge field $A_{\mu}$ and KG equations for axions $\psi^I$, then find out their energy supply vectors and work terms.

\subsubsection*{Maxwell field}
A simple ansatz for the gauge 1-form $A= - \phi(v,r) dv $ will solve the Maxwell equation without knowing details about the metric. This ansatz implies the strength 2-form $F=dA$ should be $F=-\phi' dr\wedge dv $. Then read non-vanishing componetnts $F_{rv}=-F_{vr}=-\phi'$. Notice that $F^{vr}=I^{vr}I^{rv}F_{rv}$,
the Maxwell equations implies 
\be
\frac{\partial}{\partial v}\big(  r^{D-2} \sigma \phi'  \big)=  \frac{\partial}{\partial r}\big(  r^{D-2} \sigma \phi'  \big)=0 \,,
\ee
such that the electric field strength 
\be
-\phi' = \frac{4\pi Q/\Omega}{ \sigma\,r^{D-2}} \,,
\ee
is obtained without knowing the concrete expression of functions $f$ and $\sigma$ in the metric.
On the other hand, the energy-momentum tensor for the Maxwell field in $D$ dimension is the same with Eq.\eqref{emTmunus}. We thus calculate its non-vanishing components as
\be
\begin{split} 
&T_{vv}=2\pi(\frac{Q/\Omega}{ r^{D-2}})^2 \frac{f}{\sigma^2}  \,,\;   \\&T_{vr}=T_{rv}=-2\pi(\frac{Q/\Omega}{ r^{D-2}})^2 \frac{1}{\sigma}  \,,\;   
\\& T_{ij}=2\pi\frac{(Q/\Omega)^2}{r^{2(D-3)}} \ghat_{ij} \,,
\end{split}
\ee
Then the work term for the electromagnetic field are 
\be
W_{\text{em}}= 2\pi\frac{(Q/\Omega)^2}{r^{2(D-2)}}\,, \label{wtermem}
\ee
while its energy supply vector vanishes, namely, $\Psi_{\text{em}}^a=0$. Since $W_{\text{em}}$ does not rely $x$, the averaged work term $w_{\text{em}}$ is the same.

\subsubsection*{Linear axions}
Consider ansatz $\psi^I = \alpha \delta^I_i x^i$. Scalars satisfy $\nabla^2\psi^I=0$ if coordinates $x^i$ are harmonic. To simply their energy-momentum tensor, define
\be
\Phi_{ij} = \delta_{IJ} \frac{\partial \psi^I}{\partial x^i} \frac{\partial \psi^J}{\partial x^j} \,,
\ee
and label $\Phi(x)$ as the trace of $\Phi_{ij}$, i.e., $\Phi(x)\equiv\ghat^{ij}(x)\Phi_{ij}$.
We keep the expression $\Phi(x)$ and $\ghat^{ij}(x)$ to remind us that they may depend on $x$. Hence the axions contain the following energy-momentum tensor
\be 
T_{ab} = -\frac{\Phi(x)}{2r^2}\,I_{ab} \,,\quad T_{ij}= \Phi_{ij} - \frac{\Phi(x)}{2} \ghat_{ij}(x)\,,
\ee
in which $T_{ab}$ conponents only contain trace part. Thus it also has a vanishing energy supply vector $\Psi_{\text{axions}}^a=0$. 
While its work term is
\be
W_{\text{axions}}= \frac{\Phi(x)}{2r^2}\,, \label{wtermxaxions}
\ee
which may depend on $x$. Simply smear it by integral out $x$, then we obtain the averaged work term 
\be
w_{\text{axions}}= \frac{\phi_a}{2r^2}\,, \label{wtermaxions}
\ee
in which $\phi_a\equiv\Omega^{-1}\int\Phi(x)\sqrt{\ghat} d^{D-2}x$.

The above thermodynamics method is valid since the electric field and those axion profiles lead to a vanishing total energy supply vector. Summing up all work terms as functions of $r$, Eq.\eqref{wtermem} and Eq.\eqref{wtermaxions} will contribute
\be
\begin{split}
I^{rr}&= k-\frac{16\pi G_N }{(D-2)\,r^{D-3}}\big( M/\Omega - \frac{1}{D-3}\frac{(4\pi Q/\Omega)^2}{2r^{D-3}} 
\\&\qquad\quad+ \frac{\phi_a}{2(D-3)}r^{D-3} \big)
-\frac{2\Lambda r^2}{(D-1)(D-2)} 
\\&= c-\frac{16\pi G_N }{(D-2)}\frac{M/\Omega}{r^{D-3}} + \frac{2 G_N }{(D-2)(D-3)}\frac{(4\pi Q/\Omega)^2}{r^{2(D-3)}} 
\\&\qquad\quad - \frac{2\Lambda r^2}{(D-1)(D-2)} \,.\label{highdimsolutions}
\end{split}
\ee
where $c=k-8\pi G_N\,\phi_a/((D-2)(D-3))$. Hence the metric on $\M2$ is $d\bar{s}^2=-I^{rr}dt^2+ dr^2/I^{rr}$. 
It is worth noting that the $x-$dependent unified first law gives
\be
c=\cK(x) - \frac{8\pi G_N \,\Phi(x)}{(D-2)(D-3)}\,, \label{constterminIrr}
\ee
which indicates that even though $\cK$ and $\Phi$ may depend on $x$ but $c$ is a constant.
Then we will check the constrain equation Eq.\eqref{angulareom} which leads to
\bea
&\Rhat_{ij}-\frac{(D-2)(D-3)}{2}\cK\ghat_{ij}+(D-3)(D-4)\,\frac{c}{2}\,\ghat_{ij} \nonumber\\&=8\pi G_N\,\big( \Phi_{ij} -\frac{1}{2}\Phi\ghat_{ij}\big) \,.\label{arbShape}
\eea
If $D>4$, the trace of Eq.\eqref{arbShape} matches Eq.\eqref{constterminIrr}. Up to here, the Maxwell equation is solved; KG equations are solved by choosing harmonic coordinates, and the thermodynamics method simplifies solving the $ab$-components of the Einstein equation. Thus Eq.\eqref{arbShape} is the only equation left to be solved for higher dimensional generalization of solutions Eq.\eqref{Solu1} and Eq.\eqref{Solu2}.

On the other hand, if $D=4$, there is $\Rhat_{ij}=\cK\ghat_{ij}$ because of the two-dimensional transverse space. Moreover, the LHS of Eq.\eqref{arbShape} is zero since the term with $c$ vanishes in the case of $D=4$. Therefore, $\Phi_{ij}-\Phi\ghat_{ij}/2=0$ is the constrain condiction for axions and the spatial geometry $\ghat_{ij}$. A nontrivial geometry with non-constant $\cK$ implies a non-constant $\Phi$. Hence the condition $\Phi_{ij}-\Phi\ghat_{ij}/2=0$ excludes the possibility of a single axion field. Instead, the geometry with non-constant $\cK$ requires at least two axions, like the theory described by the action Eq.\eqref{holographicAxions4dim}.

The above discussion justifies ansatzs Eq.\eqref{usualansatz} and Eq.\eqref{usualansatzalt} for specifying a concrete $\ghat_{ij}$. Eq.\eqref{usualansatz} is inspired by the unified expression $d\rho^2/(1-k \rho^2)+\rho^2d\varphi^2$ as the line-element for sphere $k=1$, plane $k=0$ and hyperbolic surface $k=-1$. Eq.\eqref{usualansatz} only introduces a deformed term $e(\rho)$ in $g_{\rho\rho}$. Though coordinate $\rho$ is not a harmonic function, it can be transformed as $p(\rho)$ such that the line-element $\ghat_{ij}dx^idx^j$ becomes $\rho^2(p)(dp^2+d\varphi^2)$, in which $\rho(p)$ is the inversed function for $p(\rho)$. Then $p$ is harmonic and $\psi^1=\alpha p$ solved the Laplace eqaution $\nabla^2\psi^1=0$. Finally, Eq.\eqref{diffeq2FromEin1st} is exactly derived from Eq.\eqref{constterminIrr}.

As for Eq.\eqref{usualansatzalt}, firstly, consider the line-elemnet $\exp(\lambda(x,y))(dx^2+dy^2)$ given by the harmonic coordinates $\{x,y\}$, then Eq.\eqref{constterminIrr} deduces to Eq.\eqref{genshapeEq}. Furthermore, we turn to the polar coordinate via $x=\rho\cos\varphi$ and $y=\rho\sin\varphi$. If one poses a rotational symmetry by requiring that $\partial/\partial\varphi$ is a Killing vector field, the constrain equation for the shape of transverse space, Eq.\eqref{genshapeEq}, further deduces to Eq.\eqref{altshapeEq}.





\subsection{Black hole thermodynamics}
Though the parameter $M$ is exactly the Arnowitt-Deser-Misner (ADM) mass in the asymptotic flat case, it seems difficult to be identified as the ``mass" for more general situations due to a lack of a suitable asymptotical structure. Nevertheless, the unified first law provides a quasi-local viewpoint without relying on the interpretation of global parameters. Relevant parameters like MS mass, work terms of electric field (Eq.\eqref{wtermem}), and axions (Eq.\eqref{wtermaxions}) are well-defined as quasi-local quantities. If let them take values on the horizon, the unified first law implies the first law of black hole thermodynamics in terms of such quasi-local parameters. For instance, consider a tiny falling energy package that goes through the trapping horizon which is defined as a hypersurface foliated by marginal surfaces. The horizon has vanishing expansion. Hence the equation $I^{rr}=0$ determines its location (see Appendix B). As shown in Fig.\ref{fallingintrappinghorizon}, during the accretion, the horizon begins as a Killing horizon and finally settles down as a new Killing horizon.
\begin{figure}[h]
\centering
\includegraphics[width=6.6cm]{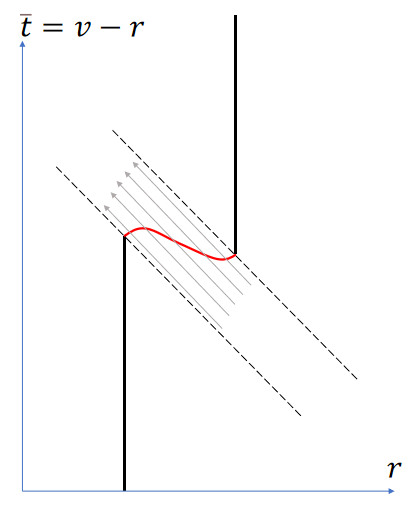}
\caption{\small The time coordinate $v$ is the null retarded time, while $\bar{t}$ is given by $v-r$. Such a sketch describes a falling process changing the size of the trapping horizon. The red curve represents the evolving tapping horizon. Gray arrows portray the in-falling energy package. Both the initial and final moment is marked by gray dotted lines while Killing horizons are indicated by black lines.} \label{fallingintrappinghorizon}
\end{figure}
To ensure the unified first law is valid, it is assumed that the $\Mco2$ part keeps unchanging. The trapping horizon only changes its size during the process.
Select a vector $z^a$ tangent to the trapping horizon, i.e., $(z^a\barnab_a I^{rr})_H=0$, and donate $\delta f=(z^a\barnab_a f)_H$, then according to Eq.\eqref{radiueomTrless}, contracting $z^a$ with the $ A \psi_a$ generates the $\kgeo \,\delta A/(8\pi G_N)$ term which is the equivalent expression of the heat flow term  $T\delta S$. The tunneling approach for Hawking radiation in dynamical spacetime confirms the relationship between $\kgeo$ at the horizon and the horizon temperature $T=\kappa_H/(2\pi)$\cite{Ren:2007xw, Ke-Xia:2009tzo, diCriscienzo:2010cp, DiCriscienzo:2009kun, Vanzo:2011wq}. Remarkably, two versions of the unified first law Eq.\eqref{x1stlaw} and Eq.\eqref{1stlaw} become
\be
\begin{split}
\delta M_{\text{MS}}-W_H\,\delta V_H &= \delta m_{\text{MS}}-w_H\,\delta V_H \
\\&= \frac{\kgeo \,\delta A}{\ivgcp} \,,  \label{1stlawonTH}
\end{split}
\ee
where $\kgeo$ is 
\be
\kgeo \equiv \frac{1}{2} \barnab^2 r \,,\label{kappageo}
\ee 
namely, the geometric surface gravity given in \cite{Hayward:1993wb, Hayward:1997jp}.
Suppose the difference between the final state and the initial state is tiny enough such that the changes for global parameters are $\delta M\simeq M_{\text{fin}} - M_{\text{ini}}$ and $\delta Q_{(i)}\simeq Q_{\text{i.fin}} - Q_{\text{i.ini}}$, etc. While $W_H$, $w_H$, and $\kgeo$ are taken values on the initial Killing horizon, especially $\kgeo=\kappa_H$ (see Appendix B). They should satisfy
\be
\delta M-\sum_i\Phi_{(i).H}\delta Q_{(i)}=\frac{\kappa_H \,\delta A}{\ivgcp} \,, \label{Global1st}
\ee
where $\Phi_{(i).H}$ should be treated as the thermo-conjugate quatity of $Q_{(i)}$.
Thus, there is
\be
\begin{split}
\delta M_{\text{MS}}-W_H\,\delta V_H &= \delta\mms(r_H)- w_H\,\delta V_{H}  \\&= \delta M-\sum_i\Phi_{(i),H}\delta Q_{(i)}\,, \label{quasiLocalGlobal}
\end{split}
\ee
which connects the quasi-local viewpoint with the global viewpoint (the rightmost) for the first law.
If apply Eq.\eqref{quasiLocalGlobal} to our solutions, one obtains
\be
\delta\Mms(r_H)- W_H\,\delta V_{H} = \delta M - \frac{4\pi Q}{\Omega r_H}\delta Q 
+\frac{\Omega r_H}{2}  \delta\Phi  \,,  \label{formal1stglb4sol}
\ee
in which $\delta\Lambda$ and $\delta\cK$ do not enter the global first law deduced from the unified first law shown by the RHS. It seems consistent with the restricted phase space formalism.
Despite the global mass parameter $M$ lacking satisfactory definitions at the present stage, the formal expression of the RHS in Eq.\eqref{formal1stglb4sol} still makes sense due to the well-defined unified first law.
Moreover, both the parameter from the curvature of the transverse space and the axionic parameter join the first law via Eq.\eqref{constterminIrr}. It indicates that we should consider the axionic charge and the new thermodynamical quantity from the curvature, no matter whether the new quantity is the topological charge, color charge, or center charge.
A more serious problem is that metrics in our solutions violate the translational symmetries along transverse directions. Thus it seems questionable to state ``homogeneity" from the first sight. However, the validity of the formal expression requires an appropriate interpretation. We expect the scaling property may be the more suitable starting point. Nevertheless, those solutions sharpen the issue of Euler homogeneity.

Nevertheless, the MS mass hints at the concept of usable energy for a black hole with a positive Ricci scalar, i.e., $\cK>0$. Suppose $r_H$ is the location of the black hole horizon.
Since the black hole horizon area does not decrease when suitable energy conditions are not violated, it is reasonable to treat the MS mass on the horizon $\Mms= (D-2)\Omega\,\cK r_H^{D-3}/(16\pi G_N) $ as the irreducible mass.
Then we further define the usable energy as $m_\text{usable}(r)=\mms(r)-m_\text{irr}$ outside the horizon $r>r_H$. The mass parameter $M$ does not appear in $m_\text{usable}(r)$. Instead, the difference of work terms between location $r$ and horizon $r_H$ determines such a new definition. Let $r$ tend to infinity, then one obtains the total usable energy.
The concept of usable energy provides an interesting understanding for Ref.\cite{Mai:2022pgf} which discussed the possibility of a Schwarzchild black hole as a battery. A charging process makes the black hole become an RN black hole. Hence the rest energy of the in-falling material is transformed into the usable energy of the black hole. While a suitable discharging process will extract such usable energy. Such an argument seems also valid when turning on the axions and cosmological constant.

\section{Summary}
In this article, we obtained two novel solutions Eq.\eqref{Solu1} and Eq.\eqref{Solu2} in the simplest holographic axion model in which the two axions serve as free scalars with canonical kinetic terms.
These novel solutions contain transverse spaces with non-constant Ricci scalars, distinguished from topological RN solutions or planar axionic solutions or solutions with a horizon geometry as an Einstein space\cite{Maeda:2016ddh}. When the cosmological constant is negative, these solutions can describe black holes with deformed horizon shapes.
We draw embedding diagrams for various situations. Their shapes usually contain a part embedded into a flat Euclidean space and some parts embedded into a Minkowski space. Solution I allows a regular transverse space, while a singular direction usually appears in other cases, including solution II.
We also calculate the DC conductivity for a topological black hole with a generally deformed shape. Compared to the planar axionic AdS black hole, the non-constant Ricci scalar cancels the direction-dependence of the axion configuration. Such a cancelation contributes to the constant that leads to the finite DC conductivity.

In addition, we re-derive the unified first law from Einstein equations in detail. The advantage of applying the unified first law is two-fold. Firstly, the unified first law extracts the crucial structure hidden in the Einstein equations. Based on such a structure, we improve the thermodynamics method proposed originally in Ref.\cite{Zhang:2013tca}  to adapt to the situations of deformed transverse spaces. It is an efficient method to construct solutions in which the metric is a warped product. The resulting constraint equation \eqref{constterminIrr} and \eqref{arbShape} hints at how to generalize Eq.\eqref{Solu1} and Eq.\eqref{Solu2}, i.e. the charged topological black holes with deformed horizon.
Secondly, the unified first law provides a quasi-local viewpoint for the horizon thermodynamics. The MS mass is a mathematically well-defined quasi-local mass. Despite the concrete meaning of the mass parameter $M$, one can always re-interpret the global first law of black hole thermodynamics as the well-defined unified first law on the horizon. In recent years, several anisotropic black holes with spacetime geometries beyond warped products have been obtained in Gauss-Bonnet gravity\cite{Peng:2021xwh, Peng:2022ttn}.
Moreover, the improved thermodynamics method requires a vanishing energy supply vector thus it is less valid for studying black holes with scalar hairs or other cases with hidden scalar degree of freedom\cite{Maeda:2015cia, Sato:2022yto, Huang:2019lsl, Huang:2020qmn}.
We expect the further generalized unified first law adapting situations including new horizon shapes, scalar hairs, and rotation (like Ref.\cite{Gundlach:2021six, Kinoshita:2021qsv}) may bring some unexpected insight into a deeper understanding of quasi-local energy\cite{Brown:1992br}, the relationship between thermodynamics and gravity\cite{Caldarelli:1999xj, Chakraborty:2015hna, Chakraborty:2015aja, Mahapatra:2020wym, Priyadarshinee:2021rch}, Kerr/CFT duality\cite{Guica:2008mu}, even quantum gravity\cite{Song:2020arr}.

Nevertheless, an understanding from the global viewpoint of our novel solutions is still lacking. The direction-dependent Ricci scalar $\Rhat(x)$ sharpens the issue of introducing new thermodynamical extended quantities to ensure the Euler homogeneity of the first law. To the best of our knowledge, this issue is attached by various approaches, including topological charge \cite{Tian:2014goa, Tian:2018hlw}, color charge \cite{Visser:2021eqk}, and center charge \cite{Wang:2022err}.
They play a similar role for topological RN black holes as the axionic charge played for planar axionic. Furthermore, our solutions show that effects from the curvature and axions can co-exist. It seems a challenge to clarify the interconnection between these charges. We left this issue for future work.

\section*{Acknowledgments}
The author thanks Ali Akil, Yusen An,  Zhongying Fan, Hyat Huang, Yuxuan Peng, Hongwei Tan, and Junlan Xian for their constructive suggestions. It is also grateful for supportments of Mayumi Aoki and Ryoko Nishikawa regarding personnel affairs when the author was at Kanazawa University, and the invitation from Yi Wang to visit HKUST Jockey Club Institute for Advanced Study. 
\setcounter{secnumdepth}{0} 

\section{Appendix A: Calculate $\Gamma^{\lambda}_{\;\mu\nu}$ and $R^{\lambda}_{\;\rho\mu\nu}$}
\setcounter{equation}{0}
\renewcommand\theequation{A.\arabic{equation}}

This appendix shows an efficient method for calculating the  Levi-Civita connection and Riemann tensor. 
Usually, the components of the Levi-Civita connection for a given metric, are calculated by
\begin{equation}
\Gamma^{\lambda}_{\mu\nu} = \frac{1}{2}g^{\lambda\sigma} (\frac{\partial g_{\mu\sigma}}{\partial x^{\nu}}+\frac{\partial g_{\nu\sigma}}{\partial x^{\mu}}-\frac{\partial g_{\mu\nu}}{\partial x^{\sigma}}) \,,\label{oldGam}
\end{equation}
The geodesic equations give hints to finding the trick. Consider the geodesic equations
\begin{equation}
\begin{split}
	&\frac{d^2x^{\lambda}}{d\tau^2}+ \Gamma^{\lambda}_{\mu\nu}\frac{dx^{\mu}}{d\tau} \frac{dx^{\nu}}{d\tau} \\&= \frac{d^2x^{\lambda}}{d\tau^2}+g^{\lambda\sigma}(\frac{d g_{\sigma\nu}}{d\tau} \frac{dx^{\nu}}{d\tau}- \frac{1}{2}\frac{\partial g_{\mu\nu}}{\partial x^{\sigma}}\frac{dx^{\mu}}{d\tau} \frac{dx^{\nu}}{d\tau}) =0  \,, 
\end{split}
\end{equation}
in which $d g_{\sigma\nu}/d\tau = (\partial g_{\sigma\nu}/\partial x^{\mu})(dx^{\mu}/d\tau)$.
If the second-order derivative term is ignored, then the structure
\begin{equation}
\Gamma^{\lambda}_{\mu\nu}dx^{\mu} dx^{\nu}= g^{\lambda\sigma}(d g_{\sigma\nu} dx^{\nu}- \frac{1}{2}\frac{\partial ds^2}{\partial x^{\sigma}})  \,, \label{trickGam}
\end{equation}
is extracted.
It serves as a coordinate-dependent rank-2 symmetric tensor, in which $\partial ds^2 / \partial x^{\sigma}$ is the short notation of $(\partial g_{\mu\nu}/\partial x^{\sigma})dx^{\mu} dx^{\nu}$. Since the coordinates frame is fixed under a particular calculation, one can simply treat $g_{\sigma\nu}$ as several functions of $x^{\mu}$ and $dg_{\sigma\nu}$ represents their differential. The trick is to calculate the structure Eq.~\eqref{trickGam} rather than to calculate components of Eq.~\eqref{oldGam} one by one. Now we use this trick to calculate the Levi-Civita connection of the metric \eqref{decom}.
The components of its inverse metric are
\be
g^{ab}=I^{ab} \,,\quad g^{ij}=\fft{\ghat^{ij}}{r^2} \,.
\ee
Thus, Eq.~\eqref{trickGam} gives
\bea
\begin{split}
\Gamma^{a}_{\;\mu\nu} dx^{\mu}dx^{\nu} =&~ \bar{\Gamma}^a_{\;bc}du^bdu^c - (r \barnab^a r) \ghat_{ij}dx^idx^j \,,\; 
\\ \Gamma^{i}_{\;\mu\nu} dx^{\mu}dx^{\nu} =&~ 2\fft{\barnab_a r}{r}du^adx^i + \hat{\Gamma}^i_{\;jk} dx^jdx^k
\,.
\end{split}
\eea
Here, the property $\partial r/\partial u^a = \barnab_a r$ is used. Noticing that product terms like $du^adx^i$ are the short notation for symmetric tensor product $(du^a dx^i \otimes dx^i du^a)/2$, every component can be correctly read as 
\bea
\begin{split}
&\Gamma^a_{\;bc}
=\bar{\Gamma}^a_{\;bc} \,,\;\;
\Gamma^a_{\;ij} = -rI^{ab}\fft{\partial r}{\partial u^b}\, \hat{g}_{ij} \,,\;\;
\\&\Gamma^i_{\;aj} = \fft{1}{r} \fft{\partial r}{\partial u^a} \delta^i_j\,,\;\;
\Gamma^i_{\;jk}  =\hat{\Gamma}^i_{\;jk}  \,. \label{dredGam}
\end{split}
\eea
It is worth noting that the components $\Gamma^a_{bc}$ and $\Gamma^i_{jk} $ are just the independent Levi-Civita connection of $\M2$ and $\Mco2$ respectively.
The author would like to introduce the covariant differential operator $\bar{\nabla}_{a}$ for $\M2$ and $\hat{\nabla}_{i}$ for $\Mco2$. The areal radius $r(u)$ can be treated as a scalar field in $\M2$. The notation $\barnab_a r$ also means $\del r/\del u^a$ while $\barnab^a r$ means $I^{ab}(\del r/\del u^b$).

Once the connection was obtained, the Riemann tensor can be calculated through $R^{\lambda}_{\;\rho\mu\nu}=\partial\Gamma^{\lambda}_{\;\nu\rho}/\partial x^{\mu} +\Gamma^{\lambda}_{\;\mu\sigma}\Gamma^{\sigma}_{\;\nu\rho} -(\mu\leftrightarrow\nu)$. It can also be treated as several 2-forms due to the anti-symmetry of exchanging $\mu$ and $\nu$, 
\be
\begin{split}
\fft{1}{2} R^{\lambda}_{\;\rho\mu\nu}dx^{\mu}\wedge dx^{\nu} 
= d\Gamma^{\lambda}_{\;\nu\rho}\wedge dx^{\nu} + (\Gamma^{\lambda}_{\;\mu\sigma}dx^{\mu})\wedge(\Gamma^{\sigma}_{\;\nu\rho}dx^{\nu})
\,. \label{cur2form_exp}
\end{split}
\ee
In order to simplify the notation, label $\fft{1}{2} R^{\lambda}_{\;\rho\mu\nu}dx^{\mu}\wedge dx^{\nu} $ as $\Omega^{\lambda}_{\;\rho}$ and $\Gamma^{\lambda}_{\;\mu\sigma}dx^{\mu}$ as $A^{\lambda}_{\;\rho}\,$, then 
\be
\Omega^{\lambda}_{\;\rho} = dA^{\lambda}_{\;\rho}+ A^{\lambda}_{\;\sigma}\wedge A^{\sigma}_{\;\rho}  \,. \label{cur2form}
\ee
These 1-forms $A^{\lambda}_{\;\rho}$ can be viewed as the connection 1-forms for the coordinates tetrad $(\partial_{\mu})^{\nu} = \delta^{\nu}_{\;\mu}$ while $\Omega^{\lambda}_{\;\rho}$ are their curvature 2-forms. Concretely, 1-forms $A^{\lambda}_{\;\rho}$ for the metric \eqref{decom} are
\be
\begin{split}
A^{a}_{\;b} =&~ \bar{\Gamma}^a_{\;cb}\, du^c =\bar{A}^a_{\;b}  \,,\\
A^{a}_{i} =& - (r \barnab^a r) \ghat_{ij}dx^j \,,\quad
A^{i}_{\;a} = \fft{\barnab_a r}{r}dx^i  \,,\\
A^{i}_{\;j} =&~ \fft{\barnab_a r}{r} \delta^{i}_{\;j} du^a + \hat{\Gamma}^i_{\;kj}dx^k = \fft{d r}{r} \delta^{i}_{\;j} + \hat{A}^i_{\;j} \,.
\end{split}
\ee
Then one obtains curvature 2-forms by applying Eq~.\eqref{cur2form}. Firstly, 
\be
\Omega^{a}_{\;b} = dA^{a}_{\;b}+ A^{a}_{\;c}\wedge A^{c}_{\;b} +A^{a}_{\;i}\wedge A^{i}_{\;b} 
= ~ \bar{\Omega}^{a}_{\;b} \,,
\ee
since term containing $\ghat_{ij} dx^i\wedge dx^j$ vanishes. Notice that $\Omega_{ai} =I_{ab}\Omega^b_{\;i} = - \Omega_{ia}= -r^2\ghat_{ij}\Omega^j_{\;a}$, calculating $\Omega^i_{\;a}$ can avoid dealing with $d\ghat_{ij}$ here:
\be
\begin{split}
\Omega^{i}_{\;a} =~ \fft{\barnab_a\barnab_b r}{r} du^b\wedge dx^i  \,,
\end{split}
\ee
therefore, 
\be
\Omega^{a}_{\;i} =-r\barnab^a\barnab_b r \,\ghat_{ij}\,du^b\wedge dx^j   \,,
\ee
The final 2-form $\Omega^i_{\;j}$ is 
\be
\begin{split}
\Omega^{i}_{\;j} &=\hat{\Omega}^{i}_{\;j} - I^{rr}\delta^i_{\;k}\,\ghat_{jl}\,dx^k\wedge dx^l 
\\&=\fft{1}{2} \big( \Rhat^{i}_{\;jkl} - I^{rr}(\delta^i_{\;k}\,\ghat_{jl}-\delta^i_{\;l}\,\ghat_{jk})\big)\,dx^k\wedge dx^l \,,
\end{split}
\ee
where the term $I^{rr}$ is the short notation for $I^{ab}\barnab_a r\barnab_b r$. Reminding $dx^{\mu}\wedge dx^{\nu}=dx^{\mu}\otimes dx^{\nu} -dx^{\nu}\otimes dx^{\mu}$, one can read the components of Riemann tensor as
\be
\begin{split}
R^a_{\;bcd}& = \bar{R}^a_{\;bcd} \,,\qquad
\\R^a_{\;ibj} &= -R^a_{\;ijb} = -r (\barnab^a \barnab_b r)\,\ghat_{ij} \,,\;
\\R^i_{\;ajb} &= -R^i_{\;abj} = -\fft{\barnab_a \barnab_b r}{r}\delta^i_j \,, \quad
\\R^i_{\;jkl} &= \hat{R}^i_{\;jkl}- I^{rr}(\delta^i_{\;k}\hat{g}_{jl}-\delta^i_{\;l}\hat{g}_{jk})\,.
\label{dimreduc}
\end{split}
\ee
The same result can be found in Ref.~\cite{Maeda:2007uu}. Contracting $\delta^i_{\;i}=D-2$, components of the Ricci tensor are
\begin{equation}
\begin{split}
	R_{ab}&=\bar{R}_{ab}-(D-2)\frac{\bar{\nabla}_a \bar{\nabla}_b r}{r} \;,\; \\
	R_{ij}&=\hat{R}_{ij}-\hat{g}_{ij}\big(r\bar{\nabla}^2 r
	+(D-3)I^{rr} \big) \;,\label{dimreducRic}
\end{split}
\end{equation}
in which the $\bar{\nabla}^2 r$ is $\bar{\nabla}^a \bar{\nabla}_a r$ for short. The Ricci scalar is
\begin{equation}
\begin{split}
	R &= I^{ab}R_{ab} + \frac{\hat{g}^{ij}}{r^2}R_{ij}
	\\&= \bar{R} -2(D-2)\frac{\bar{\nabla}^2 r}{r}
	+ \frac{\hat{R}}{r^2} -(D-2)(D-3)\frac{I^{rr}}{r^2} \,.\label{RicciSca}
\end{split}
\end{equation} 
This paper further focuses on 2-dimensional $\M2$. Since any 2-dimensional metric is conformally flat (see \cite{WaldGR, LiangGR}), the Einstein tensor of a 2-dimensional metric always vanishes, i.e., $\bar{R}_{ab}-\Rbar I_{ab}/2=0$. 
In addition, we introduce $\cK=\Rhat/((D-2)(D-3))$ for simplicity.
Therefore, the Einstein tensor $G_{\mu\nu}=R_{\mu\nu}- Rg_{\mu\nu}/2$ becomes
\begin{align}
G_{ab}=&-\frac{D-2}{r} (\barnab_a \barnab_b r - \frac{1}{2}I_{ab}\, \barnab^2 r) \nonumber \\
&+\frac{D-2}{2r} I_{ab}\,\big(\barnab^2r -\frac{D-3}{r}(\cK-I^{rr}) \big) \nonumber \;,\\
G_{ij}=&~\hat{R}_{ij} -(D-3)\cK\hat{g}_{ij}  
-\hat{g}_{ij}\big( \frac{r^2}{2} \Rbar  \nonumber \\
&\qquad -(D-3)\, r\barnab^2 r
+\frac{(D-3)(D-4)}{2}(\cK-I^{rr}) \big) \;, \label{dimreducEin} 
\end{align}
where we have separated the traceless part and trace part explicitly.  

\section{Appendix B: General Eddington-Finkelstein coordinates}
\setcounter{equation}{0}
\renewcommand\theequation{B.\arabic{equation}}
The 2-dimensional sub-spacetime must permits double null coordinates $\{u,v\}$  such that the line element becomes
\be
ds^2= -\Omega^2(u,v)dudv +r^2(u,v)\hat{g}_{ij}(x)dx^idx^j \,. \label{doublenull}
\ee
In the region where $\barnab_a r$ does not vanish, $r$ itself can be a coordinate, such that one can change to other coordinates frame like $\{v,r\}$. Since $dr=r_{,u}\,du+r_{,v}\, dv$ where $r_{,u}$, $r_{,v}$ represents $\partial r/\partial u$, $\partial r/\partial v$, the line element becomes 
\be
ds^2=\,\Omega^2\frac{ r_{,v}}{r_{,u}} dv^2 -\frac{\Omega^2}{r_{,u}}drdv +r^2\hat{g}_{ij}(x)dx^idx^j 
\,. \label{toFinEdd}
\ee
The line element \eqref{toFinEdd} is still general. The coordinates frame $\{v,r\}$ is called general Eddington-Finkelstein (GEF) coordinates in this article. Define functions $\sigma(v,r)=-2r_{,u}/\Omega^2$ and $f(v,r)=-4r_{,u}r_{,v}/\Omega^2$, metric components under the GEF coordinates are
\be
\begin{split}
&g_{vv}=I_{vv}=-\frac{f(v,r)}{\sigma^2(v,r)} \,,\quad   g_{rr}=I_{rr}=0 \,, \\& g_{vr}=g_{rv}=I_{vr}=I_{rv}=\frac{1}{\sigma(v,r)} \,,\quad   g_{ij}=r^2\ghat_{ij} \,,
\end{split}
\ee
while inverse metric components are  
\be
\begin{split}
&g^{vv}=I^{vv}=0 \,,\quad   g^{rr}=I^{rr}=f(v,r) \,, \\& g^{vr}=g^{rv}=I^{vr}=I^{rv}=\sigma(v,r) \,,\quad   g^{ij}=\frac{\ghat^{ij} }{r^2}\,.
\end{split}
\ee

In general, the shape of spacetime $\M2 \times \Mco2$ with a given metric $\ghat_{ij}$ for the unit $\Mco2$ is described by two functions. In double null coordinates, they are $\Omega(u,v)$ and $r(u,v)$, while in GEF coordinates, they are $\sigma(v,r)$ and $f(v,r)$. The usage of function $f(v,r)$ is convenient since it picks up the important function $I^{rr}=\nabla^{\mu}r\nabla_{\mu}r$. Further, the determinant of $I_{ab}$ in the GEF coordinates is simply $I=-\sigma^2$. Then $\sqrt{-I}$ equals to $\sigma^{-1}$ up to a sign. Therefore, the Laplacian of $r$ in $\M2$ is
\be
\barnab^2 r= \sigma\frac{\partial (\sigma^{-1}I^{vr})}{\partial v}+\sigma\frac{\partial(\sigma^{-1}I^{rr})}{\partial r} =f'- \frac{\sigma'}{\sigma}f\,,
\ee
where $f'=\partial f/\partial r$ and $\sigma'=\partial \sigma/\partial r$. The result leads to the following simple expression 
\be
\kgeo = \frac{1}{2}\barnab^2 r = \frac{f'}{2}- \frac{\sigma'}{\sigma}\frac{f}{2} \,,
\ee
for the geometric surface gravity~\eqref{kappageo} in terms of $f$ and $\sigma$:

Then we will calculate the expansions of null vector fields tangent to $\M2$. Specify those null fields as
\be
k^{\mu} \frac{\partial}{\partial X^{\mu} }= \frac{\partial}{\partial v} +\frac{f}{2\sigma } \frac{\partial}{\partial r}\,,\quad l^{\mu} \frac{\partial}{\partial X^{\mu} }= -\sigma\frac{\partial}{\partial r} \,,
\ee
and require the condition $k^{\mu}l_{\nu}=-1$. Assuming the increasing direction of $v$ is future, $k^{\mu}$ and $l^{\nu}$ are all future-pointed. Their expansion can be easily calculated by the trick $\theta_{(k)}=(D-2)\,k^a\barnab_ar/r$ without dealing with $g^{\mu\nu}\nabla_{\mu}k_{\nu}$  
\be
\theta_{(k)}=(D-2)\,\frac{f}{2\sigma r} \,,\quad \theta_{(l)}= -(D-2)\,\frac{\sigma}{r}\,. \label{expansionsGEF}
\ee
Such a method is also used in Ref.\cite{Cai:2006rs}. The hypersurface $I^{rr}=f=0$ leads to $\theta_{(k)}=0$, thus determining a trapping horizon, which is defined as a hypersurface foliated by marginal surfaces\cite{Hayward:1994bu, Hayward:1997jp}. A marginal surface is a two-codimensional spatial surface with vanishing expansion. One can further classify types of trapping horizons according to the behavior of $\theta_{(l)}$ and $\cL_{l}\theta_{(k)}$, see Refs.\cite{ Hayward:1994bu, Hayward:1997jp}.

\bibliography{reference}

\begin{thebibliography}{100}%
\makeatletter
\providecommand \@ifxundefined [1]{%
 \@ifx{#1\undefined}
}%
\providecommand \@ifnum [1]{%
 \ifnum #1\expandafter \@firstoftwo
 \else \expandafter \@secondoftwo
 \fi
}%
\providecommand \@ifx [1]{%
 \ifx #1\expandafter \@firstoftwo
 \else \expandafter \@secondoftwo
 \fi
}%
\providecommand \natexlab [1]{#1}%
\providecommand \enquote  [1]{``#1''}%
\providecommand \bibnamefont  [1]{#1}%
\providecommand \bibfnamefont [1]{#1}%
\providecommand \citenamefont [1]{#1}%
\providecommand \href@noop [0]{\@secondoftwo}%
\providecommand \href [0]{\begingroup \@sanitize@url \@href}%
\providecommand \@href[1]{\@@startlink{#1}\@@href}%
\providecommand \@@href[1]{\endgroup#1\@@endlink}%
\providecommand \@sanitize@url [0]{\catcode `\\12\catcode `\$12\catcode
  `\&12\catcode `\#12\catcode `\^12\catcode `\_12\catcode `\%12\relax}%
\providecommand \@@startlink[1]{}%
\providecommand \@@endlink[0]{}%
\providecommand \url  [0]{\begingroup\@sanitize@url \@url }%
\providecommand \@url [1]{\endgroup\@href {#1}{\urlprefix }}%
\providecommand \urlprefix  [0]{URL }%
\providecommand \Eprint [0]{\href }%
\providecommand \doibase [0]{https://doi.org/}%
\providecommand \selectlanguage [0]{\@gobble}%
\providecommand \bibinfo  [0]{\@secondoftwo}%
\providecommand \bibfield  [0]{\@secondoftwo}%
\providecommand \translation [1]{[#1]}%
\providecommand \BibitemOpen [0]{}%
\providecommand \bibitemStop [0]{}%
\providecommand \bibitemNoStop [0]{.\EOS\space}%
\providecommand \EOS [0]{\spacefactor3000\relax}%
\providecommand \BibitemShut  [1]{\csname bibitem#1\endcsname}%
\let\auto@bib@innerbib\@empty
\bibitem [{\citenamefont {Bekenstein}(1974)}]{Bekenstein:1974ax}%
  \BibitemOpen
  \bibfield  {author} {\bibinfo {author} {\bibfnamefont {J.~D.}\ \bibnamefont
  {Bekenstein}},\ }\bibfield  {title} {\bibinfo {title} {{Generalized second
  law of thermodynamics in black hole physics}},\ }\href
  {https://doi.org/10.1103/PhysRevD.9.3292} {\bibfield  {journal} {\bibinfo
  {journal} {Phys. Rev. D}\ }\textbf {\bibinfo {volume} {9}},\ \bibinfo {pages}
  {3292} (\bibinfo {year} {1974})}\BibitemShut {NoStop}%
\bibitem [{\citenamefont {Bardeen}\ \emph {et~al.}(1973)\citenamefont
  {Bardeen}, \citenamefont {Carter},\ and\ \citenamefont
  {Hawking}}]{Bardeen:1973gs}%
  \BibitemOpen
  \bibfield  {author} {\bibinfo {author} {\bibfnamefont {J.~M.}\ \bibnamefont
  {Bardeen}}, \bibinfo {author} {\bibfnamefont {B.}~\bibnamefont {Carter}},\
  and\ \bibinfo {author} {\bibfnamefont {S.~W.}\ \bibnamefont {Hawking}},\
  }\bibfield  {title} {\bibinfo {title} {{The Four laws of black hole
  mechanics}},\ }\href {https://doi.org/10.1007/BF01645742} {\bibfield
  {journal} {\bibinfo  {journal} {Commun. Math. Phys.}\ }\textbf {\bibinfo
  {volume} {31}},\ \bibinfo {pages} {161} (\bibinfo {year} {1973})}\BibitemShut
  {NoStop}%
\bibitem [{\citenamefont {Hawking}(1976)}]{Hawking:1976de}%
  \BibitemOpen
  \bibfield  {author} {\bibinfo {author} {\bibfnamefont {S.~W.}\ \bibnamefont
  {Hawking}},\ }\bibfield  {title} {\bibinfo {title} {{Black Holes and
  Thermodynamics}},\ }\href {https://doi.org/10.1103/PhysRevD.13.191}
  {\bibfield  {journal} {\bibinfo  {journal} {Phys. Rev. D}\ }\textbf {\bibinfo
  {volume} {13}},\ \bibinfo {pages} {191} (\bibinfo {year} {1976})}\BibitemShut
  {NoStop}%
\bibitem [{\citenamefont {Maldacena}(1998)}]{Maldacena:1997re}%
  \BibitemOpen
  \bibfield  {author} {\bibinfo {author} {\bibfnamefont {J.~M.}\ \bibnamefont
  {Maldacena}},\ }\bibfield  {title} {\bibinfo {title} {{The Large N limit of
  superconformal field theories and supergravity}},\ }\href
  {https://doi.org/10.1023/A:1026654312961} {\bibfield  {journal} {\bibinfo
  {journal} {Adv. Theor. Math. Phys.}\ }\textbf {\bibinfo {volume} {2}},\
  \bibinfo {pages} {231} (\bibinfo {year} {1998})},\ \Eprint
  {https://arxiv.org/abs/hep-th/9711200} {arXiv:hep-th/9711200} \BibitemShut
  {NoStop}%
\bibitem [{\citenamefont {Gubser}\ \emph {et~al.}(1998)\citenamefont {Gubser},
  \citenamefont {Klebanov},\ and\ \citenamefont {Polyakov}}]{Gubser:1998bc}%
  \BibitemOpen
  \bibfield  {author} {\bibinfo {author} {\bibfnamefont {S.~S.}\ \bibnamefont
  {Gubser}}, \bibinfo {author} {\bibfnamefont {I.~R.}\ \bibnamefont
  {Klebanov}},\ and\ \bibinfo {author} {\bibfnamefont {A.~M.}\ \bibnamefont
  {Polyakov}},\ }\bibfield  {title} {\bibinfo {title} {{Gauge theory
  correlators from noncritical string theory}},\ }\href
  {https://doi.org/10.1016/S0370-2693(98)00377-3} {\bibfield  {journal}
  {\bibinfo  {journal} {Phys. Lett. B}\ }\textbf {\bibinfo {volume} {428}},\
  \bibinfo {pages} {105} (\bibinfo {year} {1998})},\ \Eprint
  {https://arxiv.org/abs/hep-th/9802109} {arXiv:hep-th/9802109} \BibitemShut
  {NoStop}%
\bibitem [{\citenamefont {Witten}(1998)}]{Witten:1998qj}%
  \BibitemOpen
  \bibfield  {author} {\bibinfo {author} {\bibfnamefont {E.}~\bibnamefont
  {Witten}},\ }\bibfield  {title} {\bibinfo {title} {{Anti-de Sitter space and
  holography}},\ }\href {https://doi.org/10.4310/ATMP.1998.v2.n2.a2} {\bibfield
   {journal} {\bibinfo  {journal} {Adv. Theor. Math. Phys.}\ }\textbf {\bibinfo
  {volume} {2}},\ \bibinfo {pages} {253} (\bibinfo {year} {1998})},\ \Eprint
  {https://arxiv.org/abs/hep-th/9802150} {arXiv:hep-th/9802150} \BibitemShut
  {NoStop}%
\bibitem [{\citenamefont {Ryu}\ and\ \citenamefont
  {Takayanagi}(2006)}]{Ryu:2006bv}%
  \BibitemOpen
  \bibfield  {author} {\bibinfo {author} {\bibfnamefont {S.}~\bibnamefont
  {Ryu}}\ and\ \bibinfo {author} {\bibfnamefont {T.}~\bibnamefont
  {Takayanagi}},\ }\bibfield  {title} {\bibinfo {title} {{Holographic
  derivation of entanglement entropy from AdS/CFT}},\ }\href
  {https://doi.org/10.1103/PhysRevLett.96.181602} {\bibfield  {journal}
  {\bibinfo  {journal} {Phys. Rev. Lett.}\ }\textbf {\bibinfo {volume} {96}},\
  \bibinfo {pages} {181602} (\bibinfo {year} {2006})},\ \Eprint
  {https://arxiv.org/abs/hep-th/0603001} {arXiv:hep-th/0603001} \BibitemShut
  {NoStop}%
\bibitem [{\citenamefont {Wald}(1993)}]{Wald:1993nt}%
  \BibitemOpen
  \bibfield  {author} {\bibinfo {author} {\bibfnamefont {R.~M.}\ \bibnamefont
  {Wald}},\ }\bibfield  {title} {\bibinfo {title} {{Black hole entropy is the
  Noether charge}},\ }\href {https://doi.org/10.1103/PhysRevD.48.R3427}
  {\bibfield  {journal} {\bibinfo  {journal} {Phys. Rev. D}\ }\textbf {\bibinfo
  {volume} {48}},\ \bibinfo {pages} {R3427} (\bibinfo {year} {1993})},\ \Eprint
  {https://arxiv.org/abs/gr-qc/9307038} {arXiv:gr-qc/9307038} \BibitemShut
  {NoStop}%
\bibitem [{\citenamefont {Iyer}\ and\ \citenamefont
  {Wald}(1994)}]{PhysRevD.50.846}%
  \BibitemOpen
  \bibfield  {author} {\bibinfo {author} {\bibfnamefont {V.}~\bibnamefont
  {Iyer}}\ and\ \bibinfo {author} {\bibfnamefont {R.~M.}\ \bibnamefont
  {Wald}},\ }\bibfield  {title} {\bibinfo {title} {Some properties of the
  noether charge and a proposal for dynamical black hole entropy},\ }\href
  {https://doi.org/10.1103/PhysRevD.50.846} {\bibfield  {journal} {\bibinfo
  {journal} {Phys. Rev. D}\ }\textbf {\bibinfo {volume} {50}},\ \bibinfo
  {pages} {846} (\bibinfo {year} {1994})}\BibitemShut {NoStop}%
\bibitem [{\citenamefont {Hawking}(1972)}]{Hawking:1971vc}%
  \BibitemOpen
  \bibfield  {author} {\bibinfo {author} {\bibfnamefont {S.~W.}\ \bibnamefont
  {Hawking}},\ }\bibfield  {title} {\bibinfo {title} {{Black holes in general
  relativity}},\ }\href {https://doi.org/10.1007/BF01877517} {\bibfield
  {journal} {\bibinfo  {journal} {Commun. Math. Phys.}\ }\textbf {\bibinfo
  {volume} {25}},\ \bibinfo {pages} {152} (\bibinfo {year} {1972})}\BibitemShut
  {NoStop}%
\bibitem [{\citenamefont {Horowitz}\ and\ \citenamefont
  {Strominger}(1991)}]{Horowitz:1991cd}%
  \BibitemOpen
  \bibfield  {author} {\bibinfo {author} {\bibfnamefont {G.~T.}\ \bibnamefont
  {Horowitz}}\ and\ \bibinfo {author} {\bibfnamefont {A.}~\bibnamefont
  {Strominger}},\ }\bibfield  {title} {\bibinfo {title} {{Black strings and
  P-branes}},\ }\href {https://doi.org/10.1016/0550-3213(91)90440-9} {\bibfield
   {journal} {\bibinfo  {journal} {Nucl. Phys. B}\ }\textbf {\bibinfo {volume}
  {360}},\ \bibinfo {pages} {197} (\bibinfo {year} {1991})}\BibitemShut
  {NoStop}%
\bibitem [{\citenamefont {Vanzo}(1997)}]{Vanzo:1997gw}%
  \BibitemOpen
  \bibfield  {author} {\bibinfo {author} {\bibfnamefont {L.}~\bibnamefont
  {Vanzo}},\ }\bibfield  {title} {\bibinfo {title} {{Black holes with unusual
  topology}},\ }\href {https://doi.org/10.1103/PhysRevD.56.6475} {\bibfield
  {journal} {\bibinfo  {journal} {Phys. Rev. D}\ }\textbf {\bibinfo {volume}
  {56}},\ \bibinfo {pages} {6475} (\bibinfo {year} {1997})},\ \Eprint
  {https://arxiv.org/abs/gr-qc/9705004} {arXiv:gr-qc/9705004} \BibitemShut
  {NoStop}%
\bibitem [{\citenamefont {Galloway}\ \emph {et~al.}(1999)\citenamefont
  {Galloway}, \citenamefont {Schleich}, \citenamefont {Witt},\ and\
  \citenamefont {Woolgar}}]{Galloway:1999bp}%
  \BibitemOpen
  \bibfield  {author} {\bibinfo {author} {\bibfnamefont {G.~J.}\ \bibnamefont
  {Galloway}}, \bibinfo {author} {\bibfnamefont {K.}~\bibnamefont {Schleich}},
  \bibinfo {author} {\bibfnamefont {D.~M.}\ \bibnamefont {Witt}},\ and\
  \bibinfo {author} {\bibfnamefont {E.}~\bibnamefont {Woolgar}},\ }\bibfield
  {title} {\bibinfo {title} {{Topological censorship and higher genus black
  holes}},\ }\href {https://doi.org/10.1103/PhysRevD.60.104039} {\bibfield
  {journal} {\bibinfo  {journal} {Phys. Rev. D}\ }\textbf {\bibinfo {volume}
  {60}},\ \bibinfo {pages} {104039} (\bibinfo {year} {1999})},\ \Eprint
  {https://arxiv.org/abs/gr-qc/9902061} {arXiv:gr-qc/9902061} \BibitemShut
  {NoStop}%
\bibitem [{\citenamefont {Emparan}\ and\ \citenamefont
  {Reall}(2002)}]{Emparan:2001wn}%
  \BibitemOpen
  \bibfield  {author} {\bibinfo {author} {\bibfnamefont {R.}~\bibnamefont
  {Emparan}}\ and\ \bibinfo {author} {\bibfnamefont {H.~S.}\ \bibnamefont
  {Reall}},\ }\bibfield  {title} {\bibinfo {title} {{A Rotating black ring
  solution in five-dimensions}},\ }\href
  {https://doi.org/10.1103/PhysRevLett.88.101101} {\bibfield  {journal}
  {\bibinfo  {journal} {Phys. Rev. Lett.}\ }\textbf {\bibinfo {volume} {88}},\
  \bibinfo {pages} {101101} (\bibinfo {year} {2002})},\ \Eprint
  {https://arxiv.org/abs/hep-th/0110260} {arXiv:hep-th/0110260} \BibitemShut
  {NoStop}%
\bibitem [{\citenamefont {Elvang}\ and\ \citenamefont
  {Figueras}(2007)}]{Elvang:2007rd}%
  \BibitemOpen
  \bibfield  {author} {\bibinfo {author} {\bibfnamefont {H.}~\bibnamefont
  {Elvang}}\ and\ \bibinfo {author} {\bibfnamefont {P.}~\bibnamefont
  {Figueras}},\ }\bibfield  {title} {\bibinfo {title} {{Black Saturn}},\ }\href
  {https://doi.org/10.1088/1126-6708/2007/05/050} {\bibfield  {journal}
  {\bibinfo  {journal} {JHEP}\ }\textbf {\bibinfo {volume} {05}},\ \bibinfo
  {pages} {050}},\ \Eprint {https://arxiv.org/abs/hep-th/0701035}
  {arXiv:hep-th/0701035} \BibitemShut {NoStop}%
\bibitem [{\citenamefont {Emparan}\ and\ \citenamefont
  {Reall}(2008)}]{Emparan:2008eg}%
  \BibitemOpen
  \bibfield  {author} {\bibinfo {author} {\bibfnamefont {R.}~\bibnamefont
  {Emparan}}\ and\ \bibinfo {author} {\bibfnamefont {H.~S.}\ \bibnamefont
  {Reall}},\ }\bibfield  {title} {\bibinfo {title} {{Black Holes in Higher
  Dimensions}},\ }\href {https://doi.org/10.12942/lrr-2008-6} {\bibfield
  {journal} {\bibinfo  {journal} {Living Rev. Rel.}\ }\textbf {\bibinfo
  {volume} {11}},\ \bibinfo {pages} {6} (\bibinfo {year} {2008})},\ \Eprint
  {https://arxiv.org/abs/0801.3471} {arXiv:0801.3471 [hep-th]} \BibitemShut
  {NoStop}%
\bibitem [{\citenamefont {Mann}(1997)}]{Mann:1996gj}%
  \BibitemOpen
  \bibfield  {author} {\bibinfo {author} {\bibfnamefont {R.~B.}\ \bibnamefont
  {Mann}},\ }\bibfield  {title} {\bibinfo {title} {{Pair production of
  topological anti-de Sitter black holes}},\ }\href
  {https://doi.org/10.1088/0264-9381/14/5/007} {\bibfield  {journal} {\bibinfo
  {journal} {Class. Quant. Grav.}\ }\textbf {\bibinfo {volume} {14}},\ \bibinfo
  {pages} {L109} (\bibinfo {year} {1997})},\ \Eprint
  {https://arxiv.org/abs/gr-qc/9607071} {arXiv:gr-qc/9607071} \BibitemShut
  {NoStop}%
\bibitem [{\citenamefont {Birmingham}(1999)}]{Birmingham:1998nr}%
  \BibitemOpen
  \bibfield  {author} {\bibinfo {author} {\bibfnamefont {D.}~\bibnamefont
  {Birmingham}},\ }\bibfield  {title} {\bibinfo {title} {{Topological black
  holes in Anti-de Sitter space}},\ }\href
  {https://doi.org/10.1088/0264-9381/16/4/009} {\bibfield  {journal} {\bibinfo
  {journal} {Class. Quant. Grav.}\ }\textbf {\bibinfo {volume} {16}},\ \bibinfo
  {pages} {1197} (\bibinfo {year} {1999})},\ \Eprint
  {https://arxiv.org/abs/hep-th/9808032} {arXiv:hep-th/9808032} \BibitemShut
  {NoStop}%
\bibitem [{\citenamefont {Emparan}(1999)}]{Emparan:1999gf}%
  \BibitemOpen
  \bibfield  {author} {\bibinfo {author} {\bibfnamefont {R.}~\bibnamefont
  {Emparan}},\ }\bibfield  {title} {\bibinfo {title} {{AdS / CFT duals of
  topological black holes and the entropy of zero energy states}},\ }\href
  {https://doi.org/10.1088/1126-6708/1999/06/036} {\bibfield  {journal}
  {\bibinfo  {journal} {JHEP}\ }\textbf {\bibinfo {volume} {06}},\ \bibinfo
  {pages} {036}},\ \Eprint {https://arxiv.org/abs/hep-th/9906040}
  {arXiv:hep-th/9906040} \BibitemShut {NoStop}%
\bibitem [{\citenamefont {Birmingham}\ and\ \citenamefont
  {Mokhtari}(2007)}]{Birmingham:2007yv}%
  \BibitemOpen
  \bibfield  {author} {\bibinfo {author} {\bibfnamefont {D.}~\bibnamefont
  {Birmingham}}\ and\ \bibinfo {author} {\bibfnamefont {S.}~\bibnamefont
  {Mokhtari}},\ }\bibfield  {title} {\bibinfo {title} {{Stability of
  topological black holes}},\ }\href
  {https://doi.org/10.1103/PhysRevD.76.124039} {\bibfield  {journal} {\bibinfo
  {journal} {Phys. Rev. D}\ }\textbf {\bibinfo {volume} {76}},\ \bibinfo
  {pages} {124039} (\bibinfo {year} {2007})},\ \Eprint
  {https://arxiv.org/abs/0709.2388} {arXiv:0709.2388 [hep-th]} \BibitemShut
  {NoStop}%
\bibitem [{\citenamefont {Klemm}\ \emph {et~al.}(1998)\citenamefont {Klemm},
  \citenamefont {Moretti},\ and\ \citenamefont {Vanzo}}]{Klemm:1997ea}%
  \BibitemOpen
  \bibfield  {author} {\bibinfo {author} {\bibfnamefont {D.}~\bibnamefont
  {Klemm}}, \bibinfo {author} {\bibfnamefont {V.}~\bibnamefont {Moretti}},\
  and\ \bibinfo {author} {\bibfnamefont {L.}~\bibnamefont {Vanzo}},\ }\bibfield
   {title} {\bibinfo {title} {{Rotating topological black holes}},\ }\href
  {https://doi.org/10.1103/PhysRevD.60.109902} {\bibfield  {journal} {\bibinfo
  {journal} {Phys. Rev. D}\ }\textbf {\bibinfo {volume} {57}},\ \bibinfo
  {pages} {6127} (\bibinfo {year} {1998})},\ \bibinfo {note} {[Erratum:
  Phys.Rev.D 60, 109902 (1999)]},\ \Eprint
  {https://arxiv.org/abs/gr-qc/9710123} {arXiv:gr-qc/9710123} \BibitemShut
  {NoStop}%
\bibitem [{\citenamefont {Morley}\ \emph {et~al.}(2018)\citenamefont {Morley},
  \citenamefont {Taylor},\ and\ \citenamefont {Winstanley}}]{Morley:2018lwn}%
  \BibitemOpen
  \bibfield  {author} {\bibinfo {author} {\bibfnamefont {T.}~\bibnamefont
  {Morley}}, \bibinfo {author} {\bibfnamefont {P.}~\bibnamefont {Taylor}},\
  and\ \bibinfo {author} {\bibfnamefont {E.}~\bibnamefont {Winstanley}},\
  }\bibfield  {title} {\bibinfo {title} {{Vacuum polarization on topological
  black holes}},\ }\href {https://doi.org/10.1088/1361-6382/aae45b} {\bibfield
  {journal} {\bibinfo  {journal} {Class. Quant. Grav.}\ }\textbf {\bibinfo
  {volume} {35}},\ \bibinfo {pages} {235010} (\bibinfo {year} {2018})},\
  \Eprint {https://arxiv.org/abs/1808.04386} {arXiv:1808.04386 [gr-qc]}
  \BibitemShut {NoStop}%
\bibitem [{\citenamefont {Bai}\ and\ \citenamefont {Ren}(2022)}]{Bai:2022obp}%
  \BibitemOpen
  \bibfield  {author} {\bibinfo {author} {\bibfnamefont {X.}~\bibnamefont
  {Bai}}\ and\ \bibinfo {author} {\bibfnamefont {J.}~\bibnamefont {Ren}},\
  }\bibfield  {title} {\bibinfo {title} {{Holographic R\'enyi entropies from
  hyperbolic black holes with scalar hair}},\ }\href
  {https://doi.org/10.1007/JHEP12(2022)038} {\bibfield  {journal} {\bibinfo
  {journal} {JHEP}\ }\textbf {\bibinfo {volume} {12}},\ \bibinfo {pages}
  {038}},\ \Eprint {https://arxiv.org/abs/2210.03732} {arXiv:2210.03732
  [hep-th]} \BibitemShut {NoStop}%
\bibitem [{\citenamefont {Gutperle}\ and\ \citenamefont
  {Strominger}(2002)}]{Gutperle:2002ai}%
  \BibitemOpen
  \bibfield  {author} {\bibinfo {author} {\bibfnamefont {M.}~\bibnamefont
  {Gutperle}}\ and\ \bibinfo {author} {\bibfnamefont {A.}~\bibnamefont
  {Strominger}},\ }\bibfield  {title} {\bibinfo {title} {{Space - like
  branes}},\ }\href {https://doi.org/10.1088/1126-6708/2002/04/018} {\bibfield
  {journal} {\bibinfo  {journal} {JHEP}\ }\textbf {\bibinfo {volume} {04}},\
  \bibinfo {pages} {018}},\ \Eprint {https://arxiv.org/abs/hep-th/0202210}
  {arXiv:hep-th/0202210} \BibitemShut {NoStop}%
\bibitem [{\citenamefont {Tasinato}\ \emph {et~al.}(2004)\citenamefont
  {Tasinato}, \citenamefont {Zavala}, \citenamefont {Burgess},\ and\
  \citenamefont {Quevedo}}]{Tasinato:2004dy}%
  \BibitemOpen
  \bibfield  {author} {\bibinfo {author} {\bibfnamefont {G.}~\bibnamefont
  {Tasinato}}, \bibinfo {author} {\bibfnamefont {I.}~\bibnamefont {Zavala}},
  \bibinfo {author} {\bibfnamefont {C.~P.}\ \bibnamefont {Burgess}},\ and\
  \bibinfo {author} {\bibfnamefont {F.}~\bibnamefont {Quevedo}},\ }\bibfield
  {title} {\bibinfo {title} {{Regular S-brane backgrounds}},\ }\href
  {https://doi.org/10.1088/1126-6708/2004/04/038} {\bibfield  {journal}
  {\bibinfo  {journal} {JHEP}\ }\textbf {\bibinfo {volume} {04}},\ \bibinfo
  {pages} {038}},\ \Eprint {https://arxiv.org/abs/hep-th/0403156}
  {arXiv:hep-th/0403156} \BibitemShut {NoStop}%
\bibitem [{\citenamefont {Lu}\ and\ \citenamefont
  {Vazquez-Poritz}(2004)}]{Lu:2004ye}%
  \BibitemOpen
  \bibfield  {author} {\bibinfo {author} {\bibfnamefont {H.}~\bibnamefont
  {Lu}}\ and\ \bibinfo {author} {\bibfnamefont {J.~F.}\ \bibnamefont
  {Vazquez-Poritz}},\ }\bibfield  {title} {\bibinfo {title} {{Nonsingular
  twisted S-branes from rotating branes}},\ }\href
  {https://doi.org/10.1088/1126-6708/2004/07/050} {\bibfield  {journal}
  {\bibinfo  {journal} {JHEP}\ }\textbf {\bibinfo {volume} {07}},\ \bibinfo
  {pages} {050}},\ \Eprint {https://arxiv.org/abs/hep-th/0403248}
  {arXiv:hep-th/0403248} \BibitemShut {NoStop}%
\bibitem [{\citenamefont {Hartnoll}\ \emph
  {et~al.}(2008{\natexlab{a}})\citenamefont {Hartnoll}, \citenamefont
  {Herzog},\ and\ \citenamefont {Horowitz}}]{Hartnoll:2008vx}%
  \BibitemOpen
  \bibfield  {author} {\bibinfo {author} {\bibfnamefont {S.~A.}\ \bibnamefont
  {Hartnoll}}, \bibinfo {author} {\bibfnamefont {C.~P.}\ \bibnamefont
  {Herzog}},\ and\ \bibinfo {author} {\bibfnamefont {G.~T.}\ \bibnamefont
  {Horowitz}},\ }\bibfield  {title} {\bibinfo {title} {{Building a Holographic
  Superconductor}},\ }\href {https://doi.org/10.1103/PhysRevLett.101.031601}
  {\bibfield  {journal} {\bibinfo  {journal} {Phys. Rev. Lett.}\ }\textbf
  {\bibinfo {volume} {101}},\ \bibinfo {pages} {031601} (\bibinfo {year}
  {2008}{\natexlab{a}})},\ \Eprint {https://arxiv.org/abs/0803.3295}
  {arXiv:0803.3295 [hep-th]} \BibitemShut {NoStop}%
\bibitem [{\citenamefont {Hartnoll}\ \emph
  {et~al.}(2008{\natexlab{b}})\citenamefont {Hartnoll}, \citenamefont
  {Herzog},\ and\ \citenamefont {Horowitz}}]{Hartnoll:2008kx}%
  \BibitemOpen
  \bibfield  {author} {\bibinfo {author} {\bibfnamefont {S.~A.}\ \bibnamefont
  {Hartnoll}}, \bibinfo {author} {\bibfnamefont {C.~P.}\ \bibnamefont
  {Herzog}},\ and\ \bibinfo {author} {\bibfnamefont {G.~T.}\ \bibnamefont
  {Horowitz}},\ }\bibfield  {title} {\bibinfo {title} {{Holographic
  Superconductors}},\ }\href {https://doi.org/10.1088/1126-6708/2008/12/015}
  {\bibfield  {journal} {\bibinfo  {journal} {JHEP}\ }\textbf {\bibinfo
  {volume} {12}},\ \bibinfo {pages} {015}},\ \Eprint
  {https://arxiv.org/abs/0810.1563} {arXiv:0810.1563 [hep-th]} \BibitemShut
  {NoStop}%
\bibitem [{\citenamefont {Andrade}\ and\ \citenamefont
  {Withers}(2014)}]{Andrade:2013gsa}%
  \BibitemOpen
  \bibfield  {author} {\bibinfo {author} {\bibfnamefont {T.}~\bibnamefont
  {Andrade}}\ and\ \bibinfo {author} {\bibfnamefont {B.}~\bibnamefont
  {Withers}},\ }\bibfield  {title} {\bibinfo {title} {{A simple holographic
  model of momentum relaxation}},\ }\href
  {https://doi.org/10.1007/JHEP05(2014)101} {\bibfield  {journal} {\bibinfo
  {journal} {JHEP}\ }\textbf {\bibinfo {volume} {05}},\ \bibinfo {pages}
  {101}},\ \Eprint {https://arxiv.org/abs/1311.5157} {arXiv:1311.5157 [hep-th]}
  \BibitemShut {NoStop}%
\bibitem [{\citenamefont {Donos}\ and\ \citenamefont
  {Gauntlett}(2014)}]{Donos:2014cya}%
  \BibitemOpen
  \bibfield  {author} {\bibinfo {author} {\bibfnamefont {A.}~\bibnamefont
  {Donos}}\ and\ \bibinfo {author} {\bibfnamefont {J.~P.}\ \bibnamefont
  {Gauntlett}},\ }\bibfield  {title} {\bibinfo {title} {{Thermoelectric DC
  conductivities from black hole horizons}},\ }\href
  {https://doi.org/10.1007/JHEP11(2014)081} {\bibfield  {journal} {\bibinfo
  {journal} {JHEP}\ }\textbf {\bibinfo {volume} {11}},\ \bibinfo {pages}
  {081}},\ \Eprint {https://arxiv.org/abs/1406.4742} {arXiv:1406.4742 [hep-th]}
  \BibitemShut {NoStop}%
\bibitem [{\citenamefont {Arias}\ and\ \citenamefont
  {Salazar~Landea}(2017)}]{Arias:2017yqj}%
  \BibitemOpen
  \bibfield  {author} {\bibinfo {author} {\bibfnamefont {R.~E.}\ \bibnamefont
  {Arias}}\ and\ \bibinfo {author} {\bibfnamefont {I.}~\bibnamefont
  {Salazar~Landea}},\ }\bibfield  {title} {\bibinfo {title} {{Thermoelectric
  Transport Coefficients from Charged Solv and Nil Black Holes}},\ }\href
  {https://doi.org/10.1007/JHEP12(2017)087} {\bibfield  {journal} {\bibinfo
  {journal} {JHEP}\ }\textbf {\bibinfo {volume} {12}},\ \bibinfo {pages}
  {087}},\ \Eprint {https://arxiv.org/abs/1708.04335} {arXiv:1708.04335
  [hep-th]} \BibitemShut {NoStop}%
\bibitem [{\citenamefont {Jiang}\ \emph {et~al.}(2017)\citenamefont {Jiang},
  \citenamefont {Liu}, \citenamefont {Lu},\ and\ \citenamefont
  {Pope}}]{Jiang:2017imk}%
  \BibitemOpen
  \bibfield  {author} {\bibinfo {author} {\bibfnamefont {W.-J.}\ \bibnamefont
  {Jiang}}, \bibinfo {author} {\bibfnamefont {H.-S.}\ \bibnamefont {Liu}},
  \bibinfo {author} {\bibfnamefont {H.}~\bibnamefont {Lu}},\ and\ \bibinfo
  {author} {\bibfnamefont {C.~N.}\ \bibnamefont {Pope}},\ }\bibfield  {title}
  {\bibinfo {title} {{DC Conductivities with Momentum Dissipation in Horndeski
  Theories}},\ }\href {https://doi.org/10.1007/JHEP07(2017)084} {\bibfield
  {journal} {\bibinfo  {journal} {JHEP}\ }\textbf {\bibinfo {volume} {07}},\
  \bibinfo {pages} {084}},\ \Eprint {https://arxiv.org/abs/1703.00922}
  {arXiv:1703.00922 [hep-th]} \BibitemShut {NoStop}%
\bibitem [{\citenamefont {Esposito}\ \emph {et~al.}(2017)\citenamefont
  {Esposito}, \citenamefont {Garcia-Saenz}, \citenamefont {Nicolis},\ and\
  \citenamefont {Penco}}]{Esposito:2017qpj}%
  \BibitemOpen
  \bibfield  {author} {\bibinfo {author} {\bibfnamefont {A.}~\bibnamefont
  {Esposito}}, \bibinfo {author} {\bibfnamefont {S.}~\bibnamefont
  {Garcia-Saenz}}, \bibinfo {author} {\bibfnamefont {A.}~\bibnamefont
  {Nicolis}},\ and\ \bibinfo {author} {\bibfnamefont {R.}~\bibnamefont
  {Penco}},\ }\bibfield  {title} {\bibinfo {title} {{Conformal solids and
  holography}},\ }\href {https://doi.org/10.1007/JHEP12(2017)113} {\bibfield
  {journal} {\bibinfo  {journal} {JHEP}\ }\textbf {\bibinfo {volume} {12}},\
  \bibinfo {pages} {113}},\ \Eprint {https://arxiv.org/abs/1708.09391}
  {arXiv:1708.09391 [hep-th]} \BibitemShut {NoStop}%
\bibitem [{\citenamefont {Baggioli}\ and\ \citenamefont
  {Buchel}(2019)}]{Baggioli:2018bfa}%
  \BibitemOpen
  \bibfield  {author} {\bibinfo {author} {\bibfnamefont {M.}~\bibnamefont
  {Baggioli}}\ and\ \bibinfo {author} {\bibfnamefont {A.}~\bibnamefont
  {Buchel}},\ }\bibfield  {title} {\bibinfo {title} {{Holographic Viscoelastic
  Hydrodynamics}},\ }\href {https://doi.org/10.1007/JHEP03(2019)146} {\bibfield
   {journal} {\bibinfo  {journal} {JHEP}\ }\textbf {\bibinfo {volume} {03}},\
  \bibinfo {pages} {146}},\ \Eprint {https://arxiv.org/abs/1805.06756}
  {arXiv:1805.06756 [hep-th]} \BibitemShut {NoStop}%
\bibitem [{\citenamefont {Esposito}\ \emph {et~al.}(2020)\citenamefont
  {Esposito}, \citenamefont {Krichevsky},\ and\ \citenamefont
  {Nicolis}}]{Esposito:2020wsn}%
  \BibitemOpen
  \bibfield  {author} {\bibinfo {author} {\bibfnamefont {A.}~\bibnamefont
  {Esposito}}, \bibinfo {author} {\bibfnamefont {R.}~\bibnamefont
  {Krichevsky}},\ and\ \bibinfo {author} {\bibfnamefont {A.}~\bibnamefont
  {Nicolis}},\ }\bibfield  {title} {\bibinfo {title} {{Solidity without
  inhomogeneity: Perfectly homogeneous, weakly coupled, UV-complete solids}},\
  }\href {https://doi.org/10.1007/JHEP11(2020)021} {\bibfield  {journal}
  {\bibinfo  {journal} {JHEP}\ }\textbf {\bibinfo {volume} {11}},\ \bibinfo
  {pages} {021}},\ \Eprint {https://arxiv.org/abs/2004.11386} {arXiv:2004.11386
  [hep-th]} \BibitemShut {NoStop}%
\bibitem [{\citenamefont {Baggioli}\ \emph {et~al.}(2021)\citenamefont
  {Baggioli}, \citenamefont {Kim}, \citenamefont {Li},\ and\ \citenamefont
  {Li}}]{Baggioli:2021xuv}%
  \BibitemOpen
  \bibfield  {author} {\bibinfo {author} {\bibfnamefont {M.}~\bibnamefont
  {Baggioli}}, \bibinfo {author} {\bibfnamefont {K.-Y.}\ \bibnamefont {Kim}},
  \bibinfo {author} {\bibfnamefont {L.}~\bibnamefont {Li}},\ and\ \bibinfo
  {author} {\bibfnamefont {W.-J.}\ \bibnamefont {Li}},\ }\bibfield  {title}
  {\bibinfo {title} {{Holographic Axion Model: a simple gravitational tool for
  quantum matter}},\ }\href {https://doi.org/10.1007/s11433-021-1681-8}
  {\bibfield  {journal} {\bibinfo  {journal} {Sci. China Phys. Mech. Astron.}\
  }\textbf {\bibinfo {volume} {64}},\ \bibinfo {pages} {270001} (\bibinfo
  {year} {2021})},\ \Eprint {https://arxiv.org/abs/2101.01892}
  {arXiv:2101.01892 [hep-th]} \BibitemShut {NoStop}%
\bibitem [{\citenamefont {Liu}\ \emph {et~al.}(2023)\citenamefont {Liu},
  \citenamefont {Wang}, \citenamefont {Wu},\ and\ \citenamefont
  {Zhang}}]{Liu:2022bdu}%
  \BibitemOpen
  \bibfield  {author} {\bibinfo {author} {\bibfnamefont {Y.}~\bibnamefont
  {Liu}}, \bibinfo {author} {\bibfnamefont {X.-J.}\ \bibnamefont {Wang}},
  \bibinfo {author} {\bibfnamefont {J.-P.}\ \bibnamefont {Wu}},\ and\ \bibinfo
  {author} {\bibfnamefont {X.}~\bibnamefont {Zhang}},\ }\bibfield  {title}
  {\bibinfo {title} {{Holographic superfluid with gauge\textendash{}axion
  coupling}},\ }\href {https://doi.org/10.1140/epjc/s10052-023-11918-9}
  {\bibfield  {journal} {\bibinfo  {journal} {Eur. Phys. J. C}\ }\textbf
  {\bibinfo {volume} {83}},\ \bibinfo {pages} {748} (\bibinfo {year} {2023})},\
  \Eprint {https://arxiv.org/abs/2212.01986} {arXiv:2212.01986 [hep-th]}
  \BibitemShut {NoStop}%
\bibitem [{\citenamefont {Bardoux}\ \emph {et~al.}(2012)\citenamefont
  {Bardoux}, \citenamefont {Caldarelli},\ and\ \citenamefont
  {Charmousis}}]{Bardoux:2012aw}%
  \BibitemOpen
  \bibfield  {author} {\bibinfo {author} {\bibfnamefont {Y.}~\bibnamefont
  {Bardoux}}, \bibinfo {author} {\bibfnamefont {M.~M.}\ \bibnamefont
  {Caldarelli}},\ and\ \bibinfo {author} {\bibfnamefont {C.}~\bibnamefont
  {Charmousis}},\ }\bibfield  {title} {\bibinfo {title} {{Shaping black holes
  with free fields}},\ }\href {https://doi.org/10.1007/JHEP05(2012)054}
  {\bibfield  {journal} {\bibinfo  {journal} {JHEP}\ }\textbf {\bibinfo
  {volume} {05}},\ \bibinfo {pages} {054}},\ \Eprint
  {https://arxiv.org/abs/1202.4458} {arXiv:1202.4458 [hep-th]} \BibitemShut
  {NoStop}%
\bibitem [{\citenamefont {Tian}\ \emph {et~al.}(2014)\citenamefont {Tian},
  \citenamefont {Wu},\ and\ \citenamefont {Zhang}}]{Tian:2014goa}%
  \BibitemOpen
  \bibfield  {author} {\bibinfo {author} {\bibfnamefont {Y.}~\bibnamefont
  {Tian}}, \bibinfo {author} {\bibfnamefont {X.-N.}\ \bibnamefont {Wu}},\ and\
  \bibinfo {author} {\bibfnamefont {H.-B.}\ \bibnamefont {Zhang}},\ }\bibfield
  {title} {\bibinfo {title} {{Holographic Entropy Production}},\ }\href
  {https://doi.org/10.1007/JHEP10(2014)170} {\bibfield  {journal} {\bibinfo
  {journal} {JHEP}\ }\textbf {\bibinfo {volume} {10}},\ \bibinfo {pages}
  {170}},\ \Eprint {https://arxiv.org/abs/1407.8273} {arXiv:1407.8273 [hep-th]}
  \BibitemShut {NoStop}%
\bibitem [{\citenamefont {Tian}(2019)}]{Tian:2018hlw}%
  \BibitemOpen
  \bibfield  {author} {\bibinfo {author} {\bibfnamefont {Y.}~\bibnamefont
  {Tian}},\ }\bibfield  {title} {\bibinfo {title} {{A topological charge of
  black holes}},\ }\href {https://doi.org/10.1088/1361-6382/ab5343} {\bibfield
  {journal} {\bibinfo  {journal} {Class. Quant. Grav.}\ }\textbf {\bibinfo
  {volume} {36}},\ \bibinfo {pages} {245001} (\bibinfo {year} {2019})},\
  \Eprint {https://arxiv.org/abs/1804.00249} {arXiv:1804.00249 [gr-qc]}
  \BibitemShut {NoStop}%
\bibitem [{\citenamefont {Zeyuan}\ and\ \citenamefont
  {Zhao}(2022)}]{Zeyuan:2021uol}%
  \BibitemOpen
  \bibfield  {author} {\bibinfo {author} {\bibfnamefont {G.}~\bibnamefont
  {Zeyuan}}\ and\ \bibinfo {author} {\bibfnamefont {L.}~\bibnamefont {Zhao}},\
  }\bibfield  {title} {\bibinfo {title} {{Restricted phase space thermodynamics
  for AdS black holes via holography}},\ }\href
  {https://doi.org/10.1088/1361-6382/ac566c} {\bibfield  {journal} {\bibinfo
  {journal} {Class. Quant. Grav.}\ }\textbf {\bibinfo {volume} {39}},\ \bibinfo
  {pages} {075019} (\bibinfo {year} {2022})},\ \Eprint
  {https://arxiv.org/abs/2112.02386} {arXiv:2112.02386 [gr-qc]} \BibitemShut
  {NoStop}%
\bibitem [{\citenamefont {Wang}\ and\ \citenamefont
  {Zhao}(2022)}]{Wang:2021cmz}%
  \BibitemOpen
  \bibfield  {author} {\bibinfo {author} {\bibfnamefont {T.}~\bibnamefont
  {Wang}}\ and\ \bibinfo {author} {\bibfnamefont {L.}~\bibnamefont {Zhao}},\
  }\bibfield  {title} {\bibinfo {title} {{Black hole thermodynamics is
  extensive with variable Newton constant}},\ }\href
  {https://doi.org/10.1016/j.physletb.2022.136935} {\bibfield  {journal}
  {\bibinfo  {journal} {Phys. Lett. B}\ }\textbf {\bibinfo {volume} {827}},\
  \bibinfo {pages} {136935} (\bibinfo {year} {2022})},\ \Eprint
  {https://arxiv.org/abs/2112.11236} {arXiv:2112.11236 [hep-th]} \BibitemShut
  {NoStop}%
\bibitem [{\citenamefont {Gao}\ \emph {et~al.}(2022)\citenamefont {Gao},
  \citenamefont {Kong},\ and\ \citenamefont {Zhao}}]{Gao:2021xtt}%
  \BibitemOpen
  \bibfield  {author} {\bibinfo {author} {\bibfnamefont {Z.}~\bibnamefont
  {Gao}}, \bibinfo {author} {\bibfnamefont {X.}~\bibnamefont {Kong}},\ and\
  \bibinfo {author} {\bibfnamefont {L.}~\bibnamefont {Zhao}},\ }\bibfield
  {title} {\bibinfo {title} {{Thermodynamics of Kerr-AdS black holes in the
  restricted phase space}},\ }\href
  {https://doi.org/10.1140/epjc/s10052-022-10080-y} {\bibfield  {journal}
  {\bibinfo  {journal} {Eur. Phys. J. C}\ }\textbf {\bibinfo {volume} {82}},\
  \bibinfo {pages} {112} (\bibinfo {year} {2022})},\ \Eprint
  {https://arxiv.org/abs/2112.08672} {arXiv:2112.08672 [hep-th]} \BibitemShut
  {NoStop}%
\bibitem [{\citenamefont {Zhao}(2022)}]{Zhao:2022dgc}%
  \BibitemOpen
  \bibfield  {author} {\bibinfo {author} {\bibfnamefont {L.}~\bibnamefont
  {Zhao}},\ }\bibfield  {title} {\bibinfo {title} {{Thermodynamics for higher
  dimensional rotating black holes with variable Newton constant *}},\ }\href
  {https://doi.org/10.1088/1674-1137/ac4f4c} {\bibfield  {journal} {\bibinfo
  {journal} {Chin. Phys. C}\ }\textbf {\bibinfo {volume} {46}},\ \bibinfo
  {pages} {055105} (\bibinfo {year} {2022})},\ \Eprint
  {https://arxiv.org/abs/2201.00521} {arXiv:2201.00521 [hep-th]} \BibitemShut
  {NoStop}%
\bibitem [{\citenamefont {Kong}\ \emph
  {et~al.}(2022{\natexlab{a}})\citenamefont {Kong}, \citenamefont {Wang},
  \citenamefont {Gao},\ and\ \citenamefont {Zhao}}]{Kong:2022gwu}%
  \BibitemOpen
  \bibfield  {author} {\bibinfo {author} {\bibfnamefont {X.}~\bibnamefont
  {Kong}}, \bibinfo {author} {\bibfnamefont {T.}~\bibnamefont {Wang}}, \bibinfo
  {author} {\bibfnamefont {Z.}~\bibnamefont {Gao}},\ and\ \bibinfo {author}
  {\bibfnamefont {L.}~\bibnamefont {Zhao}},\ }\bibfield  {title} {\bibinfo
  {title} {{Restricted Phased Space Thermodynamics for Black Holes in Higher
  Dimensions and Higher Curvature Gravities}},\ }\href
  {https://doi.org/10.3390/e24081131} {\bibfield  {journal} {\bibinfo
  {journal} {Entropy}\ }\textbf {\bibinfo {volume} {24}},\ \bibinfo {pages}
  {1131} (\bibinfo {year} {2022}{\natexlab{a}})},\ \Eprint
  {https://arxiv.org/abs/2208.07748} {arXiv:2208.07748 [hep-th]} \BibitemShut
  {NoStop}%
\bibitem [{\citenamefont {Wang}\ \emph {et~al.}(2022)\citenamefont {Wang},
  \citenamefont {Zhang}, \citenamefont {Kong},\ and\ \citenamefont
  {Zhao}}]{Wang:2022err}%
  \BibitemOpen
  \bibfield  {author} {\bibinfo {author} {\bibfnamefont {T.}~\bibnamefont
  {Wang}}, \bibinfo {author} {\bibfnamefont {Z.}~\bibnamefont {Zhang}},
  \bibinfo {author} {\bibfnamefont {X.}~\bibnamefont {Kong}},\ and\ \bibinfo
  {author} {\bibfnamefont {L.}~\bibnamefont {Zhao}},\ }\bibfield  {title}
  {\bibinfo {title} {{Topological black holes in Einstein-Maxwell and 4D
  conformal gravities revisited}},\ }\href@noop {} {\  (\bibinfo {year}
  {2022})},\ \Eprint {https://arxiv.org/abs/2211.16904} {arXiv:2211.16904
  [hep-th]} \BibitemShut {NoStop}%
\bibitem [{\citenamefont {Visser}(2022)}]{Visser:2021eqk}%
  \BibitemOpen
  \bibfield  {author} {\bibinfo {author} {\bibfnamefont {M.~R.}\ \bibnamefont
  {Visser}},\ }\bibfield  {title} {\bibinfo {title} {{Holographic
  thermodynamics requires a chemical potential for color}},\ }\href
  {https://doi.org/10.1103/PhysRevD.105.106014} {\bibfield  {journal} {\bibinfo
   {journal} {Phys. Rev. D}\ }\textbf {\bibinfo {volume} {105}},\ \bibinfo
  {pages} {106014} (\bibinfo {year} {2022})},\ \Eprint
  {https://arxiv.org/abs/2101.04145} {arXiv:2101.04145 [hep-th]} \BibitemShut
  {NoStop}%
\bibitem [{\citenamefont {Misner}\ and\ \citenamefont
  {Sharp}(1964)}]{PhysRev.136.B571}%
  \BibitemOpen
  \bibfield  {author} {\bibinfo {author} {\bibfnamefont {C.~W.}\ \bibnamefont
  {Misner}}\ and\ \bibinfo {author} {\bibfnamefont {D.~H.}\ \bibnamefont
  {Sharp}},\ }\bibfield  {title} {\bibinfo {title} {Relativistic equations for
  adiabatic, spherically symmetric gravitational collapse},\ }\href
  {https://doi.org/10.1103/PhysRev.136.B571} {\bibfield  {journal} {\bibinfo
  {journal} {Phys. Rev.}\ }\textbf {\bibinfo {volume} {136}},\ \bibinfo {pages}
  {B571} (\bibinfo {year} {1964})}\BibitemShut {NoStop}%
\bibitem [{\citenamefont {Hayward}(1996)}]{Hayward:1994bu}%
  \BibitemOpen
  \bibfield  {author} {\bibinfo {author} {\bibfnamefont {S.~A.}\ \bibnamefont
  {Hayward}},\ }\bibfield  {title} {\bibinfo {title} {{Gravitational energy in
  spherical symmetry}},\ }\href {https://doi.org/10.1103/PhysRevD.53.1938}
  {\bibfield  {journal} {\bibinfo  {journal} {Phys. Rev. D}\ }\textbf {\bibinfo
  {volume} {53}},\ \bibinfo {pages} {1938} (\bibinfo {year} {1996})},\ \Eprint
  {https://arxiv.org/abs/gr-qc/9408002} {arXiv:gr-qc/9408002} \BibitemShut
  {NoStop}%
\bibitem [{\citenamefont {Kodama}(1980)}]{10.1143/PTP.63.1217}%
  \BibitemOpen
  \bibfield  {author} {\bibinfo {author} {\bibfnamefont {H.}~\bibnamefont
  {Kodama}},\ }\bibfield  {title} {\bibinfo {title} {{Conserved Energy Flux for
  the Spherically Symmetric System and the Backreaction Problem in the Black
  Hole Evaporation}},\ }\href {https://doi.org/10.1143/PTP.63.1217} {\bibfield
  {journal} {\bibinfo  {journal} {Progress of Theoretical Physics}\ }\textbf
  {\bibinfo {volume} {63}},\ \bibinfo {pages} {1217} (\bibinfo {year}
  {1980})},\ \Eprint
  {https://arxiv.org/abs/https://academic.oup.com/ptp/article-pdf/63/4/1217/5243205/63-4-1217.pdf}
  {https://academic.oup.com/ptp/article-pdf/63/4/1217/5243205/63-4-1217.pdf}
  \BibitemShut {NoStop}%
\bibitem [{\citenamefont {Nakao}\ \emph {et~al.}(2019)\citenamefont {Nakao},
  \citenamefont {Yoo},\ and\ \citenamefont {Harada}}]{Nakao:2018knn}%
  \BibitemOpen
  \bibfield  {author} {\bibinfo {author} {\bibfnamefont {K.-i.}\ \bibnamefont
  {Nakao}}, \bibinfo {author} {\bibfnamefont {C.-M.}\ \bibnamefont {Yoo}},\
  and\ \bibinfo {author} {\bibfnamefont {T.}~\bibnamefont {Harada}},\
  }\bibfield  {title} {\bibinfo {title} {{Gravastar formation: What can be the
  evidence of a black hole?}},\ }\href
  {https://doi.org/10.1103/PhysRevD.99.044027} {\bibfield  {journal} {\bibinfo
  {journal} {Phys. Rev. D}\ }\textbf {\bibinfo {volume} {99}},\ \bibinfo
  {pages} {044027} (\bibinfo {year} {2019})},\ \Eprint
  {https://arxiv.org/abs/1809.00124} {arXiv:1809.00124 [gr-qc]} \BibitemShut
  {NoStop}%
\bibitem [{\citenamefont {H\"utsi}\ \emph {et~al.}(2021)\citenamefont
  {H\"utsi}, \citenamefont {Koivisto}, \citenamefont {Raidal}, \citenamefont
  {Vaskonen},\ and\ \citenamefont {Veerm\"ae}}]{Hutsi:2021nvs}%
  \BibitemOpen
  \bibfield  {author} {\bibinfo {author} {\bibfnamefont {G.}~\bibnamefont
  {H\"utsi}}, \bibinfo {author} {\bibfnamefont {T.}~\bibnamefont {Koivisto}},
  \bibinfo {author} {\bibfnamefont {M.}~\bibnamefont {Raidal}}, \bibinfo
  {author} {\bibfnamefont {V.}~\bibnamefont {Vaskonen}},\ and\ \bibinfo
  {author} {\bibfnamefont {H.}~\bibnamefont {Veerm\"ae}},\ }\bibfield  {title}
  {\bibinfo {title} {{Cosmological black holes are not described by the
  Thakurta metric: LIGO-Virgo bounds on PBHs remain unchanged}},\ }\href
  {https://doi.org/10.1140/epjc/s10052-021-09803-4} {\bibfield  {journal}
  {\bibinfo  {journal} {Eur. Phys. J. C}\ }\textbf {\bibinfo {volume} {81}},\
  \bibinfo {pages} {999} (\bibinfo {year} {2021})},\ \Eprint
  {https://arxiv.org/abs/2105.09328} {arXiv:2105.09328 [astro-ph.CO]}
  \BibitemShut {NoStop}%
\bibitem [{\citenamefont {Harada}\ \emph {et~al.}(2022)\citenamefont {Harada},
  \citenamefont {Maeda},\ and\ \citenamefont {Sato}}]{Harada:2021xze}%
  \BibitemOpen
  \bibfield  {author} {\bibinfo {author} {\bibfnamefont {T.}~\bibnamefont
  {Harada}}, \bibinfo {author} {\bibfnamefont {H.}~\bibnamefont {Maeda}},\ and\
  \bibinfo {author} {\bibfnamefont {T.}~\bibnamefont {Sato}},\ }\bibfield
  {title} {\bibinfo {title} {{Thakurta metric does not describe a cosmological
  black hole}},\ }\href {https://doi.org/10.1016/j.physletb.2022.137332}
  {\bibfield  {journal} {\bibinfo  {journal} {Phys. Lett. B}\ }\textbf
  {\bibinfo {volume} {833}},\ \bibinfo {pages} {137332} (\bibinfo {year}
  {2022})},\ \Eprint {https://arxiv.org/abs/2106.06651} {arXiv:2106.06651
  [gr-qc]} \BibitemShut {NoStop}%
\bibitem [{\citenamefont {Yoo}\ \emph {et~al.}(2022)\citenamefont {Yoo},
  \citenamefont {Harada}, \citenamefont {Hirano}, \citenamefont {Okawa},\ and\
  \citenamefont {Sasaki}}]{Yoo:2021fxs}%
  \BibitemOpen
  \bibfield  {author} {\bibinfo {author} {\bibfnamefont {C.-M.}\ \bibnamefont
  {Yoo}}, \bibinfo {author} {\bibfnamefont {T.}~\bibnamefont {Harada}},
  \bibinfo {author} {\bibfnamefont {S.}~\bibnamefont {Hirano}}, \bibinfo
  {author} {\bibfnamefont {H.}~\bibnamefont {Okawa}},\ and\ \bibinfo {author}
  {\bibfnamefont {M.}~\bibnamefont {Sasaki}},\ }\bibfield  {title} {\bibinfo
  {title} {{Primordial black hole formation from massless scalar
  isocurvature}},\ }\href {https://doi.org/10.1103/PhysRevD.105.103538}
  {\bibfield  {journal} {\bibinfo  {journal} {Phys. Rev. D}\ }\textbf {\bibinfo
  {volume} {105}},\ \bibinfo {pages} {103538} (\bibinfo {year} {2022})},\
  \Eprint {https://arxiv.org/abs/2112.12335} {arXiv:2112.12335 [gr-qc]}
  \BibitemShut {NoStop}%
\bibitem [{\citenamefont {Sato}\ \emph {et~al.}(2022)\citenamefont {Sato},
  \citenamefont {Maeda},\ and\ \citenamefont {Harada}}]{Sato:2022yto}%
  \BibitemOpen
  \bibfield  {author} {\bibinfo {author} {\bibfnamefont {T.}~\bibnamefont
  {Sato}}, \bibinfo {author} {\bibfnamefont {H.}~\bibnamefont {Maeda}},\ and\
  \bibinfo {author} {\bibfnamefont {T.}~\bibnamefont {Harada}},\ }\bibfield
  {title} {\bibinfo {title} {{Conformally Schwarzschild cosmological black
  holes}},\ }\href {https://doi.org/10.1088/1361-6382/ac902f} {\bibfield
  {journal} {\bibinfo  {journal} {Class. Quant. Grav.}\ }\textbf {\bibinfo
  {volume} {39}},\ \bibinfo {pages} {215011} (\bibinfo {year} {2022})},\
  \Eprint {https://arxiv.org/abs/2206.10998} {arXiv:2206.10998 [gr-qc]}
  \BibitemShut {NoStop}%
\bibitem [{\citenamefont {Escriv\`a}\ \emph {et~al.}(2022)\citenamefont
  {Escriv\`a}, \citenamefont {Tada}, \citenamefont {Yokoyama},\ and\
  \citenamefont {Yoo}}]{Escriva:2022pnz}%
  \BibitemOpen
  \bibfield  {author} {\bibinfo {author} {\bibfnamefont {A.}~\bibnamefont
  {Escriv\`a}}, \bibinfo {author} {\bibfnamefont {Y.}~\bibnamefont {Tada}},
  \bibinfo {author} {\bibfnamefont {S.}~\bibnamefont {Yokoyama}},\ and\
  \bibinfo {author} {\bibfnamefont {C.-M.}\ \bibnamefont {Yoo}},\ }\bibfield
  {title} {\bibinfo {title} {{Simulation of primordial black holes with large
  negative non-Gaussianity}},\ }\href
  {https://doi.org/10.1088/1475-7516/2022/05/012} {\bibfield  {journal}
  {\bibinfo  {journal} {JCAP}\ }\textbf {\bibinfo {volume} {05}}\bibfield
  {number} {\bibinfo  {number} { (05)},\ \bibinfo {pages} {012}},\ }\Eprint
  {https://arxiv.org/abs/2202.01028} {arXiv:2202.01028 [astro-ph.CO]}
  \BibitemShut {NoStop}%
\bibitem [{\citenamefont {Ren}\ and\ \citenamefont {Li}(2008)}]{Ren:2007xw}%
  \BibitemOpen
  \bibfield  {author} {\bibinfo {author} {\bibfnamefont {J.-R.}\ \bibnamefont
  {Ren}}\ and\ \bibinfo {author} {\bibfnamefont {R.}~\bibnamefont {Li}},\
  }\bibfield  {title} {\bibinfo {title} {{Unified first law and thermodynamics
  of dynamical black hole in n-dimensional Vaidya spacetime}},\ }\href
  {https://doi.org/10.1142/S0217732308028831} {\bibfield  {journal} {\bibinfo
  {journal} {Mod. Phys. Lett. A}\ }\textbf {\bibinfo {volume} {23}},\ \bibinfo
  {pages} {3265} (\bibinfo {year} {2008})},\ \Eprint
  {https://arxiv.org/abs/0705.4339} {arXiv:0705.4339 [gr-qc]} \BibitemShut
  {NoStop}%
\bibitem [{\citenamefont {Ke-Xia}\ \emph {et~al.}(2009)\citenamefont {Ke-Xia},
  \citenamefont {San-Min},\ and\ \citenamefont {Dan-Tao}}]{Ke-Xia:2009tzo}%
  \BibitemOpen
  \bibfield  {author} {\bibinfo {author} {\bibfnamefont {J.}~\bibnamefont
  {Ke-Xia}}, \bibinfo {author} {\bibfnamefont {K.}~\bibnamefont {San-Min}},\
  and\ \bibinfo {author} {\bibfnamefont {P.}~\bibnamefont {Dan-Tao}},\
  }\bibfield  {title} {\bibinfo {title} {{Hawking Radiation as tunneling and
  the unified first law of thermodynamics for a class of dynamical black
  holes}},\ }\href {https://doi.org/10.1142/S0218271809015254} {\bibfield
  {journal} {\bibinfo  {journal} {Int. J. Mod. Phys. D}\ }\textbf {\bibinfo
  {volume} {18}},\ \bibinfo {pages} {1707} (\bibinfo {year} {2009})},\ \Eprint
  {https://arxiv.org/abs/1105.0595} {arXiv:1105.0595 [hep-th]} \BibitemShut
  {NoStop}%
\bibitem [{\citenamefont {Di~Criscienzo}\ \emph {et~al.}(2010)\citenamefont
  {Di~Criscienzo}, \citenamefont {Hayward}, \citenamefont {Nadalini},
  \citenamefont {Vanzo},\ and\ \citenamefont {Zerbini}}]{DiCriscienzo:2009kun}%
  \BibitemOpen
  \bibfield  {author} {\bibinfo {author} {\bibfnamefont {R.}~\bibnamefont
  {Di~Criscienzo}}, \bibinfo {author} {\bibfnamefont {S.~A.}\ \bibnamefont
  {Hayward}}, \bibinfo {author} {\bibfnamefont {M.}~\bibnamefont {Nadalini}},
  \bibinfo {author} {\bibfnamefont {L.}~\bibnamefont {Vanzo}},\ and\ \bibinfo
  {author} {\bibfnamefont {S.}~\bibnamefont {Zerbini}},\ }\bibfield  {title}
  {\bibinfo {title} {{Hamilton-Jacobi tunneling method for dynamical horizons
  in different coordinate gauges}},\ }\href
  {https://doi.org/10.1088/0264-9381/27/1/015006} {\bibfield  {journal}
  {\bibinfo  {journal} {Class. Quant. Grav.}\ }\textbf {\bibinfo {volume}
  {27}},\ \bibinfo {pages} {015006} (\bibinfo {year} {2010})},\ \Eprint
  {https://arxiv.org/abs/0906.1725} {arXiv:0906.1725 [gr-qc]} \BibitemShut
  {NoStop}%
\bibitem [{\citenamefont {di~Criscienzo}\ \emph {et~al.}(2010)\citenamefont
  {di~Criscienzo}, \citenamefont {Nadalini}, \citenamefont {Vanzo},
  \citenamefont {Zerbini},\ and\ \citenamefont
  {Hayward}}]{diCriscienzo:2010cp}%
  \BibitemOpen
  \bibfield  {author} {\bibinfo {author} {\bibfnamefont {R.}~\bibnamefont
  {di~Criscienzo}}, \bibinfo {author} {\bibfnamefont {M.}~\bibnamefont
  {Nadalini}}, \bibinfo {author} {\bibfnamefont {L.}~\bibnamefont {Vanzo}},
  \bibinfo {author} {\bibfnamefont {S.}~\bibnamefont {Zerbini}},\ and\ \bibinfo
  {author} {\bibfnamefont {S.}~\bibnamefont {Hayward}},\ }\bibfield  {title}
  {\bibinfo {title} {{Invariance of the Tunneling Method for Dynamical Black
  Holes}},\ }in\ \href {https://doi.org/10.1142/9789814374552_0161} {\emph
  {\bibinfo {booktitle} {{12th Marcel Grossmann Meeting on General
  Relativity}}}}\ (\bibinfo {year} {2010})\ pp.\ \bibinfo {pages}
  {1166--1168},\ \Eprint {https://arxiv.org/abs/1006.1590} {arXiv:1006.1590
  [hep-th]} \BibitemShut {NoStop}%
\bibitem [{\citenamefont {Vanzo}\ \emph {et~al.}(2011)\citenamefont {Vanzo},
  \citenamefont {Acquaviva},\ and\ \citenamefont
  {Di~Criscienzo}}]{Vanzo:2011wq}%
  \BibitemOpen
  \bibfield  {author} {\bibinfo {author} {\bibfnamefont {L.}~\bibnamefont
  {Vanzo}}, \bibinfo {author} {\bibfnamefont {G.}~\bibnamefont {Acquaviva}},\
  and\ \bibinfo {author} {\bibfnamefont {R.}~\bibnamefont {Di~Criscienzo}},\
  }\bibfield  {title} {\bibinfo {title} {{Tunnelling Methods and Hawking's
  radiation: achievements and prospects}},\ }\href
  {https://doi.org/10.1088/0264-9381/28/18/183001} {\bibfield  {journal}
  {\bibinfo  {journal} {Class. Quant. Grav.}\ }\textbf {\bibinfo {volume}
  {28}},\ \bibinfo {pages} {183001} (\bibinfo {year} {2011})},\ \Eprint
  {https://arxiv.org/abs/1106.4153} {arXiv:1106.4153 [gr-qc]} \BibitemShut
  {NoStop}%
\bibitem [{\citenamefont {Kong}\ \emph
  {et~al.}(2022{\natexlab{b}})\citenamefont {Kong}, \citenamefont {Abdusattar},
  \citenamefont {Yin}, \citenamefont {Zhang},\ and\ \citenamefont
  {Hu}}]{Kong:2021dqd}%
  \BibitemOpen
  \bibfield  {author} {\bibinfo {author} {\bibfnamefont {S.-B.}\ \bibnamefont
  {Kong}}, \bibinfo {author} {\bibfnamefont {H.}~\bibnamefont {Abdusattar}},
  \bibinfo {author} {\bibfnamefont {Y.}~\bibnamefont {Yin}}, \bibinfo {author}
  {\bibfnamefont {H.}~\bibnamefont {Zhang}},\ and\ \bibinfo {author}
  {\bibfnamefont {Y.-P.}\ \bibnamefont {Hu}},\ }\bibfield  {title} {\bibinfo
  {title} {{The $P-V$ phase transition of the FRW universe}},\ }\href
  {https://doi.org/10.1140/epjc/s10052-022-10976-9} {\bibfield  {journal}
  {\bibinfo  {journal} {Eur. Phys. J. C}\ }\textbf {\bibinfo {volume} {82}},\
  \bibinfo {pages} {1047} (\bibinfo {year} {2022}{\natexlab{b}})},\ \Eprint
  {https://arxiv.org/abs/2108.09411} {arXiv:2108.09411 [gr-qc]} \BibitemShut
  {NoStop}%
\bibitem [{\citenamefont {Abdusattar}\ \emph {et~al.}(2022)\citenamefont
  {Abdusattar}, \citenamefont {Kong}, \citenamefont {Yin},\ and\ \citenamefont
  {Hu}}]{Abdusattar:2022bpg}%
  \BibitemOpen
  \bibfield  {author} {\bibinfo {author} {\bibfnamefont {H.}~\bibnamefont
  {Abdusattar}}, \bibinfo {author} {\bibfnamefont {S.-B.}\ \bibnamefont
  {Kong}}, \bibinfo {author} {\bibfnamefont {Y.}~\bibnamefont {Yin}},\ and\
  \bibinfo {author} {\bibfnamefont {Y.-P.}\ \bibnamefont {Hu}},\ }\bibfield
  {title} {\bibinfo {title} {{The Hawking-Page-like phase transition from FRW
  spacetime to McVittie black hole}},\ }\href
  {https://doi.org/10.1088/1475-7516/2022/08/060} {\bibfield  {journal}
  {\bibinfo  {journal} {JCAP}\ }\textbf {\bibinfo {volume} {08}}\bibfield
  {number} {\bibinfo  {number} { (08)},\ \bibinfo {pages} {060}},\ }\Eprint
  {https://arxiv.org/abs/2203.10868} {arXiv:2203.10868 [gr-qc]} \BibitemShut
  {NoStop}%
\bibitem [{\citenamefont {Hayward}(1998)}]{Hayward:1997jp}%
  \BibitemOpen
  \bibfield  {author} {\bibinfo {author} {\bibfnamefont {S.~A.}\ \bibnamefont
  {Hayward}},\ }\bibfield  {title} {\bibinfo {title} {{Unified first law of
  black hole dynamics and relativistic thermodynamics}},\ }\href
  {https://doi.org/10.1088/0264-9381/15/10/017} {\bibfield  {journal} {\bibinfo
   {journal} {Class. Quant. Grav.}\ }\textbf {\bibinfo {volume} {15}},\
  \bibinfo {pages} {3147} (\bibinfo {year} {1998})},\ \Eprint
  {https://arxiv.org/abs/gr-qc/9710089} {arXiv:gr-qc/9710089} \BibitemShut
  {NoStop}%
\bibitem [{\citenamefont {Cai}\ and\ \citenamefont {Kim}(2005)}]{Cai:2005ra}%
  \BibitemOpen
  \bibfield  {author} {\bibinfo {author} {\bibfnamefont {R.-G.}\ \bibnamefont
  {Cai}}\ and\ \bibinfo {author} {\bibfnamefont {S.~P.}\ \bibnamefont {Kim}},\
  }\bibfield  {title} {\bibinfo {title} {{First law of thermodynamics and
  Friedmann equations of Friedmann-Robertson-Walker universe}},\ }\href
  {https://doi.org/10.1088/1126-6708/2005/02/050} {\bibfield  {journal}
  {\bibinfo  {journal} {JHEP}\ }\textbf {\bibinfo {volume} {02}},\ \bibinfo
  {pages} {050}},\ \Eprint {https://arxiv.org/abs/hep-th/0501055}
  {arXiv:hep-th/0501055} \BibitemShut {NoStop}%
\bibitem [{\citenamefont {Cai}\ and\ \citenamefont {Cao}(2007)}]{Cai:2006rs}%
  \BibitemOpen
  \bibfield  {author} {\bibinfo {author} {\bibfnamefont {R.-G.}\ \bibnamefont
  {Cai}}\ and\ \bibinfo {author} {\bibfnamefont {L.-M.}\ \bibnamefont {Cao}},\
  }\bibfield  {title} {\bibinfo {title} {{Unified first law and thermodynamics
  of apparent horizon in FRW universe}},\ }\href
  {https://doi.org/10.1103/PhysRevD.75.064008} {\bibfield  {journal} {\bibinfo
  {journal} {Phys. Rev. D}\ }\textbf {\bibinfo {volume} {75}},\ \bibinfo
  {pages} {064008} (\bibinfo {year} {2007})},\ \Eprint
  {https://arxiv.org/abs/gr-qc/0611071} {arXiv:gr-qc/0611071} \BibitemShut
  {NoStop}%
\bibitem [{\citenamefont {Zhang}\ \emph
  {et~al.}(2014{\natexlab{a}})\citenamefont {Zhang}, \citenamefont {Hayward},
  \citenamefont {Zhai},\ and\ \citenamefont {Li}}]{Zhang:2013tca}%
  \BibitemOpen
  \bibfield  {author} {\bibinfo {author} {\bibfnamefont {H.}~\bibnamefont
  {Zhang}}, \bibinfo {author} {\bibfnamefont {S.~A.}\ \bibnamefont {Hayward}},
  \bibinfo {author} {\bibfnamefont {X.-H.}\ \bibnamefont {Zhai}},\ and\
  \bibinfo {author} {\bibfnamefont {X.-Z.}\ \bibnamefont {Li}},\ }\bibfield
  {title} {\bibinfo {title} {{Schwarzschild solution as a result of
  thermodynamics}},\ }\href {https://doi.org/10.1103/PhysRevD.89.064052}
  {\bibfield  {journal} {\bibinfo  {journal} {Phys. Rev. D}\ }\textbf {\bibinfo
  {volume} {89}},\ \bibinfo {pages} {064052} (\bibinfo {year}
  {2014}{\natexlab{a}})},\ \Eprint {https://arxiv.org/abs/1304.3647}
  {arXiv:1304.3647 [gr-qc]} \BibitemShut {NoStop}%
\bibitem [{\citenamefont {Maeda}\ and\ \citenamefont
  {Nozawa}(2008)}]{Maeda:2007uu}%
  \BibitemOpen
  \bibfield  {author} {\bibinfo {author} {\bibfnamefont {H.}~\bibnamefont
  {Maeda}}\ and\ \bibinfo {author} {\bibfnamefont {M.}~\bibnamefont {Nozawa}},\
  }\bibfield  {title} {\bibinfo {title} {{Generalized Misner-Sharp quasi-local
  mass in Einstein-Gauss-Bonnet gravity}},\ }\href
  {https://doi.org/10.1103/PhysRevD.77.064031} {\bibfield  {journal} {\bibinfo
  {journal} {Phys. Rev. D}\ }\textbf {\bibinfo {volume} {77}},\ \bibinfo
  {pages} {064031} (\bibinfo {year} {2008})},\ \Eprint
  {https://arxiv.org/abs/0709.1199} {arXiv:0709.1199 [hep-th]} \BibitemShut
  {NoStop}%
\bibitem [{\citenamefont {Zhang}\ \emph
  {et~al.}(2014{\natexlab{b}})\citenamefont {Zhang}, \citenamefont {Hu},\ and\
  \citenamefont {Li}}]{Zhang:2014goa}%
  \BibitemOpen
  \bibfield  {author} {\bibinfo {author} {\bibfnamefont {H.}~\bibnamefont
  {Zhang}}, \bibinfo {author} {\bibfnamefont {Y.}~\bibnamefont {Hu}},\ and\
  \bibinfo {author} {\bibfnamefont {X.-Z.}\ \bibnamefont {Li}},\ }\bibfield
  {title} {\bibinfo {title} {{Misner-Sharp Mass in $N$-dimensional $f(R)$
  Gravity}},\ }\href {https://doi.org/10.1103/PhysRevD.90.024062} {\bibfield
  {journal} {\bibinfo  {journal} {Phys. Rev. D}\ }\textbf {\bibinfo {volume}
  {90}},\ \bibinfo {pages} {024062} (\bibinfo {year} {2014}{\natexlab{b}})},\
  \Eprint {https://arxiv.org/abs/1406.0577} {arXiv:1406.0577 [gr-qc]}
  \BibitemShut {NoStop}%
\bibitem [{\citenamefont {Hu}\ and\ \citenamefont {Zhang}(2015)}]{Hu:2015xva}%
  \BibitemOpen
  \bibfield  {author} {\bibinfo {author} {\bibfnamefont {Y.-P.}\ \bibnamefont
  {Hu}}\ and\ \bibinfo {author} {\bibfnamefont {H.}~\bibnamefont {Zhang}},\
  }\bibfield  {title} {\bibinfo {title} {{Misner-Sharp Mass and the Unified
  First Law in Massive Gravity}},\ }\href
  {https://doi.org/10.1103/PhysRevD.92.024006} {\bibfield  {journal} {\bibinfo
  {journal} {Phys. Rev. D}\ }\textbf {\bibinfo {volume} {92}},\ \bibinfo
  {pages} {024006} (\bibinfo {year} {2015})},\ \Eprint
  {https://arxiv.org/abs/1502.00069} {arXiv:1502.00069 [hep-th]} \BibitemShut
  {NoStop}%
\bibitem [{\citenamefont {Tan}\ \emph {et~al.}(2017)\citenamefont {Tan},
  \citenamefont {Yang}, \citenamefont {He},\ and\ \citenamefont
  {Zhang}}]{Tan:2016wkj}%
  \BibitemOpen
  \bibfield  {author} {\bibinfo {author} {\bibfnamefont {H.-W.}\ \bibnamefont
  {Tan}}, \bibinfo {author} {\bibfnamefont {J.-B.}\ \bibnamefont {Yang}},
  \bibinfo {author} {\bibfnamefont {T.-M.}\ \bibnamefont {He}},\ and\ \bibinfo
  {author} {\bibfnamefont {J.-Y.}\ \bibnamefont {Zhang}},\ }\bibfield  {title}
  {\bibinfo {title} {{A Modified Thermodynamics Method to Generate Exact
  Solutions of Einstein Equations}},\ }\href
  {https://doi.org/10.1088/0253-6102/67/1/41} {\bibfield  {journal} {\bibinfo
  {journal} {Commun. Theor. Phys.}\ }\textbf {\bibinfo {volume} {67}},\
  \bibinfo {pages} {41} (\bibinfo {year} {2017})},\ \Eprint
  {https://arxiv.org/abs/1609.04181} {arXiv:1609.04181 [gr-qc]} \BibitemShut
  {NoStop}%
\bibitem [{\citenamefont {Kinoshita}(2021)}]{Kinoshita:2021qsv}%
  \BibitemOpen
  \bibfield  {author} {\bibinfo {author} {\bibfnamefont {S.}~\bibnamefont
  {Kinoshita}},\ }\bibfield  {title} {\bibinfo {title} {{Extension of Kodama
  vector and quasilocal quantities in three-dimensional axisymmetric
  spacetimes}},\ }\href {https://doi.org/10.1103/PhysRevD.103.124042}
  {\bibfield  {journal} {\bibinfo  {journal} {Phys. Rev. D}\ }\textbf {\bibinfo
  {volume} {103}},\ \bibinfo {pages} {124042} (\bibinfo {year} {2021})},\
  \Eprint {https://arxiv.org/abs/2103.07408} {arXiv:2103.07408 [gr-qc]}
  \BibitemShut {NoStop}%
\bibitem [{\citenamefont {Dotti}\ and\ \citenamefont
  {Gleiser}(2005)}]{Dotti:2005rc}%
  \BibitemOpen
  \bibfield  {author} {\bibinfo {author} {\bibfnamefont {G.}~\bibnamefont
  {Dotti}}\ and\ \bibinfo {author} {\bibfnamefont {R.~J.}\ \bibnamefont
  {Gleiser}},\ }\bibfield  {title} {\bibinfo {title} {{Obstructions on the
  horizon geometry from string theory corrections to Einstein gravity}},\
  }\href {https://doi.org/10.1016/j.physletb.2005.08.110} {\bibfield  {journal}
  {\bibinfo  {journal} {Phys. Lett. B}\ }\textbf {\bibinfo {volume} {627}},\
  \bibinfo {pages} {174} (\bibinfo {year} {2005})},\ \Eprint
  {https://arxiv.org/abs/hep-th/0508118} {arXiv:hep-th/0508118} \BibitemShut
  {NoStop}%
\bibitem [{\citenamefont {Maeda}(2010)}]{Maeda:2010bu}%
  \BibitemOpen
  \bibfield  {author} {\bibinfo {author} {\bibfnamefont {H.}~\bibnamefont
  {Maeda}},\ }\bibfield  {title} {\bibinfo {title} {{Gauss-Bonnet black holes
  with non-constant curvature horizons}},\ }\href
  {https://doi.org/10.1103/PhysRevD.81.124007} {\bibfield  {journal} {\bibinfo
  {journal} {Phys. Rev. D}\ }\textbf {\bibinfo {volume} {81}},\ \bibinfo
  {pages} {124007} (\bibinfo {year} {2010})},\ \Eprint
  {https://arxiv.org/abs/1004.0917} {arXiv:1004.0917 [gr-qc]} \BibitemShut
  {NoStop}%
\bibitem [{\citenamefont {Ray}(2015)}]{Ray:2015ava}%
  \BibitemOpen
  \bibfield  {author} {\bibinfo {author} {\bibfnamefont {S.}~\bibnamefont
  {Ray}},\ }\bibfield  {title} {\bibinfo {title} {{Birkhoff\textquoteright{}s
  theorem in Lovelock gravity for general base manifolds}},\ }\href
  {https://doi.org/10.1088/0264-9381/32/19/195022} {\bibfield  {journal}
  {\bibinfo  {journal} {Class. Quant. Grav.}\ }\textbf {\bibinfo {volume}
  {32}},\ \bibinfo {pages} {195022} (\bibinfo {year} {2015})},\ \Eprint
  {https://arxiv.org/abs/1505.03830} {arXiv:1505.03830 [gr-qc]} \BibitemShut
  {NoStop}%
\bibitem [{\citenamefont {Ohashi}\ and\ \citenamefont
  {Nozawa}(2015)}]{Ohashi:2015xaa}%
  \BibitemOpen
  \bibfield  {author} {\bibinfo {author} {\bibfnamefont {S.}~\bibnamefont
  {Ohashi}}\ and\ \bibinfo {author} {\bibfnamefont {M.}~\bibnamefont
  {Nozawa}},\ }\bibfield  {title} {\bibinfo {title} {{Lovelock black holes with
  a nonconstant curvature horizon}},\ }\href
  {https://doi.org/10.1103/PhysRevD.92.064020} {\bibfield  {journal} {\bibinfo
  {journal} {Phys. Rev. D}\ }\textbf {\bibinfo {volume} {92}},\ \bibinfo
  {pages} {064020} (\bibinfo {year} {2015})},\ \Eprint
  {https://arxiv.org/abs/1507.04496} {arXiv:1507.04496 [gr-qc]} \BibitemShut
  {NoStop}%
\bibitem [{\citenamefont {Zhang}\ and\ \citenamefont
  {Li}(2014)}]{Zhang:2014ala}%
  \BibitemOpen
  \bibfield  {author} {\bibinfo {author} {\bibfnamefont {H.}~\bibnamefont
  {Zhang}}\ and\ \bibinfo {author} {\bibfnamefont {X.-Z.}\ \bibnamefont {Li}},\
  }\bibfield  {title} {\bibinfo {title} {{From thermodynamics to the solutions
  in gravity theory}},\ }\href {https://doi.org/10.1016/j.physletb.2014.09.010}
  {\bibfield  {journal} {\bibinfo  {journal} {Phys. Lett. B}\ }\textbf
  {\bibinfo {volume} {737}},\ \bibinfo {pages} {395} (\bibinfo {year}
  {2014})},\ \Eprint {https://arxiv.org/abs/1406.1553} {arXiv:1406.1553
  [gr-qc]} \BibitemShut {NoStop}%
\bibitem [{\citenamefont {Hu}\ \emph {et~al.}(2017)\citenamefont {Hu},
  \citenamefont {Wu},\ and\ \citenamefont {Zhang}}]{Hu:2016hpm}%
  \BibitemOpen
  \bibfield  {author} {\bibinfo {author} {\bibfnamefont {Y.-P.}\ \bibnamefont
  {Hu}}, \bibinfo {author} {\bibfnamefont {X.-M.}\ \bibnamefont {Wu}},\ and\
  \bibinfo {author} {\bibfnamefont {H.}~\bibnamefont {Zhang}},\ }\bibfield
  {title} {\bibinfo {title} {{Generalized Vaidya Solutions and Misner-Sharp
  mass for $n$-dimensional massive gravity}},\ }\href
  {https://doi.org/10.1103/PhysRevD.95.084002} {\bibfield  {journal} {\bibinfo
  {journal} {Phys. Rev. D}\ }\textbf {\bibinfo {volume} {95}},\ \bibinfo
  {pages} {084002} (\bibinfo {year} {2017})},\ \Eprint
  {https://arxiv.org/abs/1611.09042} {arXiv:1611.09042 [gr-qc]} \BibitemShut
  {NoStop}%
\bibitem [{\citenamefont {Brill}\ \emph {et~al.}(1997)\citenamefont {Brill},
  \citenamefont {Louko},\ and\ \citenamefont {Peldan}}]{Brill:1997mf}%
  \BibitemOpen
  \bibfield  {author} {\bibinfo {author} {\bibfnamefont {D.~R.}\ \bibnamefont
  {Brill}}, \bibinfo {author} {\bibfnamefont {J.}~\bibnamefont {Louko}},\ and\
  \bibinfo {author} {\bibfnamefont {P.}~\bibnamefont {Peldan}},\ }\bibfield
  {title} {\bibinfo {title} {{Thermodynamics of (3+1)-dimensional black holes
  with toroidal or higher genus horizons}},\ }\href
  {https://doi.org/10.1103/PhysRevD.56.3600} {\bibfield  {journal} {\bibinfo
  {journal} {Phys. Rev. D}\ }\textbf {\bibinfo {volume} {56}},\ \bibinfo
  {pages} {3600} (\bibinfo {year} {1997})},\ \Eprint
  {https://arxiv.org/abs/gr-qc/9705012} {arXiv:gr-qc/9705012} \BibitemShut
  {NoStop}%
\bibitem [{\citenamefont {Song}\ \emph {et~al.}(2021)\citenamefont {Song},
  \citenamefont {Li}, \citenamefont {Ma},\ and\ \citenamefont
  {Zhang}}]{Song:2020arr}%
  \BibitemOpen
  \bibfield  {author} {\bibinfo {author} {\bibfnamefont {S.}~\bibnamefont
  {Song}}, \bibinfo {author} {\bibfnamefont {H.}~\bibnamefont {Li}}, \bibinfo
  {author} {\bibfnamefont {Y.}~\bibnamefont {Ma}},\ and\ \bibinfo {author}
  {\bibfnamefont {C.}~\bibnamefont {Zhang}},\ }\bibfield  {title} {\bibinfo
  {title} {{Entropy of black holes with arbitrary shapes in loop quantum
  gravity}},\ }\href {https://doi.org/10.1007/s11433-021-1770-3} {\bibfield
  {journal} {\bibinfo  {journal} {Sci. China Phys. Mech. Astron.}\ }\textbf
  {\bibinfo {volume} {64}},\ \bibinfo {pages} {120411} (\bibinfo {year}
  {2021})},\ \Eprint {https://arxiv.org/abs/2002.08869} {arXiv:2002.08869
  [gr-qc]} \BibitemShut {NoStop}%
\bibitem [{\citenamefont {Ren}(2022)}]{Ren:2019lgw}%
  \BibitemOpen
  \bibfield  {author} {\bibinfo {author} {\bibfnamefont {J.}~\bibnamefont
  {Ren}},\ }\bibfield  {title} {\bibinfo {title} {{Analytic solutions of
  neutral hyperbolic black holes with scalar hair}},\ }\href
  {https://doi.org/10.1103/PhysRevD.106.086023} {\bibfield  {journal} {\bibinfo
   {journal} {Phys. Rev. D}\ }\textbf {\bibinfo {volume} {106}},\ \bibinfo
  {pages} {086023} (\bibinfo {year} {2022})},\ \Eprint
  {https://arxiv.org/abs/1910.06344} {arXiv:1910.06344 [hep-th]} \BibitemShut
  {NoStop}%
\bibitem [{\citenamefont {Hawking}\ and\ \citenamefont
  {Ellis}(1973)}]{HawkingEllis}%
  \BibitemOpen
  \bibfield  {author} {\bibinfo {author} {\bibfnamefont {S.}~\bibnamefont
  {Hawking}}\ and\ \bibinfo {author} {\bibfnamefont {G.}~\bibnamefont
  {Ellis}},\ }\href@noop {} {\emph {\bibinfo {title} {{The large scale
  structure of space-time}}}}\ (\bibinfo {year} {1973})\BibitemShut {NoStop}%
\bibitem [{\citenamefont {Hayward}(1994)}]{Hayward:1993wb}%
  \BibitemOpen
  \bibfield  {author} {\bibinfo {author} {\bibfnamefont {S.~A.}\ \bibnamefont
  {Hayward}},\ }\bibfield  {title} {\bibinfo {title} {{General laws of black
  hole dynamics}},\ }\href {https://doi.org/10.1103/PhysRevD.49.6467}
  {\bibfield  {journal} {\bibinfo  {journal} {Phys. Rev. D}\ }\textbf {\bibinfo
  {volume} {49}},\ \bibinfo {pages} {6467} (\bibinfo {year}
  {1994})}\BibitemShut {NoStop}%
\bibitem [{\citenamefont {Mai}\ and\ \citenamefont {Yang}(2022)}]{Mai:2022pgf}%
  \BibitemOpen
  \bibfield  {author} {\bibinfo {author} {\bibfnamefont {Z.-F.}\ \bibnamefont
  {Mai}}\ and\ \bibinfo {author} {\bibfnamefont {R.-Q.}\ \bibnamefont {Yang}},\
  }\bibfield  {title} {\bibinfo {title} {{Rechargeable black hole battery}},\
  }\href@noop {} {\  (\bibinfo {year} {2022})},\ \Eprint
  {https://arxiv.org/abs/2210.10587} {arXiv:2210.10587 [gr-qc]} \BibitemShut
  {NoStop}%
\bibitem [{\citenamefont {Maeda}\ and\ \citenamefont
  {Martinez}(2018)}]{Maeda:2016ddh}%
  \BibitemOpen
  \bibfield  {author} {\bibinfo {author} {\bibfnamefont {H.}~\bibnamefont
  {Maeda}}\ and\ \bibinfo {author} {\bibfnamefont {C.}~\bibnamefont
  {Martinez}},\ }\bibfield  {title} {\bibinfo {title} {{All static and
  electrically charged solutions with Einstein base manifold in the
  arbitrary-dimensional Einstein-Maxwell system with a massless scalar
  field}},\ }\href {https://doi.org/10.1140/epjc/s10052-018-6334-7} {\bibfield
  {journal} {\bibinfo  {journal} {Eur. Phys. J. C}\ }\textbf {\bibinfo {volume}
  {78}},\ \bibinfo {pages} {860} (\bibinfo {year} {2018})},\ \Eprint
  {https://arxiv.org/abs/1603.03436} {arXiv:1603.03436 [gr-qc]} \BibitemShut
  {NoStop}%
\bibitem [{\citenamefont {Peng}(2021)}]{Peng:2021xwh}%
  \BibitemOpen
  \bibfield  {author} {\bibinfo {author} {\bibfnamefont {Y.}~\bibnamefont
  {Peng}},\ }\bibfield  {title} {\bibinfo {title} {{New topological
  Gauss-Bonnet black holes in five dimensions}},\ }\href
  {https://doi.org/10.1103/PhysRevD.104.084004} {\bibfield  {journal} {\bibinfo
   {journal} {Phys. Rev. D}\ }\textbf {\bibinfo {volume} {104}},\ \bibinfo
  {pages} {084004} (\bibinfo {year} {2021})},\ \Eprint
  {https://arxiv.org/abs/2105.08482} {arXiv:2105.08482 [gr-qc]} \BibitemShut
  {NoStop}%
\bibitem [{\citenamefont {Peng}(2022)}]{Peng:2022ttn}%
  \BibitemOpen
  \bibfield  {author} {\bibinfo {author} {\bibfnamefont {Y.}~\bibnamefont
  {Peng}},\ }\bibfield  {title} {\bibinfo {title} {{New Anisotropic
  Gauss-Bonnet Black Holes in Five Dimensions at the Critical Point}},\
  }\href@noop {} {\  (\bibinfo {year} {2022})},\ \Eprint
  {https://arxiv.org/abs/2205.03830} {arXiv:2205.03830 [gr-qc]} \BibitemShut
  {NoStop}%
\bibitem [{\citenamefont {Maeda}(2015)}]{Maeda:2015cia}%
  \BibitemOpen
  \bibfield  {author} {\bibinfo {author} {\bibfnamefont {H.}~\bibnamefont
  {Maeda}},\ }\bibfield  {title} {\bibinfo {title} {{The
  Roberts\textendash{}(A)dS spacetime}},\ }\href
  {https://doi.org/10.1088/0264-9381/32/13/135025} {\bibfield  {journal}
  {\bibinfo  {journal} {Class. Quant. Grav.}\ }\textbf {\bibinfo {volume}
  {32}},\ \bibinfo {pages} {135025} (\bibinfo {year} {2015})},\ \Eprint
  {https://arxiv.org/abs/1501.03524} {arXiv:1501.03524 [gr-qc]} \BibitemShut
  {NoStop}%
\bibitem [{\citenamefont {Huang}\ \emph {et~al.}(2019)\citenamefont {Huang},
  \citenamefont {Fan},\ and\ \citenamefont {L\"u}}]{Huang:2019lsl}%
  \BibitemOpen
  \bibfield  {author} {\bibinfo {author} {\bibfnamefont {H.}~\bibnamefont
  {Huang}}, \bibinfo {author} {\bibfnamefont {Z.-Y.}\ \bibnamefont {Fan}},\
  and\ \bibinfo {author} {\bibfnamefont {H.}~\bibnamefont {L\"u}},\ }\bibfield
  {title} {\bibinfo {title} {{Static and Dynamic Charged Black Holes}},\ }\href
  {https://doi.org/10.1140/epjc/s10052-019-7477-x} {\bibfield  {journal}
  {\bibinfo  {journal} {Eur. Phys. J. C}\ }\textbf {\bibinfo {volume} {79}},\
  \bibinfo {pages} {975} (\bibinfo {year} {2019})},\ \Eprint
  {https://arxiv.org/abs/1908.07970} {arXiv:1908.07970 [hep-th]} \BibitemShut
  {NoStop}%
\bibitem [{\citenamefont {Huang}\ \emph {et~al.}(2020)\citenamefont {Huang},
  \citenamefont {L\"u},\ and\ \citenamefont {Yang}}]{Huang:2020qmn}%
  \BibitemOpen
  \bibfield  {author} {\bibinfo {author} {\bibfnamefont {H.}~\bibnamefont
  {Huang}}, \bibinfo {author} {\bibfnamefont {H.}~\bibnamefont {L\"u}},\ and\
  \bibinfo {author} {\bibfnamefont {J.}~\bibnamefont {Yang}},\ }\bibfield
  {title} {\bibinfo {title} {{Bronnikov-like Wormholes in Einstein-Scalar
  Gravity}},\ }\href@noop {} {\  (\bibinfo {year} {2020})},\ \Eprint
  {https://arxiv.org/abs/2010.00197} {arXiv:2010.00197 [gr-qc]} \BibitemShut
  {NoStop}%
\bibitem [{\citenamefont {Gundlach}\ \emph {et~al.}(2021)\citenamefont
  {Gundlach}, \citenamefont {Bourg},\ and\ \citenamefont
  {Davey}}]{Gundlach:2021six}%
  \BibitemOpen
  \bibfield  {author} {\bibinfo {author} {\bibfnamefont {C.}~\bibnamefont
  {Gundlach}}, \bibinfo {author} {\bibfnamefont {P.}~\bibnamefont {Bourg}},\
  and\ \bibinfo {author} {\bibfnamefont {A.}~\bibnamefont {Davey}},\ }\bibfield
   {title} {\bibinfo {title} {{Fully constrained, high-resolution
  shock-capturing, formulation of the Einstein-fluid equations in 2+1
  dimensions}},\ }\href {https://doi.org/10.1103/PhysRevD.104.024061}
  {\bibfield  {journal} {\bibinfo  {journal} {Phys. Rev. D}\ }\textbf {\bibinfo
  {volume} {104}},\ \bibinfo {pages} {024061} (\bibinfo {year} {2021})},\
  \Eprint {https://arxiv.org/abs/2103.04435} {arXiv:2103.04435 [gr-qc]}
  \BibitemShut {NoStop}%
\bibitem [{\citenamefont {Brown}\ and\ \citenamefont
  {York}(1993)}]{Brown:1992br}%
  \BibitemOpen
  \bibfield  {author} {\bibinfo {author} {\bibfnamefont {J.~D.}\ \bibnamefont
  {Brown}}\ and\ \bibinfo {author} {\bibfnamefont {J.~W.}\ \bibnamefont {York},
  \bibfnamefont {Jr.}},\ }\bibfield  {title} {\bibinfo {title} {{Quasilocal
  energy and conserved charges derived from the gravitational action}},\ }\href
  {https://doi.org/10.1103/PhysRevD.47.1407} {\bibfield  {journal} {\bibinfo
  {journal} {Phys. Rev. D}\ }\textbf {\bibinfo {volume} {47}},\ \bibinfo
  {pages} {1407} (\bibinfo {year} {1993})},\ \Eprint
  {https://arxiv.org/abs/gr-qc/9209012} {arXiv:gr-qc/9209012} \BibitemShut
  {NoStop}%
\bibitem [{\citenamefont {Caldarelli}\ \emph {et~al.}(2000)\citenamefont
  {Caldarelli}, \citenamefont {Cognola},\ and\ \citenamefont
  {Klemm}}]{Caldarelli:1999xj}%
  \BibitemOpen
  \bibfield  {author} {\bibinfo {author} {\bibfnamefont {M.~M.}\ \bibnamefont
  {Caldarelli}}, \bibinfo {author} {\bibfnamefont {G.}~\bibnamefont
  {Cognola}},\ and\ \bibinfo {author} {\bibfnamefont {D.}~\bibnamefont
  {Klemm}},\ }\bibfield  {title} {\bibinfo {title} {{Thermodynamics of
  Kerr-Newman-AdS black holes and conformal field theories}},\ }\href
  {https://doi.org/10.1088/0264-9381/17/2/310} {\bibfield  {journal} {\bibinfo
  {journal} {Class. Quant. Grav.}\ }\textbf {\bibinfo {volume} {17}},\ \bibinfo
  {pages} {399} (\bibinfo {year} {2000})},\ \Eprint
  {https://arxiv.org/abs/hep-th/9908022} {arXiv:hep-th/9908022} \BibitemShut
  {NoStop}%
\bibitem [{\citenamefont {Chakraborty}\ and\ \citenamefont
  {Padmanabhan}(2015)}]{Chakraborty:2015hna}%
  \BibitemOpen
  \bibfield  {author} {\bibinfo {author} {\bibfnamefont {S.}~\bibnamefont
  {Chakraborty}}\ and\ \bibinfo {author} {\bibfnamefont {T.}~\bibnamefont
  {Padmanabhan}},\ }\bibfield  {title} {\bibinfo {title} {{Thermodynamical
  interpretation of the geometrical variables associated with null surfaces}},\
  }\href {https://doi.org/10.1103/PhysRevD.92.104011} {\bibfield  {journal}
  {\bibinfo  {journal} {Phys. Rev. D}\ }\textbf {\bibinfo {volume} {92}},\
  \bibinfo {pages} {104011} (\bibinfo {year} {2015})},\ \Eprint
  {https://arxiv.org/abs/1508.04060} {arXiv:1508.04060 [gr-qc]} \BibitemShut
  {NoStop}%
\bibitem [{\citenamefont {Chakraborty}\ \emph {et~al.}(2015)\citenamefont
  {Chakraborty}, \citenamefont {Parattu},\ and\ \citenamefont
  {Padmanabhan}}]{Chakraborty:2015aja}%
  \BibitemOpen
  \bibfield  {author} {\bibinfo {author} {\bibfnamefont {S.}~\bibnamefont
  {Chakraborty}}, \bibinfo {author} {\bibfnamefont {K.}~\bibnamefont
  {Parattu}},\ and\ \bibinfo {author} {\bibfnamefont {T.}~\bibnamefont
  {Padmanabhan}},\ }\bibfield  {title} {\bibinfo {title} {{Gravitational field
  equations near an arbitrary null surface expressed as a thermodynamic
  identity}},\ }\href {https://doi.org/10.1007/JHEP10(2015)097} {\bibfield
  {journal} {\bibinfo  {journal} {JHEP}\ }\textbf {\bibinfo {volume} {10}},\
  \bibinfo {pages} {097}},\ \Eprint {https://arxiv.org/abs/1505.05297}
  {arXiv:1505.05297 [gr-qc]} \BibitemShut {NoStop}%
\bibitem [{\citenamefont {Mahapatra}\ \emph {et~al.}(2020)\citenamefont
  {Mahapatra}, \citenamefont {Priyadarshinee}, \citenamefont {Reddy},\ and\
  \citenamefont {Shukla}}]{Mahapatra:2020wym}%
  \BibitemOpen
  \bibfield  {author} {\bibinfo {author} {\bibfnamefont {S.}~\bibnamefont
  {Mahapatra}}, \bibinfo {author} {\bibfnamefont {S.}~\bibnamefont
  {Priyadarshinee}}, \bibinfo {author} {\bibfnamefont {G.~N.}\ \bibnamefont
  {Reddy}},\ and\ \bibinfo {author} {\bibfnamefont {B.}~\bibnamefont
  {Shukla}},\ }\bibfield  {title} {\bibinfo {title} {{Exact topological charged
  hairy black holes in AdS Space in $D$-dimensions}},\ }\href
  {https://doi.org/10.1103/PhysRevD.102.024042} {\bibfield  {journal} {\bibinfo
   {journal} {Phys. Rev. D}\ }\textbf {\bibinfo {volume} {102}},\ \bibinfo
  {pages} {024042} (\bibinfo {year} {2020})},\ \Eprint
  {https://arxiv.org/abs/2004.00921} {arXiv:2004.00921 [hep-th]} \BibitemShut
  {NoStop}%
\bibitem [{\citenamefont {Priyadarshinee}\ \emph {et~al.}(2021)\citenamefont
  {Priyadarshinee}, \citenamefont {Mahapatra},\ and\ \citenamefont
  {Banerjee}}]{Priyadarshinee:2021rch}%
  \BibitemOpen
  \bibfield  {author} {\bibinfo {author} {\bibfnamefont {S.}~\bibnamefont
  {Priyadarshinee}}, \bibinfo {author} {\bibfnamefont {S.}~\bibnamefont
  {Mahapatra}},\ and\ \bibinfo {author} {\bibfnamefont {I.}~\bibnamefont
  {Banerjee}},\ }\bibfield  {title} {\bibinfo {title} {{Analytic topological
  hairy dyonic black holes and thermodynamics}},\ }\href
  {https://doi.org/10.1103/PhysRevD.104.084023} {\bibfield  {journal} {\bibinfo
   {journal} {Phys. Rev. D}\ }\textbf {\bibinfo {volume} {104}},\ \bibinfo
  {pages} {084023} (\bibinfo {year} {2021})},\ \Eprint
  {https://arxiv.org/abs/2108.02514} {arXiv:2108.02514 [hep-th]} \BibitemShut
  {NoStop}%
\bibitem [{\citenamefont {Guica}\ \emph {et~al.}(2009)\citenamefont {Guica},
  \citenamefont {Hartman}, \citenamefont {Song},\ and\ \citenamefont
  {Strominger}}]{Guica:2008mu}%
  \BibitemOpen
  \bibfield  {author} {\bibinfo {author} {\bibfnamefont {M.}~\bibnamefont
  {Guica}}, \bibinfo {author} {\bibfnamefont {T.}~\bibnamefont {Hartman}},
  \bibinfo {author} {\bibfnamefont {W.}~\bibnamefont {Song}},\ and\ \bibinfo
  {author} {\bibfnamefont {A.}~\bibnamefont {Strominger}},\ }\bibfield  {title}
  {\bibinfo {title} {{The Kerr/CFT Correspondence}},\ }\href
  {https://doi.org/10.1103/PhysRevD.80.124008} {\bibfield  {journal} {\bibinfo
  {journal} {Phys. Rev. D}\ }\textbf {\bibinfo {volume} {80}},\ \bibinfo
  {pages} {124008} (\bibinfo {year} {2009})},\ \Eprint
  {https://arxiv.org/abs/0809.4266} {arXiv:0809.4266 [hep-th]} \BibitemShut
  {NoStop}%
\bibitem [{\citenamefont {Wald}(1984)}]{WaldGR}%
  \BibitemOpen
  \bibfield  {author} {\bibinfo {author} {\bibfnamefont {R.}~\bibnamefont
  {Wald}},\ }\href@noop {} {\emph {\bibinfo {title} {{General relativity}}}}\
  (\bibinfo {year} {1984})\BibitemShut {NoStop}%
\bibitem [{\citenamefont {Liang}\ and\ \citenamefont {Zhou}(2023)}]{LiangGR}%
  \BibitemOpen
  \bibfield  {author} {\bibinfo {author} {\bibfnamefont {C.-B.}\ \bibnamefont
  {Liang}}\ and\ \bibinfo {author} {\bibfnamefont {B.}~\bibnamefont {Zhou}},\
  }\href@noop {} {\emph {\bibinfo {title} {{Differential geometry and general
  relativity}}}}\ (\bibinfo {year} {2023})\BibitemShut {NoStop}%
\end{thebibliography}%

\end{document}